%% file: slr.tex
\PassOptionsToPackage{table}{xcolor}
\documentclass[format=acmsmall, review=false, screen=true, anonymous=false]{acmart}

\usepackage{booktabs} 
\usepackage[ruled]{algorithm2e} 
\usepackage{natbib}
\usepackage{enumitem}
\usepackage{tabularx}
\usepackage{siunitx}
\usepackage{multirow}
\usepackage{subcaption}
\usepackage[justification=centering]{caption}
\usepackage{titlesec}
\usepackage{textcomp}

\definecolor{lightblue}{HTML}{DEEAF6}
\newcolumntype{Y}{>{\centering\arraybackslash}X}

\acmJournal{TOCHI}
\acmVolume{TBD}
\acmNumber{TBD}
\acmArticle{1}
\acmYear{2022}
\acmMonth{6}
\acmPrice{15.00}
\acmDOI{10.1145/3534929}

\setcopyright{acmlicensed}

\begin{document}
\title{A Trade-off-centered Framework of Content Moderation}

\author{Jialun Aaron Jiang}
\affiliation{
  \institution{University of Colorado Boulder}
  \department{Department of Information Science}
  \streetaddress{1045 18th St.}
  \city{Boulder}
  \state{CO}
  \postcode{80309}
  \country{USA}}
\email{aaron.jiang@colorado.edu}

\author{Peipei Nie}
\affiliation{
  \institution{University of Washington}
  \department{Paul G. Allen School of Computer Science and Engineering}
  \streetaddress{185 E Stevens Way NE}
  \city{Seattle}
  \state{WA}
  \postcode{98195}
  \country{USA}}
\email{niep@cs.washington.edu}

\author{Jed R. Brubaker}
\affiliation{
  \institution{University of Colorado Boulder}
  \department{Department of Information Science}
  \streetaddress{1045 18th St.}
  \city{Boulder}
  \state{CO}
  \postcode{80309}
  \country{USA}}
\email{jed.brubaker@colorado.edu}

\author{Casey Fiesler}
\affiliation{
  \institution{University of Colorado Boulder}
  \department{Department of Information Science}
  \streetaddress{1045 18th St.}
  \city{Boulder}
  \state{CO}
  \postcode{80309}
  \country{USA}}
\email{casey.fiesler@colorado.edu}

\renewcommand{\shortauthors}{J. A. Jiang et al.}

\begin{abstract}
Content moderation research typically prioritizes representing and addressing challenges for one group of stakeholders or communities in one type of context. While taking a focused approach is reasonable or even favorable for empirical case studies, it does not address how content moderation works in multiple contexts. Through a systematic literature review of 86 content moderation papers that document empirical studies, we seek to uncover patterns and tensions within past content moderation research. We find that content moderation can be characterized as a series of trade-offs around moderation actions, styles, philosophies, and values. We discuss how facilitating cooperation and preventing abuse, two key elements in Grimmelmann’s definition of moderation, are inherently dialectical in practice. We close by showing how researchers, designers, and moderators can use our framework of trade-offs in their own work, and arguing that trade-offs should be of central importance in investigating and designing content moderation.
\end{abstract}

%
%


%
%



\maketitle

\input{slr-body-rev}

\end{document}

%% file: slr-body-rev.tex
\section{Introduction}

In recent years, HCI and social computing research has shown an increasing interest in content moderation, investigating its critical role in online community building and engagement, social media user safety, and broader social issues. While the philosophical exploration of handling contentious matters in online communities dates back to as early as 1978 when Gengle~\cite{gengle_communitree_1981} conceptualized ``Fairwitness'' who were impartial citizens serving as guides of online conversations, modern scholarship of content moderation has widely accepted Grimmelmann's definition: a mechanism to facilitate cooperation and prevent abuse \cite{grimmelmann_virtues_2015}. The topic of content moderation has been attracting an abundance of research ever since usually focusing on how different types of moderation can help various communities. Existing content moderation studies typically prioritize representing and addressing challenges for one type of people or communities in one particular type of context. While such a focused approach is reasonable and sometimes favorable for individual studies, what remains hidden are the broader insights into how content moderation works.

A benefit of taking a holistic view of moderation research is that it will uncover how moderation would work differently in different contexts, and existing research has already shown evidence that such difference exists: For example, automated moderation can work consistently and quickly in a way that is required by large-scale moderation, but it lacks nuanced understandings needed by individual cases that often fall into the gray area of policies and rules \cite{jhaver_human-machine_2019}. Moderators in text-based online spaces rely heavily on removing and editing content, but the same methods completely break down in communities where voice chat or virtual reality is the dominant mode of interaction \cite{jiang_moderation_2019, blackwell_harassment_2019}. These two examples, and many more like them, demonstrate the complexity and difficulty of moderation in practice when it needs to satisfy multiple needs risen from multiple contexts. An examination of the broader picture of moderation will reveal patterns and distinctions in moderation's success and challenges, as well as the reasons that contributed to them. The insights will also help researchers, designers, moderators, and regular internet users to reflect on moderation by considering factors that they may not have considered before.

Through a systematic literature review, we present a trade-off-centered framework of moderation that synthesizes findings of individual papers into four major, interrelated trade-offs at increasing levels of abstraction. Our framework also further highlights the subtle and evasive trade-offs in moderation philosophies and values, and surfaces the dialectical tension between facilitating cooperation and preventing abuse, the two elements in Grimmelmann's \cite{grimmelmann_virtues_2015} definition of moderation. Finally, we show how researchers, designers, and moderators can be benefited from our framework, and argue that trade-offs should be of central importance in investigating and designing content moderation.

\section{Method: Systematic Literature Review}
To understand patterns and trends in existing literature about content moderation, we conducted a systematic literature review, following best practices established in different fields \cite{liberati_prisma_2009}, as well as rigorous review studies in the HCI and CSCW literature \cite{chancellor_who_2019, disalvo_mapping_2010, seering_applications_2018}. This section will describe our search strategy to identify candidate papers, inclusion criteria to filter the candidate papers into a corpus for analysis, and analysis techniques.

\subsection{Search Strategy}

Prior work in content moderation shows that there is not one field that completely covers all content moderation research. A published reading list of content moderation \cite{gillespie_content_2019} shows this line of research primarily happens in social computing, human-computer interaction, computational social science, and communication, spanning a wide range of ACM and AAAI conferences (e.g., ACM CHI, ACM CSCW, AAAI ICWSM) and journals (e.g., Social Media + Society, New Media \& Society, International Journal of Communication). 

Therefore, to ensure a robust coverage across venues, we used a combination of search databases. Following the method used by Chancellor et al. who did a similarly interdisciplinary meta-review \cite{chancellor_who_2019}, we used the ACM Digital Library to search ACM journals and conferences, the AAAI Digital Library (implemented with Google custom search) to search AAAI publications, and Web of Science for other journal publications. 

Using a keyword search within the above databases, we identified an initial set of candidate papers published between 2004\footnote{While we did not arbitrarily limit the starting point in our initial search, our search keywords only yielded results dating back to 2004 across several databases.}, the earliest time that we could find content moderation research, and the time our search concluded (\textcolor{black}{October 2020}). Based on keywords used in all published papers about content moderation in CSCW 2018 and 2019, two conferences with a relatively high amount of empirical content moderation work, as well as keywords that they have used to describe content moderation, supplemented with our domain knowledge, the final list of keywords included:

\begin{quote}
    content moderation, platform moderation, community moderation, platform governance, community governance, internet governance. 
\end{quote}

In order to check the validity of these keywords, we manually went through every paper (regardless of whether it related to content moderation) published in one conference (ICWSM 2019) and one journal (Social Media + Society papers published in 2019)---which constituted a total of 158 published papers---and created a subset of papers about moderation. We selected these two venues because they contained a reasonably small total number of papers for us to go through while not sacrificing the quality of our validity check. We then performed a keyword search of that conference and journal, and ensured that the keyword search did not result in any false negatives. False positives were retained, since they could be filtered out by our inclusion criteria (described below). Our search strategy finally yielded 1,074 papers in total (309 from the ACM Digital Library, 35 from the AAAI Library, and 730 from Web of Science).

\subsection{Inclusion Criteria}
Each paper identified with our keyword search needs to meet the following criteria to be included in the corpus:

\begin{enumerate}
    \item \textbf{Archival \& peer-reviewed:} A paper needs to be archival and peer-reviewed for inclusion, because these papers have been scrutinized by experts to ensure their validity and rigorousness and thus meet the publication criteria of the chosen venues. We did not include non-archival papers such as late-breaking work or workshop papers because they often include work that is incomplete and ongoing.
    
    \item \textbf{Empirical study:} The paper needs to describe at least one empirical study to be included for analysis. An ``empirical study'' here means a study that collects data from people. This definition means:
        \begin{enumerate}
            \item Since we focused on real-world moderation practices and challenges validated by real moderators, we did not include essays or papers that are purely theoretical analysis. However, studies that use empirical evidence to validate social science theories would meet this criterion.
            
            \item Papers that describe systems would only qualify if they also describe user studies, which includes formative studies before building the system, and evaluative studies of how people use the system.
            
            \item We also did not include papers that only summarize or evaluate existing studies, because the qualifying studies that these papers build on would already be included in the keyword search. 
        \end{enumerate}
        
    \item \textbf{Moderation practices, challenges, impacts, or recommendations:} For this study we only focus on these four facets of moderation. Therefore, for a paper to be included, it needs to document at least one of the following:
        \begin{enumerate}
            \item Existing moderation practices or approaches;
            \item Existing moderation challenges or problems;
            \item Impacts and consequences of existing moderation practices;
            \item Recommendations, implications, or future directions for designing and implementing content moderation.
        \end{enumerate}

\end{enumerate}

Since these details may not be included in the paper titles or keywords, we also read the abstract of each paper. 
After manually filtering and deduplicating using the inclusion criteria above, \textcolor{black}{we retained 74 papers (35 from the ACM Digital Library, 7 from the AAAI Library, and 32 from Web of Science). A number of papers did not pass the inclusion criteria because they focused on the governance of \emph{internet infrastructure}, such as DNS servers and IP addresses, a result of our search keyword ``internet governance.'' Additionally, we noted a lot of papers that provided valuable insights about moderation but did not pass criteria (2) or (3), such as legal scholarships, technical reports, thought-pieces and essays, and papers that described systems but did not have any component listed in (3).} We further added 12 papers from reading the bibliographies in these papers, resulting in a total of 86 papers in our corpus. 

\subsection{Characterizing the Corpus}

\begin{figure}
    \centering
    \includegraphics[width=\textwidth]{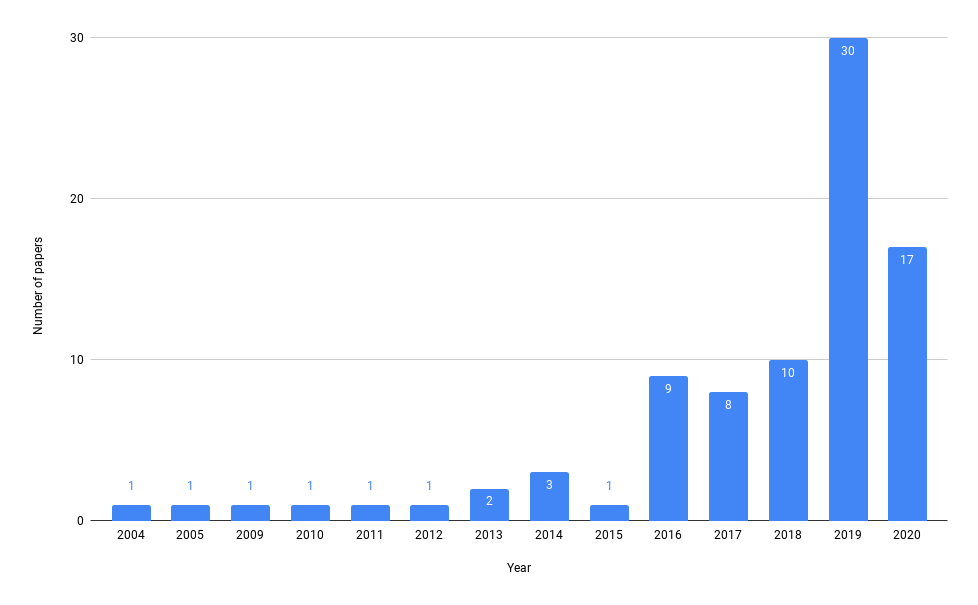}
    \caption{Number of papers in our corpus by year.}
    \label{fig:barchart}
\end{figure}

Within our corpus, Reddit was the most commonly studied platform and was the focus of 29 papers, far exceeding other platforms. Other platforms with multiple occurrences included: Facebook (including Facebook Groups) (8), Instagram (5), YouTube (5), Twitter (4), Twitch (4), Slashdot (3), Discord (2), Tumblr (2), Wikipedia (2). Platforms that only appeared in one paper include: Flickr, Tinder, Vine, Pinterest, Yelp, Yik Yak, Mayo Clinic Connect, Weibo, CNN.com, New York Times, StreamPlus.com., FutureLearn, iFixIt, League of Legends Tribunal, Whisper, and Everything 2. The interview studies and survey studies in our corpus had an average of 21.7 and 503.8 participants respectively. Appendix \ref{sec:appendix} shows a complete list of papers in our corpus, and Fig. \ref{fig:barchart} shows the number of papers in our corpus by year. As Fig. \ref{fig:barchart} shows, there is a significant uptick in the number of papers from 2016, suggesting that content moderation is a nascent field of study. 

Given that content moderation research frequently addresses negative or harmful experiences, we paid particular attention to research ethics in our corpus. 11 out of the 86 papers in our corpus had explicit discussions of research ethics considerations, beyond statements that the study was approved by an ethics review board. This relatively rare occurrence of explicit research ethics discussions is in line with the findings from a recent systematic literature review of Reddit-based research studies~\cite{proferes_studying_2021}. Papers that did discuss research ethics focused on the ethics of studying sensitive issues such as eating disorder and handling removed data, and assigning participants in experiment conditions that might incur additional harm. 

\subsection{Analysis Techniques}
We conducted a thematic analysis of the papers in our corpus, \textcolor{black}{following the six phases outlined by Braun and Clarke \cite{braun_using_2006}}. The first two authors first engaged in one round of open coding by closely reading \textcolor{black}{and coding} a sample of the corpus. Specifically, we each sampled one paper per year for every year with any publication, with an eye toward a breadth of research paradigms (e.g., qualitative and/or quantitative) and topics (e.g., volunteer moderation and/or commercial moderation), and open coded these papers. During this round of open coding, we regularly came together to discuss emergent code groups such as ``removing content'' and ``automated moderation bots.'' Then, we open coded the rest of the corpus with an eye toward the preliminary code groups identified in the sample. Two more rounds of iterative coding led us to identifying higher-level categories such as ``moderation actions'' and ``rules and norms.'' The first author used these categories to produce a set of descriptive theme memos that described each category grounded in the quotes from the papers. All authors then discussed the theme memo and developed the relationships between the categories. 

While we observed recurring themes such as ``moderation transparency'' or ''educational approaches,'' none of these themes applied across all contexts. Furthermore, themes almost always existed in opposite pairs---for example, there were papers arguing for both providing explanations in moderation and not doing so. This pattern made us realize that the recurring theme \emph{across} contexts was the competing choices---or \emph{trade-offs}---in content moderation, and the resulting trade-off-centered framework we discuss below represents one way of thinking about and analyzing content moderation practices across a variety of domains. \textcolor{black}{Further, while we did not analyze the papers with Grimmelmann's categorization of moderation \cite{grimmelmann_virtues_2015} in mind or seek to expand upon it, we found that the more concrete trade-offs around moderation actions and styles broadly aligned with his categorization. However, the same did not apply for the more abstract trade-offs that we identified. Therefore, in describing the trade-offs, we only discuss Grimmelmann's categorization where it applies well in order to connect the trade-offs to this key piece of literature. }

\subsection{Limitations}
We would like to note three major limitations in our analysis. First, our analysis does not differentiate between platform-wide commercial moderation and volunteer moderation in smaller communities, because this distinction was not salient in our corpus. We suspect that the absence of such distinction was due to an overall lack of empirical research of commercial moderation practices. However, we do note that there is research related to user \emph{perceptions} of commercial moderation, which is included in our corpus. As such, the framework we present is also likely more representative of volunteer moderation.

\textcolor{black}{Second, given the focus on empirical research in our systematic literature review, we acknowledge that our framework is inadequate in addressing numerous other perspectives in content moderation, such as legal perspectives and critical perspectives. These alternative perspectives have provided valuable insights into underrepresented topics in our analysis, such as commercial moderation and algorithmic approaches to moderation (e.g., \cite{gorwa_algorithmic_2020}). Similarly, while the papers included in our analysis only date back to 2004 due to the empirical research criteria, there have been discussions about content moderation that predate our corpus (e.g., \cite{reid_communities_1999, herring_searching_2002, kollock_managing_1996, noauthor_problems_2002}). While beyond the scope of this paper, we encourage others to examine and expand on our framework through diverse perspectives. }

Last but not least, the fact that content moderation is a nascent field of study inherently bears limitations of the work presented in this paper. We acknowledge that our corpus is biased toward one platform (Reddit), and while we tried to use diverse examples, references to papers are disproportionately skewed toward several descriptive papers that provide summaries of a broad range of practices. Therefore, the framework presented by this paper is by no means a perfect one, but we hope this paper will nonetheless contribute to the literature by providing a new perspective for examining content moderation.

\begin{figure}[htbp!]
    \centering
    \includegraphics[width=\textwidth]{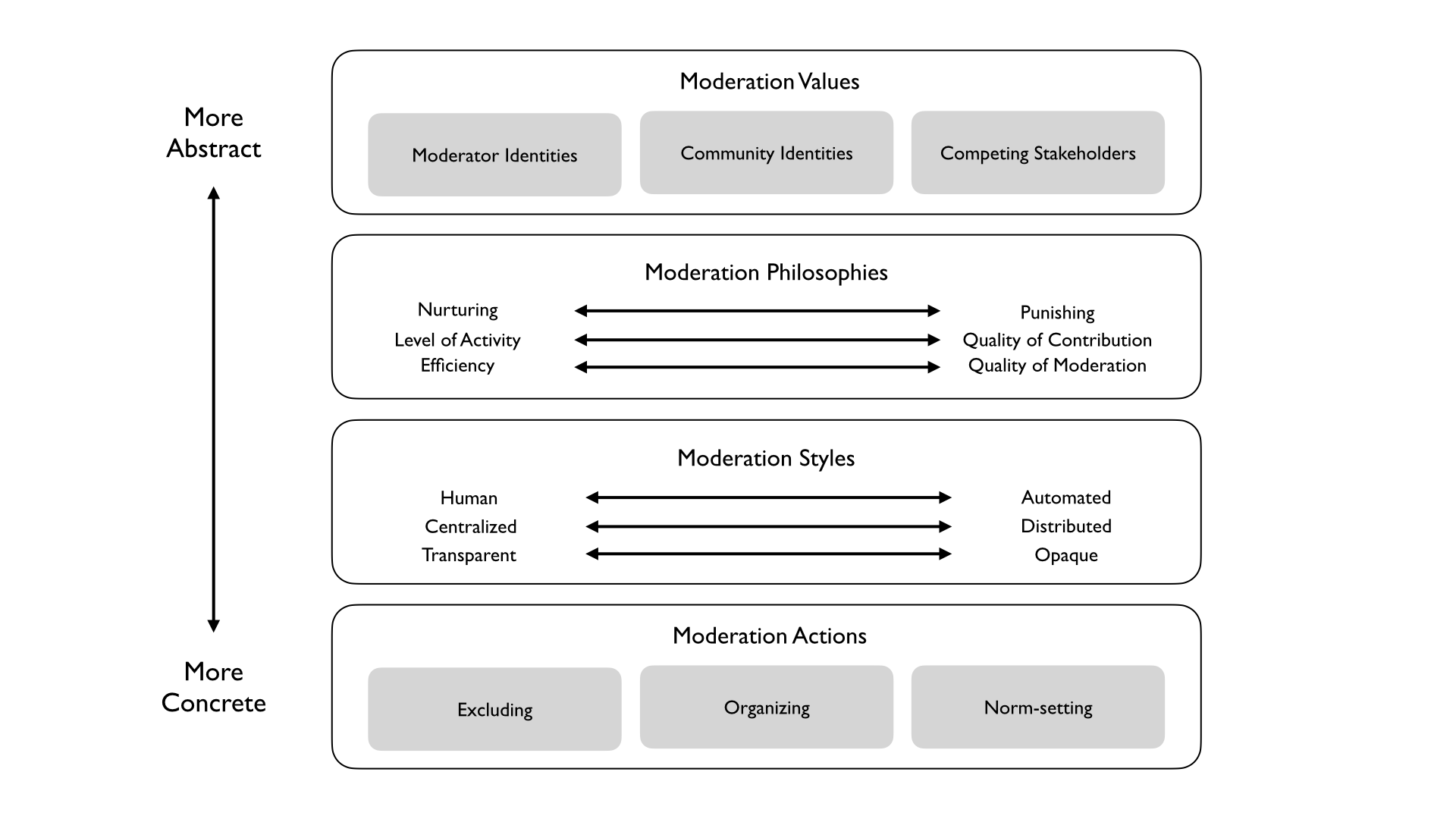}
    \caption{Diagram of our trade-off-centered framework of content moderation. The level of abstraction increases from moderation actions to moderation values. Note that elements and arrows within a single layer do not vary in the levels of abstraction.}
    \label{fig:framework}
\end{figure}

\section{A Trade-off-centered Framework of Content Moderation}

Our trade-off-centered framework of content moderation consists of four interrelated layers of trade-offs in increasing levels of abstraction: Moderation actions, moderation styles, moderation philosophies, and moderation values. Fig. \ref{fig:framework} shows a diagram that visualizes this framework. Trade-offs in the more abstract layers impact those in the more concrete layers. We envision our framework to be an analytical tool that helps people examine and make sense of content moderation practices, rather than a mental model that prescribes moderators' thought processes.

The moderation actions layer represents multiple concrete moderation techniques that moderators can use to manage their communities, such as issuing warnings, removing content, and banning people. \textcolor{black}{While we did not analyze the moderation actions with Grimmelmann's definition of moderation \cite{grimmelmann_virtues_2015} in mind, we found that these actions broadly aligned with Grimmelmann's categorization of moderation ``verbs''---excluding, organizing, norm-setting, and pricing---and described them as such.} These actions \textcolor{black}{have} varying levels of harshness, and reveal trade-offs between stifling the community and exposing the community to harm, as well as in forgoing the opportunity to educate community members about acceptable behavior by immediately removing violating behavior.

The moderation styles layer goes up one level of abstraction by addressing how moderators can carry out the actions in the moderation actions layer, \textcolor{black}{which we found to be similar to some of the ``adverbs''} in Grimmelmann's moderation framework. The styles layer consists of three specific trade-offs representing competing choices in the ways in which any of the moderation actions could be taken: Human vs. automated, centralized vs. distributed, and transparent vs. opaque. 

The moderation philosophies layer is one level more abstract than the moderation styles layer, and describes the philosophies that guide tendencies toward specific choices in moderation styles and moderation actions. The moderation philosophies layer consists of three trade-offs representing competing needs in content moderation: Nurturing vs. punishing, level of activity vs. quality of contribution, and efficiency vs. quality of moderation.

The moderation values layer is the topmost layer, representing the competing values that impact decisions in the trade-offs in moderation philosophies, styles, and actions. We broadly classify the values into three categories: Moderator identities, community identities, and competing stakeholders.

Our analysis, which we describe later in the following sections, revealed increasing levels of abstraction in our framework: While prior literature often describes moderation actions and styles as concrete findings (to the extent that there are existing categorizations such as Grimmelmann's \cite{grimmelmann_virtues_2015}), moderation philosophies and values are more evasive, which are often discussed only as speculations, if discussed at all. For the same reason, while we could find individual papers detailing the trade-offs in moderation actions and styles, our discussion of the trade-offs in moderation philosophies and values required a synthesis of multiple papers.

How prior literature discussed the trade-offs also influenced our organization within the layers. The trade-offs in moderation styles and philosophies existed in clear, opposing binaries in our corpus, so we used arrows to represent them. However, trade-offs in moderation actions and values involved multiple possible options, and as a result, here we directly listed the categories of options instead. While we vertically order the layers to represent different levels of abstraction, it does not apply to the trade-offs within the moderation styles and philosophies layers; these trade-offs are equal and do not vary in the levels of abstraction.

It is important to note that none of the options in any of the four layers are mutually exclusive. Real content moderation practices are almost always a mixture of different options, with different actions, styles, philosophies, and values existing at the same time. Therefore, the arrows in moderation styles and philosophies are ``slider scales'' where the decision could fall anywhere in the middle, instead of at one extreme or the other. Similarly, choices in moderation actions and values are also not mutually exclusive. Our notion of a trade-off is not a one-vs-all choice, but a balance to achieve among many legitimate alternatives. 

In the remainder of this paper, we explain each of the four layers of trade-offs in detail, and close by discussing how different people can use our framework in their own work.

\section{Trade-offs in Moderation Actions}
\label{sec:actions}

The first trade-off that we identified was around the moderation actions against rule-breaking behaviors, similar to the techniques, or ``verbs,'' in Grimmelmann's \cite{grimmelmann_virtues_2015} framework of content moderation. We found that moderators took different actions (for example, removing content or issuing warnings) to enforce content moderation. These actions had various levels of harshness, associated with different, sometimes competing outcomes and consequences. 

Grimmelmann \cite{grimmelmann_virtues_2015} categorized techniques into four broad categories: Excluding, organizing, norm-setting, and pricing. In our corpus, we also found more granular moderation actions that correspond to these \textcolor{black}{four} categories. 

One of the common actions was \textbf{exclusion}, which means to deprive certain people of access to an online community, and often takes the form of banning and the less harsh version of it---timeouts (i.e., ban from participation for a certain period of time). \textcolor{black}{55 out of 86} papers in our corpus mentioned some type of exclusion. Sometimes whole communities may be excluded by platforms, such as the ban of several subreddits in 2015 due to their violation of Reddit's anti-harassment policy \cite{chandrasekharan_you_2017}. In communities with voice chatting functionalities, moderators also practiced muting, which excludes people from participating in voice chats but not necessarily text chats \cite{jiang_moderation_2019}. The widespread use of exclusion was captured by Seering et al.~\cite{seering_moderator_2019}:

\begin{quote}
    \textcolor{black}{Nearly all moderators mentioned using timeouts, bans, or equivalents, though eagerness to use them varied. Communities with more laissez-faire ideologies used these only for egregious offenses, while communities intended to be safe spaces were usually quicker to use them. \cite[p. 14]{seering_moderator_2019}}
\end{quote}

\textbf{Organizing}, appearing in \textcolor{black}{68} papers, was the most common type of action that focuses on content rather than people. It ``shapes the flow of content from authors to readers'' \cite{grimmelmann_virtues_2015}, which, in our corpus, includes removing and annotating content. While removal often intends to solely get rid of content that violates the community rules, annotating can serve a multitude of purposes. For example, post annotations in Reddit, called ``flairs,'' are used as labels that categorize posts, whereas annotations in Wikipedia such as the Neutral Point of View (NPOV) tag are meant to notify readers that an article may be violating certain Wikipedia guidelines. In the case where the organized content is violating, prior research indicates differences between removing and annotating in the efficacy of helping community members adhere to norms. In a study of r/ChangeMyView, Srinivasan et al. \cite{srinivasan_content_2019} showed the causal effect that post removals indeed improve norm adherence. In contrast, Pavalanathan et al. \cite{pavalanathan_mind_2018} found that NPOV tags in Wikipedia did not help the editors to adopt the desired writing style, but did improve the overall quality of tagged articles, likely because of the contribution of other editors who edited upon the NPOV tags.

In addition to direct sanctions taken on people or content, moderators also widely used warnings (mentioned by \textcolor{black}{27} papers), which are less harsh, and fall into the premise of \textbf{norm-setting} by denouncing bad behavior \cite{grimmelmann_virtues_2015}. Moderators issued warnings to tell rule violators that they did something wrong, and sometimes also did so publicly to educate the community more broadly. Seering et al. \cite{seering_moderator_2019} also noted that warnings ranged from light to strong, the latter often accompanied by temporary sanctions mentioned above. Skousen et al.  \cite{skousen_successful_2020} in their study of an online health community also documented ``indirect policing'' practices similar to warnings to deescalate conflicts.

\textcolor{black}{We did not find any direct evidence of monetary \textbf{pricing}, perhaps due to social media platforms' overall pursuit of a high level of user engagement and lack of incentive to inhibit participation. However, by adopting a broader view of pricing as frictions to users, Vaccaro et al. \cite{vaccaro_at_2020} documented how existing appeal process on social media platforms shifted the cost of understanding documentations and formulating cohesive arguments to the users, effectively inducing a ``price'' on overturning enforcement.}

While it might seem that moderators were able to choose freely from a suite of possible moderation actions---excluding, organizing, norm-setting\textcolor{black}{, and pricing}---against rule-violating content or people, an underlying tension across these \textcolor{black}{four} types of actions was whether or not to \emph{remove} certain content or people. For example, while several studies documented moderators prioritizing warnings over direct punishments such as removal or banning (e.g., \cite{jiang_moderation_2019, skousen_successful_2020}), we also saw communities that were less hesitant to employ these harsher sanctions \cite{seering_moderator_2019}. Furthermore, with any of these actions, moderators had an additional choice to make: Whether or not to provide explanations. These different prioritizations reveal two immediate trade-offs. The first trade-off is one that is well-documented by prior research: Too much leniency may expose the community to harm, while too much harshness may stifle the community \cite{gurzick_view_2009, kraut_building_2011}. The second is more subtle: Removing violating content or people prevents them from staying in the community, but it also forgoes the opportunity to educate the community about acceptable behavior. Behind different competing choices in these trade-offs around moderation actions are differences in moderators' philosophies and values, which we discuss later in this paper.

\section{Trade-offs in Moderation Styles}
In addition to moderation actions, trade-offs are also present in how the moderators carry out these actions, which we term moderation styles. These moderation styles resemble ``distinctions'' in Grimmelmann's \cite{grimmelmann_virtues_2015} moderation framework (though the identified styles here do not cover all of them), serving as ``adverbs'' that describe the actions (``verbs'') mentioned in the previous section. In our analysis, we identified three major trade-offs around moderation styles in our corpus: Human vs. automated, centralized vs. distributed, and transparent vs. opaque.

\subsection{Human vs. Automated}
The trade-off between human and automated moderation refers to whether a moderation action was performed by a human or some type of automated system. It is important to note that current moderation systems are rarely purely human or purely automated, nor did any study in our corpus argue for a move toward either of these extremes. Moderation systems that we saw are always a hybrid of human and automated moderation, but the degrees to which they rely on humans or automation vary.

Arguments for more human moderation most commonly appeared when the moderation decisions were difficult and required a nuanced understanding of contexts:

\begin{quote}
Moderators we interviewed were happy to have tools that deal with the most obviously unwanted content, such as links to malware or pornography, but they have a strong preference to make the hard decisions themselves. \cite[p. 14]{seering_moderator_2019}
\end{quote}

One example of such unwanted content that was not obvious was memes, which derives their meanings from multiple layers of contexts \cite{jiang_perfect_2018}. Therefore, in response to Facebook's image recognition tool, Proch\'{a}zka \cite{prochazka_making_2019} questioned technology's ability to understand memes whose meanings were fluid and context-dependent, and argued for the necessity of human moderation of them.

Another thread of cases that warranted more human moderation was community-building. Seering et al. \cite{seering_moderator_2019} found that negotiation of acceptable and unacceptable behaviors was critical for community growth, and such negotiation necessarily requires human involvement.

However, negotiation of community norms can take on many shapes and forms. Jhaver et al. \cite{jhaver_does_2019} focused on a particular one of them: Providing removal explanations, and argued that subpar explanations can have detrimental effects on the community:

\begin{quote}
In cases where the removal reasons are unclear, human moderators should continue to provide such explanations. ... We expect that inaccurate removal explanations are likely to increase resentment among the moderated users rather than improve their attitudes about the community. \cite[pp. 22-23]{jhaver_does_2019}
\end{quote}

Despite the benefits of human moderation, moderation research also described the pressing need for automated moderation. As online communities quickly grew into sizes that humans could not reasonably handle (e.g, millions of users), automated moderation provided a solution for moderation at scale \cite{chandrasekharan_crossmod_2019}. In addition to the ability to moderate large volumes of content, speed was also an advantage of automated moderation that humans struggled to achieve. As the prerequisite for human moderation was that a human had to be online and see the potentially violating content, automated moderation triumphed in timeliness by offering 24/7 monitoring \cite{jhaver_human-machine_2019}. Beyond scalability, Jhaver et al. also noted that automated moderation offered a high level of consistency, since moderation rules were hard-coded into the automated systems \cite{jhaver_human-machine_2019}. However, such consistency presented a trade-off when facing the unique adaptability to contexts offered by humans, which Jhaver et al. also acknowledged.

Humans' ability to understand nuanced contexts became important in complex, high-stake situations such as when distinguishing hate speech from newsworthiness \cite{caplan_content_2018}, where the line between violating and non-violating was critical but blurry. Prior research extensively documented the trade-off between automated tools' ability to handle massive scale of content and human's ability to tell the subtle difference between whether certain content is violating rules (i.e. to reduce false positives), in various cases such as pro-eating disorder communities \cite{chancellor_multimodal_2017, chancellor_norms_2018}, crowdsourced blocklists  \cite{jhaver_online_2018}, copyright infringement detection \cite{gray_playing_2020}, and even country-wide ethnic violence \cite{jhaver_human-machine_2019}. Chancellor et al. \cite{chancellor_multimodal_2017} specifically pointed out that automated tools could magnify any errors they made, as well as the remedy required to correct these errors, precisely because of their ability to scale.

To summarize, studies reveal benefits in both human and automated moderation: Humans are capable of handling complex nuances, while automated systems offer the kind of moderation required by the massive scale of today's online community. However, these very benefits can become drawbacks in different situations, and the trade-offs between human and automated moderation remain a persistent challenge to content moderation.

\subsection{Centralized vs. Distributed}

The trade-off between centralized and distributed moderation refers to whether moderation decisions are made by designated moderators or regular users and community members. Similar to human vs. automated moderation, the configuration of centralized vs. distributed moderation is often a hybrid one in today's online communities, landing somewhere between purely centralized and distributed. For example, Facebook has centralized moderation teams around the world to enforce their community guidelines, as well as volunteer moderators in Facebook Groups to make their own rules and enforce their own moderation \cite{gillespie_custodians_2018}. Likewise, Reddit also has platform-wide moderators as well as volunteer moderators in individual subreddits that form the moderation system on Reddit that we see today \cite{fiesler_reddit_2018}. Even within the premise of a single subreddit, many subreddits also allow regular members to contribute to moderation decisions such as rule making in addition to the moderators. Furthermore, Reddit users also have the ability to upvote or downvote posts, which impacts the visibility of these posts \cite{fiesler_reddit_2018}.

Many papers pointed out drawbacks of distributed moderation that indicate the advantages of a centralized fashion. The arguments against distributed moderation focused on the lack of expertise from regular users, as well as their personal biases which made them incapable of making decisions representative of the community ideal: 

\begin{quote}
r/AskHistorians moderators described a variety of reasons why they opposed using the karma system as an indication of quality. First, the majority of those who upvote responses do not have the requisite expertise to evaluate quality; second, voting reflects user bias; and third, earlier comments tend to receive more upvotes, regardless of quality. \cite[p. 15]{gilbert_i_2020}
\end{quote}

These drawbacks of distributed moderation suggested that centralized moderation would be more consistent, standardized, and made by qualified experts. Some participants in Fan and Zhang's \cite{fan_digital_2020} digital jury experiment expressed a similar lack of confidence in the quality of user input. Furthermore, Duguay et al. \cite{duguay_queer_2018} found that distributed moderation could harm minority users disproportionately:

\begin{quote}
Co-moderation works against minority user groups on two levels. First, the majority of users on such a mainstream platform as Instagram are statistically more likely to be heterosexual and may have difficulty understanding the aims and culturally specific aesthetics of queer women's photos. Secondly, those who are compelled to flag others' photos do so because they feel strongly about the content, usually because they are offended by its violation of their personal norms, which may be sexist or homophobic. \cite[p. 245]{duguay_queer_2018}
\end{quote}

Here, Duguay et al. suggested that decisions from distributed moderation could favor majority norms against marginalized groups, a finding echoed by Park et al. \cite{park_supporting_2016} in pointing out the ``undesirable popularity bias'' in crowdsourced moderation of news comments. 

However, distributed moderation also has desirable advantages. \textcolor{black}{Not only was distributed moderation a feasible model \cite{lampe_crowdsourcing_2014}, we also} saw many cases where users had higher confidence in distributed moderation over centralized moderation (e.g., \cite{ehrett_e-judiciaries_2016, seering_moderator_2019}), with one study \cite{draper_distributed_2019} specifically arguing that distributed deliberation practices could foster a positive digital environment. In their study, Fan and Zhang \cite{fan_digital_2020} found that compared to distributed moderation, centralized moderation was less democratically legitimate in the framework of procedural justice, characterized by a lack of accountability to the public.

Furthermore, since centralized moderation converged moderation to a small team of moderators, they had to ``spend countless hours in order to maintain the community'' \cite{chandrasekharan_internets_2018}, which suggested distributed moderation's potential ability to diffuse moderators' workload. The ability to reduce workload, however, was at odds with the desire of expertise in moderation, which was the major advantage of centralized moderation and typically only the moderators possessed. Lampe and Resnick \cite{lampe_slashdot_2004}, in one of earliest studies of content moderation, summarized this inevitable trade-off between improving efficiency and seeking expertise:

\begin{quote}
These findings highlight tensions among timeliness, accuracy, limiting the influence of individual moderators, and minimizing the effort required of individual moderators. We believe any system of distributed moderation will eventually have to make trade-offs among these goals. \cite[p. 8]{lampe_slashdot_2004}
\end{quote}

In addition to the moderation work itself, the expertise desired in centralized moderation and the public accountability desired in distributed moderation also highlight another trade-off: Is the credibility derived from the experts or that derived from the public more desirable? Kayhan et al. \cite{kayhan_content_2013} rightfully pointed out this trade-off in perceived credibility, and came to the conclusion: It depends.

\begin{quote}
[G]overnance credibility is a contextual variable that varies from one situation to the next. Governance mechanisms implemented in two different organizations may not be equally credible if the governors are different. In a given context, expert-governance may be perceived as being more credible than community-governance if users trust the experts more than the community members (or vice versa).  \cite[p. 75]{kayhan_content_2013}
\end{quote}

In summary, we found that the trade-off between centralized and distributed moderation was one that revolved around perceived expertise, efficiency, and credibility. Just like the case of human vs. automated moderation, our analysis indicates that the centralized vs distributed trade-off may be inevitable.

\subsection{Transparent vs. Opaque}

The trade-off between transparent and opaque moderation is prominent in our dataset. While this trade-off is similar to the distinction of transparently vs. secretly in Grimmelmann's \cite{grimmelmann_virtues_2015} moderation framework, Grimmelmann's distinction focuses more on whether the fact that some kind of moderation had happened is explicit and public. However, the distinction between transparency and opacity here in our dataset focuses more on whether explanations are provided with sanctions, and the visibility of the act of moderation is less of a concern. 

We saw an undeniable push for transparency in our analysis, with ample discussion of the benefits of providing explanations. Studies found that transparency enhanced legitimacy, perceived consistency  \cite{witt_rule_2019}, and accountability \cite{fan_digital_2020}, and could prevent confusion and frustration that breeded the often incorrect folk theories for why certain content was sanctioned \cite{jhaver_did_2019, jiang_moderation_2019, suzor_what_2019, west_censored_2018}. Providing explanations also helped community members adhere to norms and improve their future behaviors \cite{jhaver_does_2019, tyler_social_2019}, and educated users about community rules \cite{jhaver_did_2019}.

Despite a multitude of benefits of being transparent, we also saw valid reasons for not providing explanations. Many studies \cite{jhaver_did_2019, jhaver_human-machine_2019, juneja_through_2020} reported that providing explanations of actions by automated moderation tools enabled malicious actors to game the rules: 

\begin{quote}
We found that moderators do not reveal the details of exactly how AutoMod[erator]\footnote{AutoModerator is a system built into Reddit that allows moderators to define rules to be automatically applied to posts in their subreddit \cite{noauthor_automoderator_nodate}.} works to their users. ... Our participants told us that although Reddit provides them the ability to make this wiki page public, they choose not to do so to avoid additional work and to ensure that bad actors do not game the Automod rules and post undesirable content that AutoMod cannot detect. \cite[p. 21]{jhaver_human-machine_2019}
\end{quote}

Chancellor et al. \cite{chancellor_thyghgapp_2016} explored such circumvention of hard-coded rules in detail through a case study of how pro-eating disorder communities used lexical variation to avoid hashtag-based moderation on Instagram, which did not even publicize how it moderated hashtags. The need to prevent rule circumvention extended beyond tool configuration to community rule making itself: Many moderators chose to phrase their rules vaguely and broadly so that they could have the necessary interpretative flexibility when it came the time to enforce these rules \cite{jiang_moderation_2019, juneja_through_2020}.

Explanations provided by humans had different problems. Contrary to recent findings, Petrič and Petrovčweč \cite{petric_elements_2014} found that providing explanations did not increase users' sense of community. Furthermore, Seering et al. \cite{seering_moderator_2019} found that transparency could be a source of conflict within communities, because community members often would not notice unannounced moderation decisions. Possible disagreements and conflicts resulting from transparency could escalate to harms against moderators, as Gilbert \cite{gilbert_i_2020} suggested in her study of r/AskHistorians:

\begin{quote}
While the stickied [explanation] comment may have reduced the total number of questions and comments than the question would have received without the stickied comment, it did not solve the problem entirely and resulted in additional emotional labor as users responded to the stickied post with insults. \cite[p. 22]{gilbert_i_2020}
\end{quote}

Several other studies echoed the emotional labor associated with moderation (e.g., \cite{wohn_volunteer_2019, dosono_moderation_2019}), but the physical labor as well. Providing explanations is a nontrivial amount of work. Jhaver et al. \cite{jhaver_does_2019} advocated the use of automated tools to provide explanations to handle the enormous traffic that online communities often experience today. However, as we mentioned previously, automated tools have the potential to magnify their errors, and tools mistakenly providing the wrong explanations could exacerbate the conflict and hostility toward moderators.

The trade-off between transparency and opacity is difficult, with no benefits of one side clearly outweighing those of the other. In an in-depth study of Reddit's moderation transparency, Juneja et al. \cite{juneja_through_2020} made this trade-off prominent by showing that moderators had divided opinions on almost every issue related to moderation transparency, including whether or not to make removals obvious, to provide explanations for sanctions, to share details of AutoModerator implementations, and to make moderation logs public, for the same reasons we discuss above. The tug of war between improving behaviors, legitimacy, and accountability, and preventing rule circumvention, conflict, and attack toward moderators remained a subtle balance to achieve in content moderation.

Overall, these three trade-offs in moderation styles, together with the trade-offs in moderation actions that we discussed in Section \ref{sec:actions}, reflect deeper decision making rationales in content moderation, which we discuss in the next section.

\section{Trade-offs in Moderation Philosophies}
The moderation actions and styles above reflect moderators' varying moderation philosophies, which are prioritizations of competing needs that led to the actions and styles that the moderators chose to employ. In our dataset, we identified three major trade-offs in moderation philosophies: Nurturing vs. punishing, efficiency vs. quality of moderation, and level of activity vs. quality of contribution. 

\subsection{Nurturing vs. Punishing}

Nurturing and punishing both aim to create a positive online environment, but reflect different ideals in moderation's purposes, which Ruckenstein and Turunen \cite{ruckenstein_re-humanizing_2020} conceptualized as ``the logic of choice'' and ``the logic of care.'' Nurturing takes an educational approach that aims to improve or reform community members' behavior, while punishing focuses on removing the rule-violating content from the community, and making sure that the rule-violating person receives consequences for their behavior. 

We saw nurturing typically associated with less harsh and more educational actions like providing warnings, offering explanations, and actively diffusing conflicts in the community. Seering et al. \cite{seering_moderator_2019} noted that moderators who took a nurturing approach saw misbehaviors as something to be reformed rather than to be eliminated:

\begin{quote}
    Rather than seeing misbehavior as something that could be ``cleaned up'' by algorithms or bans, many moderators choose to engage personally during incidents to set an example for future interactions. \cite[p. 2]{seering_moderator_2019}
\end{quote}

A reformative approach can be desirable especially because not all misbehaviors come from malicious perpetrators who intentionally disrupt communities. Jhaver et al. \cite{jhaver_did_2019} found that some people broke rules simply because they misinterpreted or unintentionally overlooked the rules, and argued that it was worthwhile to nurture these sincere users by offering explanations so as to not drive them away. Furthermore, as we discussed in the transparent vs. opaque trade-off, providing explanations to educate users could improve their behavior as well as their perceptions of content moderation in their communities. These benefits had prompted researchers to argue for a nurturing rather than punitive approach in content moderation \cite{jhaver_did_2019, west_censored_2018}.  

However, punishing can also be valuable in community maintenance. While arguing for a general nurturing approach to moderation, Jhaver et al. \cite{jhaver_did_2019} also highlighted the necessity of punishment: 

\begin{quote}
We note that although supporting users who have the potential to be valuable contributors is a worthy goal, there are other constraints and trade-offs that need to be considered. For example, moderator teams, particularly on platforms like Reddit where voluntary users regulate content, often have limited human resources. Such teams may prioritize removing offensive or violent content to keep their online spaces usable. \cite[p. 26]{jhaver_did_2019}
\end{quote}

While suggesting differential treatments of rule violation between well-meaning and malicious people, Jhaver et al. rightfully pointed out the limitation in human moderation resources---providing detailed, customized nurturing requires human work, a point we have reiterated in discussing the transparency vs. opacity trade-off. Furthermore, more human resources invested in nurturing meant less in punitive actions such as removal, which was necessary to remove harmful content to prevent them from overwhelming legitimate content. Einwiller and Kim \cite{einwiller_how_2020}, through a study of online content providers in four countries, extended Jhaver et al.'s \cite{jhaver_did_2019} volunteer moderation-based arguments to commercially-moderated platforms, highlighting the heightened difficulty of a nurturing approach when the scale was much larger:

\begin{quote}
    [Interviewees] stated that decisively pointing out publicly where and why comments violated the policy and referring to the respective policy could help educate the poster and those observing. When the volume of [harmful online content] is large, however, doing so is often impossible. It is also a challenge to do this when a user is clearly trolling or posts are severely harming others so that they have to be removed immediately. \cite[p. 198]{einwiller_how_2020}
\end{quote}

Einwiller and Kim identified severity as another key reason for taking a punitive approach to prevent exposing platform users to harm. Jiang et al. \cite{jiang_characterizing_2020} found that platform moderation had to face a wide range of harmful content, from insensitive jokes to coordination of mass murder. The latter obviously requires immediate removal and possibly an account ban, rather than a kind, educational message saying that mass murder does not contribute to a positive online environment. The severity-based moderation philosophy applied not only to platforms, but to smaller communities as well \cite{blackwell_harassment_2019, jiang_moderation_2019, seering_moderator_2019}. Therefore, the configuration in the nurturing vs. punishing trade-off, like in all other trade-offs, is a hybrid one in practice, with differing tendencies toward one or the other depending on the specific community context.

\subsection{Level of Activity vs. Quality of Contributions}
The trade-off between level of activity and quality of contributions is related to content in the community. It represents competing desires of a large amount of traffic in a community (e.g., a large number of members, a high amount of daily posts), and high quality contributions in the community (e.g., correct categorization, minimum low-effort posting\footnote{Often called ``shitposting'' in online communities.}).

The trade-off between level of activity and quality of contribution relates to how strictly moderators enforce the community rules, which represents a trade-off that we have discussed in moderation actions: Loose moderation retains community members but may also retain low quality or even harmful content, whereas strict moderation promotes high quality content but may stifle the community \cite{gurzick_view_2009, kraut_building_2011}. Srinivasan et al. \cite{srinivasan_content_2019}, for example, concluded that strict moderation through removal contributed to a high quality of community content, but also acknowledged the possibility that authors of the moderated posts might get discouraged and leave the community. Furthermore, research \cite{jhaver_does_2019} found that providing explanations, the more nurturing and less punitive approach than mere removal, also had the potential to alienate users and drive them away, noting ``moderators may need to consider whether having high traffic is more important to them than having quality content on their community.'' 

The battle between traffic and quality was also one that community members realized. Jhaver et al. \cite{jhaver_did_2019} found that community members would intentionally break rules that ensure clean organization of community content, which in their case, was a rule that mandated that questions are only posted in designated threads. While community members acknowledged the purpose and necessity of that rule, they believed that it made individual questions invisible and ``stifled community interactions,'' and chose to break the rule with speculations that their posts would subsequently be removed. 

Here, it is clear that making their own questions visible was more important to these community members, and the potential benefits outweighed the risks of breaking the rule. However, considering the scale of today's online community, having questions scattered in the community without a centralized repository (e.g., a question thread) may overwhelm other members in the community. Similarly, members of the r/NoSleep subreddit noted that while strict regulation helped them survive the surge of newcomers as a result of becoming one of the default subreddits, it also deprived old members of the kind of freedom they used to enjoy \cite{kiene_surviving_2016}. Therefore, having to face different types of community members, the answer to the level of activity vs. quality of contribution trade-off may not be obvious to the moderators.

\subsection{Efficiency vs. Quality of Moderation}

While quality of community contribution was an important consideration, so was another kind of quality---the quality of moderation. The trade-off between efficiency and quality represent two competing characteristics of content moderation work. On the one hand, moderation needs to be efficient in order to monitor and handle content in a timely manner. On the other hand, moderation also needs to fulfill goals related to quality, which converge to the central goal of making sure all content receives appropriate treatment. There is a reason for the overly-broad definition of quality. Later we will show that the meaning of quality is complex, and consists of multiple, sometimes competing factors.

The reason for the need for efficiency is straightforward: Undesirable contents should not stay up for too long. Undesirable contents range from unuseful to harmful, and the longer they exist, the more impact they have on the community. Moderation research from the earliest time has expressed the desire for efficiency \cite{lampe_slashdot_2004, wang_whispers_2014}, which became one of the primary reasons for the widespread use of automated moderation tools \cite{jhaver_human-machine_2019}. 

Minimizing delay in moderation has become even more important as online communities gain variety. The most prominent examples are communities with real-time interactions. Seering et al. \cite{seering_shaping_2017} noted that on the live streaming platform Twitch, ``due to the synchronous nature of conversations ... moderation decisions need[ed] to happen immediately.'' In voice-based communities on Discord, interactions were not only in real-time but also ephemeral. Therefore, unless moderation could happen with virtually no delay, moderators needed to seek evidence of rule breaking to make sure it even happened, a problem Discord moderators were constantly facing \cite{jiang_moderation_2019}. Furthermore, the need for efficiency did not only exist for identifying the misbehavior, but also for deciding what actions to take on the misbehavior:

\begin{quote}
    However, participants also described aspects they did not like about deliberation. Eight people mentioned lower efficacy. One user identified a trade-off between efficacy and richer user input.  \cite[p. 9]{fan_digital_2020}
\end{quote}

In Fan and Zhang's digital jury experiment where they recruited users to serve as ``jurors'' that make moderation decisions, they found that deliberation between the jurors delayed the moderation decision in sacrifice to careful, democratic decision making. The digital jury example is exemplary of the trade-off between efficiency and quality in question, with quality represented by ``richer user input.'' 

The meaning of moderation quality, however, is more complex. Different studies conceptualized ``quality'' differently, as already shown by previous sections. For example, Fan and Zhang \cite{fan_digital_2020} considered ``quality'' to be democratic legitimacy and accountability. Schoenebeck et al. \cite{schoenebeck_drawing_2020} argued that ``quality'' should be customized moderation that did not fail some people while privileging others. Jhaver et al. \cite{jhaver_human-machine_2019} believed ``quality'' of moderation to be minimal incorrect decisions (i.e., ``false negatives'' and ``false positives''), though the concept of ``correct'' might be just as complex as ``quality.'' In a tricky case of r/AskHistorian where moderators had to choose between directly removing a post and explaining why that post was subtly harmful, Gilbert et al. \cite{gilbert_i_2020} presented an example of almost complete surrender of efficiency for the pursuit of high-quality moderation, where a moderator biked to a nearby, paywalled library to find answers to a community member's question.

While these examples are by no means comprehensive, all of them require human deliberation and thus, a sacrifice of efficiency to various extents. As the trade-off between efficiency and quality pertains to every moving part in what we have described in the previous sections about moderation actions and styles, it might be the case that decisions in different parts will compete with each other and impact the overall efficiency vs. quality trade-off as a whole.

Overall, these three trade-offs in moderation philosophies presented in this section represent competing desires for the purpose of moderation, the process of moderation, and the community content shaped by moderation. These subtle decisions in philosophies reflect values that different stakeholders in online communities hold, which we discuss in the next section.

\section{Trade-offs in Moderation Values}
So far, we have discussed many trade-offs in moderation actions, styles, and philosophies. These trade-offs show competing needs that are all legitimate, have pros and cons, and do not have clear, ``right'' answers. However, facing these trade-offs, moderators must make decisions, and we found that these decisions were impacted by trade-offs in the values that they might hold. While these trade-offs were not characterized by pairs of polar opposites, any value position could come in tension with other alternatives. In our dataset, we identified three facets in the trade-offs in moderation values: Moderator identities, community identities, and competing stakeholders.

\subsection{Moderator Identities}

Moderator identities are what moderators see themselves as in their communities, such as governors, teachers, and gardeners, to give a few examples. Prior research (e.g., \cite{wohn_volunteer_2019}) often referred to these identities as social roles characterized by designated tasks, but here we use the term ``identity'' to emphasize moderators' self-perceived high-level responsibilities that transcend specific tasks. The differences in moderator identities have been a prominent theme since the earliest of moderation research in our corpus:

\begin{quote}
    One admin saw his role as being particularly centred on careful management of people in ``keeping the peace'' and maximising the potential of others, while another saw his role as being more based around the filtering of discussions and the group pool. \cite[p. 12]{holmes_every_2011}
\end{quote}

In their study of Flickr administrators, Holmes and Cox found moderator identities that correspond to the nurturing vs. punishing trade-off in moderation philosophies. As online communities evolve, the perceived identities also start to vary more. For example, Matias \cite{matias_civic_2019} listed a range of identities that his participants self-identified with, along with different corresponding duties. For example, dictators ``make all the decisions,'' janitors ``clean up,'' and martyrs ``give hell to anyone who dared to ... threaten [their] communities.'' Similarly, Seering et al. \cite{seering_moderator_2019} in their study of 56 moderators also found several identities, including arbiters, community managers, role models, etc.

While we do not get into the details of the subtle differences and overlaps between the identities listed here (which deserves its own research), moderators used them to justify the moderation decisions they had made. While prior research has not always made explicit connections between these identities and philosophies, styles, and actions, it is reasonable to speculate that moderators who identify as arbiters would prefer to adopt centralized moderation, and those who identify as curators would care about the quality of contributions more than the level of community activity. Overall, taking up a certain identity means to serve certain responsibilities and purposes, and to take actions accordingly \cite{wohn_volunteer_2019}.

Wohn \cite{wohn_volunteer_2019} also pointed out that moderator identities were not mutually exclusive. The co-presence of these different, sometimes competing identities showed that there was a need for many of them---for example, moderation may need to be nurturing and punishing, instead of nurturing or punishing. Another line of research on social roles (e.g., \cite{yang_seekers_2019}) also echoed these simultaneously existing identities. Gurzick et al. \cite{gurzick_view_2009} described how moderators were aware of the need to balance identities, and that moderators ``debated the proper role that they should take and negotiated the amount of activity that would be reasonable.'' The negotiation of these identities shows that making decisions in the trade-offs in actions, styles, and philosophies may extend beyond their own pros and cons, to a deliberation of value differences.

\subsection{Community Identities}

In addition to how moderators see themselves, how moderators see their communities also has an impact on how they moderate them. We call communities' self-conception of what kind of communities they are as ``community identities.'' The perceived identity of a community determines who and what is welcome or unwelcome in the community, and what purpose the community is supposed to serve.

A prominent trade-off in community identities is that between, as Gibson \cite{gibson_free_2019} named, ``free speech,'' and ``safe spaces.'' The former referred to online spaces that promote free expression of opinions, while the latter emphasized mitigating potential harm that speech could cause. Gibson found that compared to ``free speech'' spaces, in ``safe spaces'' moderators removed significantly more content, indicating a punitive tendency that focused more on the quality of community content (in this case, content that did not harm marginalized communities). Like Gibson, other research \cite{grover_detecting_2019, phadke_many_2020} also revealed these two often competing identities, highlighting that it is a difficult trade-off to balance:

\begin{quote}
   As political and ideological stratification in society continues to grow, and online communities focused on ideological commitments become more numerous, moderators of online  platforms ... face  difficult  challenges  in  how  to  balance the right to free expression, with broader concerns of public safety and wellbeing. \cite[p. 203]{grover_detecting_2019}
\end{quote}

Trade-offs in community identities also existed for communities committed to certain topics, where moderators struggled to balance competing conceptualizations of the topic. For example, in r/Paleo, a subreddit for the paleo diet\footnote{Wikipedia explains paleo diet as ``a modern fad diet consisting of foods thought to mirror those eaten during the Paleolithic era'' \cite{noauthor_paleolithic_2020}.}, moderators struggled to maintain a balance between a consistent community conception of paleo diet and individualized understandings of what paleo diet is:

\begin{quote}
    Paleo faces a tension between the need to maintain some kind of coherent concept of the diet while also allowing flexibility for adherents to pursue a diet that accounts for individual differences. One way of negotiating this tension comes through the rules of the subreddit. One of the only rules that r/paleo moderators actively enforce is not to ``[a]ct like your One True Paleo™ is the be-all, end-all and is perfect for every human on Earth.'' \cite[p. 1919]{squirrell_platform_2019}
\end{quote}

Here, the moderators did not decide on one particular identity to pursue as a community, but simply required that members keep an open mind toward all versions of the paleo diet. 

The differences in how certain topics are conceptualized also exists in research of these communities, with pro-eating disorder (pro-ED) communities the most prominent. A long line of research \cite{chancellor_multimodal_2017, chancellor_norms_2018, chancellor_thyghgapp_2016, gerrard_beyond_2018, feuston_conformity_2020} on moderating pro-ED communities shows a clear trajectory of how pro-ED communities are viewed: From communities that promote eating disorders as a legitimate lifestyle, to those that support and help people with eating disorders. The co-presence of competing conceptualizations meant that the same content could be treated differently due to (1) how they were perceived, and (2) whether that perception matched the community identity:

\begin{quote}
    Harm reduction provides resources for individuals who have an eating disorder, but cannot or will not recover, to stay safe and informed. Despite benefits, harm reduction resources are treated differently across eating disorder spaces online. While some communities freely permit them, others, such as one of the active subreddits in our digital ethnography, have moderation teams dedicated to removing posts related to tips or advice and carefully overseeing content related to harm reduction. \cite[p. 16]{feuston_conformity_2020}
\end{quote}

Feuston et al. \cite{feuston_conformity_2020} argued that content moderation should consider the full complexity of marginalized experiences such as eating disorders, and not cast negative stereotypes on content like harm reduction that might help those in need. 

While Feuston et al. provided an example of how fulfilling stereotypical community identities could be harmful, Gilbert \cite{gilbert_i_2020} further complicated the issue by demonstrating how fulfilling seemingly innocent identities could also cause unintentional harm. In the same example we discussed in the efficiency vs. quality of moderation trade-off, where an r/AskHistorians member posted a question about the background of a historical photo featuring naked women in military, fulfilling the community identity became at odds with the need of being contextually sensitive:

\begin{quote}
    In circumstances in which biased or insensitive questions are asked, moderators are tasked with making the decision to let the question stand or remove it, and experts with the decision to respond to the question or ignore it. ... During our interview, moderator, Mark Evans described deliberating whether or not to remove the question: ``We had a discussion about removing it because the pictures are incredibly ... exploitative ... And we just felt so shitty as moderators, because here was our community, which is meant to be giving people answers about the past, but what it's doing is providing Redditors with porn. And that's what it ended up doing. And that's why people have ended up looking at it and it's become a platform for these poor women to become humiliated again, like 80 years after the event. Again.''  \cite[pp. 11-12]{gilbert_i_2020}
\end{quote}

As Gilbert later pointed out, the issue of whether or not to provide people with answers about an exploitative past raised questions about trade-offs between centralized and distributed moderation, as well as ``free speech'' and ``safe spaces.'' While prior research argued that community identity might not be as salient in the moderation of platforms due to the lack of strong ties \cite{fiesler_creativity_2019}, the discrepancy of perceived platform identity could still be a source of conflict, like when Yelp users left one-star reviews for a merchant that employed someone who had contentious political beliefs on immigration, many of which Yelp removed \cite{medeiros_picketing_2019}. While Yelp intended the reviews to be about the commercial services of merchants, the users found them as ``symbolically significant means of signaling social disapproval.'' Medeiros \cite{medeiros_picketing_2019} characterized the unintended use of reviews as ``a genuinely vexing moderation challenge for Yelp, suggesting a limit to the site's ability to enforce rules that dichotomize political and commercial content.'' 

In both examples above, core to the problems is the different prioritization of community identities across different stakeholders. We explore the impacts of different stakeholders in the next section.

\subsection{Competing Stakeholders}

Moderation is often expected to satisfy multiple stakeholders and their often different needs, which presents a difficult task for moderators who often have to make decisions that serve some over others. Matias \cite{matias_civic_2019}, for example, summarized volunteer moderators' work of serving different stakeholders as their ``civic labor'':

\begin{quote}
    This ``civic labor'' requires moderators to serve three masters with whom they negotiate the idea of moderation: the platform, reddit participants, and other moderators. Moderators differ in the pressure they receive from these parties and the weight they give them. Some face further stakeholders outside the platform. Yet attempts to make sense of moderation by focusing on any one of these relationships can bring the other actors out of focus. \cite[p. 8]{matias_civic_2019}
\end{quote}

While we have shown in previous sections the impact of community members and other moderators, platforms are also a significant factor. Volunteer moderators' power cannot reach beyond the purview of the platform where their communities are hosted, and consequences could be severe when negotiations with platforms fail. One such example is the Reddit blackout, where many moderators shut down their communities in response to Reddit's dismissal of an employee who routinely offered support to volunteer moderators. Matias \cite{matias_going_2016} showed that such protest against the platform was still a negotiation among moderators, users, and the platform:

\begin{quote}
   Reddit employees played a key role in these negotiations [with Reddit]. ... Across subreddits of all sizes, relations among moderators were also associated with participation in the blackout. ... Community members also played an important role in action against the platform by pressuring moderators to join the blackout, discussing and voting in decisions, and sometimes even punishing moderators who disagreed.  \cite[pp. 1146-1147]{matias_going_2016}
\end{quote}

Like volunteer moderation, commercial moderation faces the same trade-off between multiple stakeholders. The common factor was users---for example, differently politically affiliated users also perceived content moderation differently \cite{hua_characterizing_2020, shen_discourse_2019}. Schoenebeck et al. \cite{schoenebeck_drawing_2020} also found that people with different backgrounds had significantly different preferences for the kinds of remedy social media sites could offer for online harassment.

However, platform moderation also needed to satisfy a new set of stakeholders. First, unsurprisingly, platforms have to operate under the requirement of local law, which often ban severely harmful content on a statutory level such as child pornography and terrorism \cite{einwiller_how_2020, gillespie_custodians_2018, zeng_how_2017}. However, these content may still provide value to someone else:

\begin{quote}
    Often disturbing, graphic, and controversial, human rights-related media like the Werfalli and Syrian war videos pose a dilemma for platforms hosting them, involving difficult tradeoffs between their perceived social value and their possible harms. \cite[p. 2]{banchik_disappearing_2020}
\end{quote}

Banchik \cite{banchik_disappearing_2020} found that even graphically abusive content may prove to be valuable documentations to various human rights workers, adding that:

\begin{quote}
     Practitioners I spoke with expressed added concern that biased or merely ill-informed human reviewers ``without the requisite knowledge'' would decide the fate of vital documentation. Moreover, most practitioners did not blame platforms alone for the removal of content, but instead saw the topography of takedowns as far more complex. \cite[p. 7]{banchik_disappearing_2020}
\end{quote}

Platforms are also aware of the complexity of harmful content given their potential public value. Facebook's Community Standards \cite{facebook_community_nodate}, for example, states:

\begin{quote}
    In some cases, we allow content for public awareness which would otherwise go against our Community Standards---if it is newsworthy and in the public interest. We do this only after weighing the public interest value against the risk of harm and we look to international human rights standards to make these judgments.
\end{quote}

However, Facebook's decision to not remove some violence-inciting message on the same ground provoked heated debate among users and various experts \cite{shieber_zuckerberg_2020}. 

Furthermore, for platform designers, the fundamental need to moderate content for users becomes a trade-off to consider with the psychological health of moderators. Both academic research \cite{karunakaran_testing_2019, luo_emotional_2020, riedl_downsides_2020} and journalistic coverage \cite{newton_secret_2019} revealed the emotional impact of moderating disturbing content. As platform technologies evolve into new forms like live-streaming video, produced content are more likely to provoke intensified emotional reactions, and therefore what is asked from moderators, both logistically and emotionally, can also escalate \cite{luo_emotional_2020}.

Above are only some of the examples of the full complexity of content moderation in a multi-stakeholder environment. Realization of the needs of multiple stakeholders has prompted many studies to call out against a one-size-fits-all approach to content moderation \cite{blackwell_classification_2017, gallagher_comparing_2016, jiang_moderation_2019, schoenebeck_drawing_2020}. However, as desirable as customized moderation might be, it may not be entirely feasible due to constraints in human and technological resources. Then, whose needs are prioritized, and what downstream impacts it has on various trade-offs, are critical problems to consider in content moderation.

\section{How Different Stakeholders Can Use Our Framework}
Our framework offers a way to examine content moderation that posits trade-offs in the front and center. As an example, Seering et al.'s \cite{seering_moderator_2019} findings on the differences in actions taken by moderators toward misbehaviors indicate that values impact moderator actions. However, if we examine their findings through the lens of our framework, we can reveal several additional research questions related to the trade-offs that could have happened: While communities with more laissez-faire ideologies use fewer bans than communities that intended to be ``safe spaces,'' what prompted the communities to side with certain ideologies over others? Do moderators' perceived identities differ between Reddit and Facebook, and does that have an impact on differences in the level of reliance on automated tools? These are only a few examples of the questions we can ask from the application of our framework, and Seering et al.'s \cite{seering_moderator_2019} speculation of the answers to the latter question testified to the value of our framework---``The difference [in the preference of automated tools] likely results from the importance of continuously evolving community values in decisions made by moderators.'' Answers to these questions will offer a deep, rich understanding of the inner-working of content moderation from a new angle.

The above example is only one way \textbf{researchers} of content moderation can use this framework as an analytical tool in their own research. For example, \textcolor{black}{beyond identifying moderation actions in the community, }a researcher can use our framework to \textcolor{black}{go one step further and} identify key trade-offs in moderators' decision-making, investigate why moderators took certain actions instead of other actions they could have taken, and trace back to their philosophies and values behind these decisions. Furthermore, researchers can also use our framework to identify potential value tensions behind certain philosophies, and potential caveats of recommendations they might make. For example, \textcolor{black}{do certain platform affordances favor certain philosophies and values?} When recommending that the moderation of a community or a platform should be more transparent, what are the potential stakeholder tensions that may prevent it from doing so? How can it resolve such tensions to get closer to the researchers' ideal?

\textbf{Designers} of content moderation can use our framework as a heuristic for their design, either to improve an existing content moderation system, or to build a new one. Designers who wish to improve an existing moderation system can use our framework to identify key decision points that moderators may struggle with and to be critically aware of the trade-offs and tensions involved. While their designs may inevitably favor one side of a trade-off, designers can consciously find their ideal balance in the trade-off so that their designs can be more considerate of the other side. Similar to the case of researchers, some trade-offs may not be applicable or salient to some communities or platforms. While designers should focus on the trade-offs as appropriate, with our framework they can also consider making some previously invisible trade-offs more salient as a potential form of improvement. 

Designers who wish to build new content moderation systems can use our framework as a guide to support moderators in key decision points. For example, designers may consider explicitly showing the available actions and decisions to moderators as trade-offs instead of a simple listing, as well as the potential consequences of making different decisions. Designers can present these trade-offs not only in manuals or training materials, but also in the interface of moderators' day-to-day work, so that moderators can be more informed when making decisions. 

\textbf{Moderators} may also benefit from our framework as a way to encourage reflexivity in their own work. For example, our framework will allow moderators to realize that when they make a decision on doing something, they are also making decisions on not doing something else. Therefore, moderators will be able to make more conscious trade-offs in their work, and have elaborate justifications for past decisions that may be valuable for revising or improving their workflow.

Finally, \textbf{users}, or people who are moderated, may find our framework informative when participating in content moderation in various ways. As a key element in content moderation, users will be able to learn the full complexity of moderation from the trade-off-centered framework, and therefore be more informed when disputing moderation decisions, contributing to rule making, or engaging in conversations about content moderation in general.

\section{Trade-offs Define Content Moderation}
Our framework characterizes content moderation in terms of trade-offs on multiple levels. First, we found many competing choices in trade-offs in moderation actions and styles. Each choice has its own pros and cons that, as we have shown, relate to trade-offs in moderation philosophies. For example, the trade-off between leniency and harshness and that between immediately removing harm and long-term education in moderation actions demonstrate clear connections to the level of activity vs. quality of contribution and the nurturing vs. punishing trade-offs respectively. The different pros and cons of competing moderation styles also find their way to trade-offs in philosophies. Overall, moderation philosophies reflect the fundamental needs and purposes that moderation actions and styles aim to serve.

In trade-offs in moderation philosophies, many options are often believed to (or at least be supposed to) go hand in hand with each other: Moderation should be both educational for sincere community members and punishing for malicious actors. Moderation should be both efficient and of high quality. Moderation should maintain community members' engagement and activities while ensuring a high quality of contribution. While these goals often seem to be congruent, in our analysis of moderation literature, we found that they were often at odds with each other. As ideal as it would be to achieve both sides of the trade-offs, we saw evidence that a tendency toward one side may necessarily be at the cost of the other.

Furthermore, these philosophies trace back to \textcolor{black}{Grimmelmann's commonly-cited definition of content moderation as ``the governance mechanisms that structure participation in a community to facilitate cooperation and prevent abuse'' \cite{grimmelmann_virtues_2015}.} The trade-offs in moderation philosophies echo the goals of moderation in Grimmelmann's definition: Nurturing, moderation quality, and level of activity are different facets of facilitating cooperation, while punishing, moderation efficiency, and quality of contribution represent different dimensions of preventing abuse. However, while Grimmelmann indicates that these two goals are to be achieved at the same time, our trade-off centered analysis shows a different relationship: Facilitating cooperation and preventing abuse may be at tension with each other in practice. If the two definitional components of content moderation constitute a trade-off, then we argue that content moderation as a whole can be conceptualized as a series of trade-offs, and that moderation work is making choices and striking balances between simultaneously desirable goals.

Then, how can moderators balance facilitating cooperation and preventing abuse? The trade-offs in values may provide answers to this question. Our findings suggest that the driving force behind which component is favored more is dependent on the moderators' perceived identities of themselves and their visions for their communities, both of which are also shaped by various stakeholders including other moderators, community members, platforms, legal requirements, etc. These forces work together and converge toward a unique decision point between facilitating cooperation and preventing abuse. 

While we have summarized the major trade-offs that we have identified in our corpus, there may be other, likely more granular trade-offs that we have not listed in the paper. Therefore, while using the trade-offs identified in this paper as a preliminary checklist may prove useful, we believe that a trade-off-centered perspective in content moderation will be more valuable. Therefore, when examining a decision in content moderation, we urge researchers and designers to consider them as a trade-off instead: What is the other side of the trade-off? Does it privilege someone's perspective and disadvantage someone else's?

\section{Conclusion}
In this paper, we propose a trade-off-centered framework of content moderation. We describe four major layers of trade-offs in our framework at increasing levels of abstractions---in moderation actions, styles, philosophies, and values---and how they are related to each other. These trade-offs are pervasive in content moderation practices, and reveal the dialectic nature of content moderation. While existing literature largely conceptualizes content moderation as being built on a series of standalone actions, we believe a trade-off centered framework provides a more holistic perspective: What are pros and cons of taking a certain moderation action? What do stakeholders gain and give up by taking up certain philosophies? What does it mean for the community if any trade-off rationale becomes normative and codified and enforced at scale? We believe that moderation researchers, designers, moderators, and users will all find value in taking a trade-off-centered approach to content moderation, and we hope this paper will provide a fresh agenda for content moderation research.

\begin{acks}
We thank Jes Feuston, Joseph Seering, and anonymous reviewers for their valuable feedback.
\end{acks}

\bibliographystyle{ACM-Reference-Format}
\bibliography{slr}

\newpage

\appendix

\section{Papers Included in the Systematic Literature Review}
\label{sec:appendix}
\input{papers_table}

\clearpage

\section{Components of Trade-offs and Relevant Papers}
\label{sec:appendix2}
\input{category_table}


%% file: papers_table.tex
\begin{table}[h!]
\centering
\rowcolors{2}{white}{lightblue}
\renewcommand{\arraystretch}{1.2}
\caption{List of papers included in our systematic literature review.}
\begin{tabularx}{\textwidth}{YYYYY}
\rowcolor{white}
\textbf{Paper}                      & \textbf{Qualitative} & \textbf{Quantitative} & \textbf{Volunteer Moderation} & \textbf{Commercial Moderation}  \\
\midrule
(Seering et al., 2020)~\cite{seering_metaphors_2020}        &  \textbullet           &             & \textbullet                    &                     \\
(Vaccaro et al., 2020)~\cite{vaccaro_at_2020}        &  \textbullet             & \textbullet            &                    &  \textbullet                   \\
(Rajadesingan et al. 2020)~\cite{rajadesingan_quick_2020}  &\textbullet &  & \textbullet&\\
(Feuston et al., 2020)~\cite{feuston_conformity_2020}             & \textbullet           &              & \textbullet                    & \textbullet                     \\
(Fan \& Zhang, 2020)~\cite{fan_digital_2020}               & \textbullet           & \textbullet            &                      &                    \\
(Juneja et al., 2020)~\cite{juneja_through_2020}              & \textbullet           & \textbullet            & \textbullet                    &                    \\
(Hua et al., 2020)~\cite{hua_characterizing_2020}                 & \textbullet           & \textbullet            &                      & \textbullet                   \\
(Luo et al., 2020)~\cite{luo_emotional_2020}                 &             & \textbullet            & \textbullet                    & \textbullet                    \\
(Phadke \& Mitra, 2020)~\cite{phadke_many_2020}            & \textbullet           & \textbullet            &                      & \textbullet                     \\
(Gilbert, 2020)~\cite{gilbert_i_2020}                    & \textbullet           &              & \textbullet                    &                     \\
(Obar \& Oeldorf-Hirsch, 2020)~\cite{obar_biggest_2020}     &             & \textbullet            &                      & \textbullet                   \\
(Riedl et al., 2020)~\cite{riedl_downsides_2020}               &             & \textbullet            &                      & \textbullet                   \\
(Einwiller \& Kim, 2020)~\cite{einwiller_how_2020}           & \textbullet           & \textbullet            &                      & \textbullet                     \\
(Banchik, 2020)~\cite{banchik_disappearing_2020}                    & \textbullet           &              &                      & \textbullet                    \\
(Schoenebeck et al., 2020)~\cite{schoenebeck_drawing_2020}         &             & \textbullet            &                      & \textbullet                     \\
(Skousen et al., 2020)~\cite{skousen_successful_2020}             & \textbullet           &              & \textbullet                    & \textbullet                     \\
(Gray \& Suzor, 2020)~\cite{gray_playing_2020}              & \textbullet           & \textbullet            &                      & \textbullet

\end{tabularx}
\end{table}

\begin{table}[]
\centering
\rowcolors{2}{white}{lightblue}
\renewcommand{\arraystretch}{1.2}
\begin{tabularx}{\textwidth}{YYYYY}

(Datta \& Adar, 2019)~\cite{datta_extracting_2019}              &             & \textbullet            &                      & \textbullet                      \\
(Grover \& Mark, 2019)~\cite{grover_detecting_2019}             &             & \textbullet            & \textbullet                    & \textbullet       \\
(S. Jiang et al., 2019)~\cite{jiang_bias_2019}            &             & \textbullet            &                      & \textbullet                      \\
(Redmiles et al., 2019)~\cite{redmiles_i_2019}            & \textbullet           & \textbullet            &                      & \textbullet   \\

(Karunakaran \& Ramakrishan, 2019)~\cite{karunakaran_testing_2019} & \textbullet           & \textbullet            &                      & \textbullet        \\
(Kiene et al., 2019)~\cite{kiene_technological_2019}               & \textbullet           &              & \textbullet                    &                         \\
(Blackwell et al., 2019)~\cite{blackwell_harassment_2019}           & \textbullet           &              & \textbullet                    & \textbullet  \\
(Jhaver, Bruckman, et al., 2019)~\cite{jhaver_does_2019}   &             & \textbullet            & \textbullet                    &                         \\
(Jhaver, Birman, et al., 2019)~\cite{jhaver_human-machine_2019}     & \textbullet           &              & \textbullet                    &                         \\
(Jhaver, Appling, et al., 2019)~\cite{jhaver_did_2019}    & \textbullet           & \textbullet            & \textbullet                    &                         \\
(Srinivasan et al., 2019)~\cite{srinivasan_content_2019}          &             & \textbullet            & \textbullet                    &                         \\
(J. A. Jiang et al., 2019)~\cite{jiang_moderation_2019}         & \textbullet           &              & \textbullet                    &                         \\
(Chandrasekharan et al., 2019)~\cite{chandrasekharan_crossmod_2019}     & \textbullet           & \textbullet            & \textbullet                    &                         \\
(Dosono et al., 2019)~\cite{dosono_moderation_2019}                       & \textbullet           &              & \textbullet                    &                         \\
(Wohn, 2019)~\cite{wohn_volunteer_2019}                       & \textbullet           &              & \textbullet                    &                         \\
(Fiesler \& Bruckman, 2019)~\cite{fiesler_creativity_2019}        & \textbullet           &              & \textbullet                    &                         \\
(Gibson, 2019)~\cite{gibson_free_2019}                     &             & \textbullet            & \textbullet                    &                         \\
(Tyler et al., 2019)~\cite{tyler_social_2019}               &             & \textbullet            &                      & \textbullet                       \\
(Potts et al., 2019)~\cite{potts_boycotting_2019}               &             & \textbullet            & \textbullet                    &                         \\
(Procházka, 2019)~\cite{prochazka_making_2019}                  & \textbullet           &              &                      & \textbullet

\end{tabularx}
\end{table}

\begin{table}[]
\centering
\rowcolors{2}{white}{lightblue}
\renewcommand{\arraystretch}{1.2}
\begin{tabularx}{\textwidth}{YYYYYY}
(Squirrell, 2019)~\cite{squirrell_platform_2019}                  & \textbullet           &              & \textbullet                    &                         \\
(Seering et al., 2019)~\cite{seering_moderator_2019}             & \textbullet           &              & \textbullet                    &                         \\
(Witt et al., 2019)~\cite{witt_rule_2019}                &             & \textbullet            &                      & \textbullet                       \\
(Matias, 2019)~\cite{matias_civic_2019}                    & \textbullet           &              & \textbullet                    & \textbullet                       \\
(Juneström, 2019)~\cite{junestrom_online_2019}                  & \textbullet           & \textbullet            &                      & \textbullet                       \\
(Shen \& Rose, 2019)~\cite{shen_discourse_2019}               &             & \textbullet            &                      & \textbullet                       \\
(Medeiros, 2019)~\cite{medeiros_picketing_2019}                   & \textbullet           &              &                      & \textbullet                       \\
(Draper, 2019)~\cite{draper_distributed_2019}                     & \textbullet           &              & \textbullet                    &                         \\
(Suzor et al., 2019)~\cite{suzor_what_2019}               & \textbullet           &              &                      & \textbullet                       \\
(Nurik, 2019)~\cite{nurik_men_2019}                      & \textbullet           &              &                      & \textbullet \\
(Duguay et al., 2018)~\cite{duguay_queer_2018}              & \textbullet           &              &                      & \textbullet                       \\
(Fiesler et al., 2018)~\cite{fiesler_reddit_2018}             & \textbullet           & \textbullet            & \textbullet                    & \textbullet                       \\
(Blackwell et al., 2018)~\cite{blackwell_when_2018}           & \textbullet           & \textbullet            & \textbullet                    &             \\
(Jhaver et al., 2018)~\cite{jhaver_online_2018}              & \textbullet           & \textbullet            & \textbullet                    &                         \\
(Matias \& Mou, 2018)~\cite{matias_civilservant_2018}              & \textbullet           &              & \textbullet                    &                         \\
(Chancellor et al., 2018)~\cite{chancellor_norms_2018}          &             & \textbullet            & \textbullet                    &                         \\
(Pavalanathan et al., 2018)~\cite{pavalanathan_mind_2018}        &             & \textbullet            & \textbullet                    &                         \\
(Chandrasekharan et al., 2018)~\cite{chandrasekharan_internets_2018}     & \textbullet           & \textbullet            & \textbullet                    &                         \\
(Gerrard, 2018)~\cite{gerrard_beyond_2018}                    & \textbullet           &              &                      & \textbullet                       \\
(West, 2018)~\cite{west_censored_2018}                       & \textbullet           & \textbullet            &                      & \textbullet                       \\
(Keegan \& Fiesler, 2017)~\cite{keegan_evolution_2017}          &             & \textbullet            & \textbullet                    &                         \\

\end{tabularx}
\end{table}

\begin{table}[]
\centering
\rowcolors{2}{lightblue}{white}
\renewcommand{\arraystretch}{1.2}
\begin{tabularx}{\textwidth}{YYYYY}
(Chancellor et al., 2017)~\cite{chancellor_multimodal_2017}          &             & \textbullet            &                      & \textbullet                       \\
(Pellicone \& Ahn, 2017)~\cite{pellicone_game_2017}           & \textbullet           &              & \textbullet                    &                         \\
(Blackwell et al., 2017)~\cite{blackwell_classification_2017}           & \textbullet           &              &                      & \textbullet                       \\
(Chandrasekharan et al., 2017)~\cite{chandrasekharan_you_2017}     &             & \textbullet            &                      & \textbullet                       \\
(Seering et al., 2017)~\cite{seering_shaping_2017}             &             & \textbullet            & \textbullet                    &                         \\
(Zeng et al., 2017)~\cite{zeng_how_2017}                &             & \textbullet            &                      & \textbullet                       \\
(Cheng et al., 2017)~\cite{cheng_anyone_2017}               &             & \textbullet            &                      & \textbullet                       \\
(Newell et al., 2016)~\cite{newell_user_2016}              & \textbullet           & \textbullet            &                      & \textbullet                       \\
(Chancellor, Pater, et al., 2016)~\cite{chancellor_thyghgapp_2016}  &             & \textbullet            &                      & \textbullet                       \\
(Park et al., 2016)~\cite{park_supporting_2016}                & \textbullet           &              &                      & \textbullet                       \\
(Centivany \& Glushko, 2016)~\cite{centivany_popcorn_2016}       & \textbullet           &              & \textbullet                    &                         \\
(Matias, 2016)~\cite{matias_going_2016}                    & \textbullet           & \textbullet            &                      & \textbullet                       \\
(Gallagher \& Savage, 2016)~\cite{gallagher_comparing_2016}        &             & \textbullet            &                      & \textbullet \\
(Getto \& Labriola, 2016)~\cite{getto_ifixit_2016}          & \textbullet           &              & \textbullet                    &                         \\
(Kiene et al., 2016)~\cite{kiene_surviving_2016}               & \textbullet           &              & \textbullet                    &                         \\
(Ehrett, 2016)~\cite{ehrett_e-judiciaries_2016}                     & \textbullet           & \textbullet            & \textbullet                    & \textbullet       \\
(Vashistha et al., 2015)~\cite{vashistha_sangeet_2015}           & \textbullet           & \textbullet            &                      & \textbullet                       \\
(Wang et al., 2014)~\cite{wang_whispers_2014}                &             & \textbullet            &                      & \textbullet                       \\
(Petrič \& Petrovčič, 2014)~\cite{petric_elements_2014}        &             & \textbullet            & \textbullet                    & \textbullet                       \\
(Lampe et al., 2014)~\cite{lampe_crowdsourcing_2014}               & \textbullet           & \textbullet            & \textbullet                    &                         
\end{tabularx}
\end{table}

\begin{table}[]
\centering
\rowcolors{2}{lightblue}{white}
\renewcommand{\arraystretch}{1.2}
\begin{tabularx}{\textwidth}{YYYYY}

(Kayhan \& Bhattacherjee, 2013)~\cite{kayhan_content_2013}    &             & \textbullet            &                      & \textbullet                       \\
(Heinze et al., 2013)~\cite{heinze_ideal_2013}              & \textbullet           & \textbullet            &                      & \textbullet                       \\
(Sarkar et al., 2012)~\cite{sarkar_quantitative_2012}              &             & \textbullet            &                      & \textbullet                       \\
(Holmes \& Cox, 2011)~\cite{holmes_every_2011}              & \textbullet           & \textbullet            & \textbullet                    &                         \\
(Liao et al., 2010)~\cite{liao_chinese_2010}                & \textbullet           & \textbullet            & \textbullet                    & \textbullet                       \\
(Gurzick et al., 2009)~\cite{gurzick_view_2009}             & \textbullet           &              &                      & \textbullet                       \\
(Lampe \& Johnston, 2005)~\cite{lampe_follow_2005}          &             & \textbullet            & \textbullet                    & \textbullet                       \\
(Lampe \& Resnick, 2004)~\cite{lampe_slashdot_2004}           & \textbullet           & \textbullet            & \textbullet                    & \textbullet                       \\

\bottomrule
\end{tabularx}
\end{table}

%% file: category_table.tex
\begin{table}[h!]
\centering
\renewcommand{\arraystretch}{1.2}
\caption{Components of trade-offs and relevant papers.}
\begin{tabularx}{\textwidth}{YY}
\textbf{Trade-off} & \textbf{Papers} \\

\midrule
\multicolumn{2}{c}{\textbf{Moderation Actions}} \\ \midrule
Exclusion & \cite{seering_metaphors_2020, vaccaro_at_2020, feuston_conformity_2020, fan_digital_2020, juneja_through_2020, hua_characterizing_2020, phadke_many_2020, gilbert_i_2020, einwiller_how_2020, schoenebeck_drawing_2020, skousen_successful_2020, datta_extracting_2019, kiene_technological_2019, blackwell_harassment_2019, jhaver_does_2019, jhaver_human-machine_2019, srinivasan_content_2019, jiang_moderation_2019, dosono_moderation_2019, wohn_volunteer_2019, gibson_free_2019, potts_boycotting_2019, prochazka_making_2019, squirrell_platform_2019, seering_moderator_2019, witt_rule_2019, matias_civic_2019, shen_discourse_2019, draper_distributed_2019, suzor_what_2019, nurik_men_2019, duguay_queer_2018, fiesler_reddit_2018, jhaver_online_2018, matias_civilservant_2018, chancellor_norms_2018, pavalanathan_mind_2018, chandrasekharan_internets_2018, gerrard_beyond_2018, west_censored_2018, chancellor_multimodal_2017, pellicone_game_2017, chandrasekharan_you_2017, seering_shaping_2017, cheng_anyone_2017, newell_user_2016, chancellor_thyghgapp_2016, park_supporting_2016, centivany_popcorn_2016, matias_going_2016, kiene_surviving_2016, ehrett_e-judiciaries_2016, liao_chinese_2010, gurzick_view_2009, chandrasekharan_crossmod_2019} \\ 
Organizing & \cite{seering_metaphors_2020, vaccaro_at_2020, rajadesingan_quick_2020, feuston_conformity_2020, fan_digital_2020, juneja_through_2020, gilbert_i_2020, riedl_downsides_2020, einwiller_how_2020, banchik_disappearing_2020, schoenebeck_drawing_2020, skousen_successful_2020, gray_playing_2020, datta_extracting_2019, redmiles_i_2019,kiene_technological_2019, blackwell_harassment_2019, jhaver_does_2019, jhaver_human-machine_2019, jhaver_did_2019,srinivasan_content_2019, jiang_moderation_2019, dosono_moderation_2019, wohn_volunteer_2019, fiesler_creativity_2019, gibson_free_2019, tyler_social_2019, potts_boycotting_2019, prochazka_making_2019, squirrell_platform_2019, seering_moderator_2019, witt_rule_2019, matias_civic_2019, junestrom_online_2019, shen_discourse_2019, medeiros_picketing_2019, draper_distributed_2019, suzor_what_2019, nurik_men_2019, duguay_queer_2018, fiesler_reddit_2018, jhaver_online_2018, matias_civilservant_2018, chancellor_norms_2018, pavalanathan_mind_2018, chandrasekharan_internets_2018, gerrard_beyond_2018, west_censored_2018, keegan_evolution_2017, chancellor_multimodal_2017, blackwell_classification_2017, chandrasekharan_you_2017, seering_shaping_2017, zeng_how_2017, cheng_anyone_2017,chancellor_thyghgapp_2016, park_supporting_2016, matias_going_2016, kiene_surviving_2016, vashistha_sangeet_2015, wang_whispers_2014, petric_elements_2014, heinze_ideal_2013, sarkar_quantitative_2012, holmes_every_2011, liao_chinese_2010, gurzick_view_2009, chandrasekharan_crossmod_2019} \\
Norm setting & \cite{seering_metaphors_2020, vaccaro_at_2020, feuston_conformity_2020, fan_digital_2020, phadke_many_2020, gilbert_i_2020, einwiller_how_2020, datta_extracting_2019, grover_detecting_2019, kiene_technological_2019, jhaver_did_2019, srinivasan_content_2019, jiang_moderation_2019, wohn_volunteer_2019, gibson_free_2019, squirrell_platform_2019, seering_moderator_2019, draper_distributed_2019, nurik_men_2019, duguay_queer_2018, jhaver_online_2018, chandrasekharan_internets_2018, seering_shaping_2017, newell_user_2016, centivany_popcorn_2016, ehrett_e-judiciaries_2016, petric_elements_2014} \\
Pricing & \cite{vaccaro_at_2020} \\ \midrule

 \multicolumn{2}{c}{\textbf{Moderation Styles}} \\ \midrule
 Human vs. Automated & \cite{seering_metaphors_2020, vaccaro_at_2020, gray_playing_2020, jhaver_does_2019, jhaver_human-machine_2019, prochazka_making_2019, seering_moderator_2019, jhaver_online_2018, chancellor_norms_2018, chancellor_multimodal_2017, chandrasekharan_crossmod_2019} \\

Centralized vs. Distributed & \cite{fan_digital_2020, gilbert_i_2020, seering_moderator_2019, draper_distributed_2019, duguay_queer_2018, chandrasekharan_internets_2018, keegan_evolution_2017, park_supporting_2016, ehrett_e-judiciaries_2016, lampe_crowdsourcing_2014, kayhan_content_2013, lampe_slashdot_2004} \\
Transparent vs. Opaque & \cite{seering_metaphors_2020, vaccaro_at_2020, rajadesingan_quick_2020, fan_digital_2020, juneja_through_2020, gilbert_i_2020, jiang_bias_2019, jhaver_does_2019, jhaver_human-machine_2019, tyler_social_2019, seering_moderator_2019, witt_rule_2019, suzor_what_2019, jhaver_online_2018, chancellor_thyghgapp_2016, petric_elements_2014, sarkar_quantitative_2012} \\ \midrule
 \multicolumn{2}{c}{\textbf{Moderation Philosophies}} \\ \midrule
Nurturing vs. Punishing & \cite{seering_metaphors_2020, vaccaro_at_2020, einwiller_how_2020, skousen_successful_2020, jhaver_does_2019, seering_moderator_2019, west_censored_2018} \\
Level of Activity vs. Quality of Contributions & \cite{vaccaro_at_2020, skousen_successful_2020, jhaver_does_2019, srinivasan_content_2019, keegan_evolution_2017, kiene_surviving_2016} \\
Efficiency vs. Quality of Moderation & \cite{seering_metaphors_2020, vaccaro_at_2020, fan_digital_2020, gilbert_i_2020, jhaver_human-machine_2019, squirrell_platform_2019, seering_moderator_2019, wang_whispers_2014, lampe_slashdot_2004} \\ \midrule
 \multicolumn{2}{c}{\textbf{Moderation Values}} \\ \midrule
Moderator Identities & \cite{seering_metaphors_2020, wohn_volunteer_2019, seering_moderator_2019, matias_civic_2019, jhaver_online_2018, keegan_evolution_2017, holmes_every_2011, gurzick_view_2009} \\ 
Community Identities & \cite{seering_metaphors_2020, feuston_conformity_2020, phadke_many_2020, gilbert_i_2020, grover_detecting_2019, fiesler_creativity_2019, gibson_free_2019, squirrell_platform_2019, seering_moderator_2019, medeiros_picketing_2019, keegan_evolution_2017} \\
Competing Stakeholders & \cite{seering_metaphors_2020, vaccaro_at_2020, hua_characterizing_2020, luo_emotional_2020, riedl_downsides_2020, einwiller_how_2020, banchik_disappearing_2020, schoenebeck_drawing_2020, karunakaran_testing_2019, matias_civic_2019, shen_discourse_2019, blackwell_classification_2017, zeng_how_2017, centivany_popcorn_2016, matias_going_2016, gallagher_comparing_2016, ehrett_e-judiciaries_2016} \\ \bottomrule
\end{tabularx}
\end{table}

%% file: slr.bbl

\begin{thebibliography}{112}


\ifx \showCODEN    \undefined \def \showCODEN     #1{\unskip}     \fi
\ifx \showDOI      \undefined \def \showDOI       #1{#1}\fi
\ifx \showISBNx    \undefined \def \showISBNx     #1{\unskip}     \fi
\ifx \showISBNxiii \undefined \def \showISBNxiii  #1{\unskip}     \fi
\ifx \showISSN     \undefined \def \showISSN      #1{\unskip}     \fi
\ifx \showLCCN     \undefined \def \showLCCN      #1{\unskip}     \fi
\ifx \shownote     \undefined \def \shownote      #1{#1}          \fi
\ifx \showarticletitle \undefined \def \showarticletitle #1{#1}   \fi
\ifx \showURL      \undefined \def \showURL       {\relax}        \fi
\providecommand\bibfield[2]{#2}
\providecommand\bibinfo[2]{#2}
\providecommand\natexlab[1]{#1}
\providecommand\showeprint[2][]{arXiv:#2}

\bibitem[\protect\citeauthoryear{??}{noa}{[n.d.]}]%
        {noauthor_automoderator_nodate}
 \bibinfo{year}{[n.d.]}\natexlab{}.
\newblock \bibinfo{title}{automoderator - reddit.com}.
\newblock
\newblock
\urldef\tempurl%
\url{https://www.reddit.com/wiki/automoderator}
\showURL{%
\tempurl}


\bibitem[\protect\citeauthoryear{??}{noa}{2002}]%
        {noauthor_problems_2002}
 \bibinfo{year}{2002}\natexlab{}.
\newblock \bibinfo{booktitle}{\emph{Problems of conflict management in virtual
  communities}}.
\newblock \bibinfo{publisher}{Routledge}.
\newblock
\showISBNx{978-0-203-19495-9}
\urldef\tempurl%
\url{https://doi.org/10.4324/9780203194959-16}
\showDOI{\tempurl}
\newblock
\shownote{Pages: 145-174 Publication Title: Communities in Cyberspace.}


\bibitem[\protect\citeauthoryear{??}{noa}{2020}]%
        {noauthor_paleolithic_2020}
 \bibinfo{year}{2020}\natexlab{}.
\newblock \bibinfo{title}{Paleolithic diet}.
\newblock
\newblock
\urldef\tempurl%
\url{https://en.wikipedia.org/w/index.php?title=Paleolithic_diet&oldid=979608774}
\showURL{%
\tempurl}
\newblock
\shownote{Page Version ID: 979608774.}


\bibitem[\protect\citeauthoryear{Banchik}{Banchik}{2020}]%
        {banchik_disappearing_2020}
\bibfield{author}{\bibinfo{person}{Anna~Veronica Banchik}.}
  \bibinfo{year}{2020}\natexlab{}.
\newblock \showarticletitle{Disappearing acts: {Content} moderation and
  emergent practices to preserve at-risk human rights{\textendash}related
  content}.
\newblock \bibinfo{journal}{\emph{New Media \& Society}} (\bibinfo{date}{March}
  \bibinfo{year}{2020}), \bibinfo{pages}{1461444820912724}.
\newblock
\showISSN{1461-4448}
\urldef\tempurl%
\url{https://doi.org/10.1177/1461444820912724}
\showDOI{\tempurl}
\newblock
\shownote{Publisher: SAGE Publications.}


\bibitem[\protect\citeauthoryear{Blackwell, Chen, Schoenebeck, and
  Lampe}{Blackwell et~al\mbox{.}}{2018}]%
        {blackwell_when_2018}
\bibfield{author}{\bibinfo{person}{Lindsay Blackwell},
  \bibinfo{person}{Tianying Chen}, \bibinfo{person}{Sarita Schoenebeck}, {and}
  \bibinfo{person}{Cliff Lampe}.} \bibinfo{year}{2018}\natexlab{}.
\newblock \showarticletitle{When {Online} {Harassment} {Is} {Perceived} as
  {Justified}}. In \bibinfo{booktitle}{\emph{Twelfth {International} {AAAI}
  {Conference} on {Web} and {Social} {Media}}}.
\newblock
\urldef\tempurl%
\url{https://www.aaai.org/ocs/index.php/ICWSM/ICWSM18/paper/view/17902}
\showURL{%
\tempurl}


\bibitem[\protect\citeauthoryear{Blackwell, Dimond, Schoenebeck, and
  Lampe}{Blackwell et~al\mbox{.}}{2017}]%
        {blackwell_classification_2017}
\bibfield{author}{\bibinfo{person}{Lindsay Blackwell}, \bibinfo{person}{Jill
  Dimond}, \bibinfo{person}{Sarita Schoenebeck}, {and} \bibinfo{person}{Cliff
  Lampe}.} \bibinfo{year}{2017}\natexlab{}.
\newblock \showarticletitle{Classification and {Its} {Consequences} for
  {Online} {Harassment}: {Design} {Insights} from {HeartMob}}.
\newblock \bibinfo{journal}{\emph{Proceedings of the ACM on Human-Computer
  Interaction}} \bibinfo{volume}{1}, \bibinfo{number}{CSCW}
  (\bibinfo{date}{Dec.} \bibinfo{year}{2017}), \bibinfo{pages}{24:1--24:19}.
\newblock
\urldef\tempurl%
\url{https://doi.org/10.1145/3134659}
\showDOI{\tempurl}


\bibitem[\protect\citeauthoryear{Blackwell, Ellison, Elliott-Deflo, and
  Schwartz}{Blackwell et~al\mbox{.}}{2019}]%
        {blackwell_harassment_2019}
\bibfield{author}{\bibinfo{person}{Lindsay Blackwell}, \bibinfo{person}{Nicole
  Ellison}, \bibinfo{person}{Natasha Elliott-Deflo}, {and} \bibinfo{person}{Raz
  Schwartz}.} \bibinfo{year}{2019}\natexlab{}.
\newblock \showarticletitle{Harassment in {Social} {Virtual} {Reality}:
  {Challenges} for {Platform} {Governance}}.
\newblock \bibinfo{journal}{\emph{Proceedings of the ACM on Human-Computer
  Interaction}} \bibinfo{volume}{3}, \bibinfo{number}{CSCW}
  (\bibinfo{date}{Nov.} \bibinfo{year}{2019}), \bibinfo{pages}{100:1--100:25}.
\newblock
\urldef\tempurl%
\url{https://doi.org/10.1145/3359202}
\showDOI{\tempurl}


\bibitem[\protect\citeauthoryear{Braun and Clarke}{Braun and Clarke}{2006}]%
        {braun_using_2006}
\bibfield{author}{\bibinfo{person}{Virginia Braun} {and}
  \bibinfo{person}{Victoria Clarke}.} \bibinfo{year}{2006}\natexlab{}.
\newblock \showarticletitle{Using thematic analysis in psychology}.
\newblock \bibinfo{journal}{\emph{Qualitative Research in Psychology}}
  \bibinfo{volume}{3}, \bibinfo{number}{2} (\bibinfo{date}{Jan.}
  \bibinfo{year}{2006}), \bibinfo{pages}{77--101}.
\newblock
\showISSN{1478-0887}
\urldef\tempurl%
\url{https://doi.org/10.1191/1478088706qp063oa}
\showDOI{\tempurl}


\bibitem[\protect\citeauthoryear{Caplan}{Caplan}{2018}]%
        {caplan_content_2018}
\bibfield{author}{\bibinfo{person}{Robyn Caplan}.}
  \bibinfo{year}{2018}\natexlab{}.
\newblock \bibinfo{title}{Content or {Context} {Moderation}?}
\newblock
\newblock
\urldef\tempurl%
\url{https://datasociety.net/output/content-or-context-moderation/}
\showURL{%
\tempurl}


\bibitem[\protect\citeauthoryear{Centivany and Glushko}{Centivany and
  Glushko}{2016}]%
        {centivany_popcorn_2016}
\bibfield{author}{\bibinfo{person}{Alissa Centivany} {and}
  \bibinfo{person}{Bobby Glushko}.} \bibinfo{year}{2016}\natexlab{}.
\newblock \showarticletitle{"{Popcorn} {Tastes} {Good}": {Participatory}
  {Policymaking} and {Reddit}'s}. In \bibinfo{booktitle}{\emph{Proceedings of
  the 2016 {CHI} {Conference} on {Human} {Factors} in {Computing} {Systems}}}
  \emph{(\bibinfo{series}{{CHI} '16})}. \bibinfo{publisher}{Association for
  Computing Machinery}, \bibinfo{address}{New York, NY, USA},
  \bibinfo{pages}{1126--1137}.
\newblock
\showISBNx{978-1-4503-3362-7}
\urldef\tempurl%
\url{https://doi.org/10.1145/2858036.2858516}
\showDOI{\tempurl}


\bibitem[\protect\citeauthoryear{Chancellor, Baumer, and
  De~Choudhury}{Chancellor et~al\mbox{.}}{2019}]%
        {chancellor_who_2019}
\bibfield{author}{\bibinfo{person}{Stevie Chancellor}, \bibinfo{person}{Eric
  P.~S. Baumer}, {and} \bibinfo{person}{Munmun De~Choudhury}.}
  \bibinfo{year}{2019}\natexlab{}.
\newblock \showarticletitle{Who is the "{Human}" in {Human}-{Centered}
  {Machine} {Learning}: {The} {Case} of {Predicting} {Mental} {Health} from
  {Social} {Media}}.
\newblock \bibinfo{journal}{\emph{Proc. ACM Hum.-Comput. Interact.}}
  \bibinfo{volume}{3}, \bibinfo{number}{CSCW} (\bibinfo{date}{Nov.}
  \bibinfo{year}{2019}), \bibinfo{pages}{147:1--147:32}.
\newblock
\showISSN{2573-0142}
\urldef\tempurl%
\url{https://doi.org/10.1145/3359249}
\showDOI{\tempurl}


\bibitem[\protect\citeauthoryear{Chancellor, Hu, and De~Choudhury}{Chancellor
  et~al\mbox{.}}{2018}]%
        {chancellor_norms_2018}
\bibfield{author}{\bibinfo{person}{Stevie Chancellor}, \bibinfo{person}{Andrea
  Hu}, {and} \bibinfo{person}{Munmun De~Choudhury}.}
  \bibinfo{year}{2018}\natexlab{}.
\newblock \showarticletitle{Norms {Matter}: {Contrasting} {Social} {Support}
  {Around} {Behavior} {Change} in {Online} {Weight} {Loss} {Communities}}. In
  \bibinfo{booktitle}{\emph{Proceedings of the 2018 {CHI} {Conference} on
  {Human} {Factors} in {Computing} {Systems}}} \emph{(\bibinfo{series}{{CHI}
  '18})}. \bibinfo{publisher}{Association for Computing Machinery},
  \bibinfo{address}{Montreal QC, Canada}, \bibinfo{pages}{1--14}.
\newblock
\showISBNx{978-1-4503-5620-6}
\urldef\tempurl%
\url{https://doi.org/10.1145/3173574.3174240}
\showDOI{\tempurl}


\bibitem[\protect\citeauthoryear{Chancellor, Kalantidis, Pater, De~Choudhury,
  and Shamma}{Chancellor et~al\mbox{.}}{2017}]%
        {chancellor_multimodal_2017}
\bibfield{author}{\bibinfo{person}{Stevie Chancellor}, \bibinfo{person}{Yannis
  Kalantidis}, \bibinfo{person}{Jessica~A. Pater}, \bibinfo{person}{Munmun
  De~Choudhury}, {and} \bibinfo{person}{David~A. Shamma}.}
  \bibinfo{year}{2017}\natexlab{}.
\newblock \showarticletitle{Multimodal {Classification} of {Moderated} {Online}
  {Pro}-{Eating} {Disorder} {Content}}. In
  \bibinfo{booktitle}{\emph{Proceedings of the 2017 {CHI} {Conference} on
  {Human} {Factors} in {Computing} {Systems}}} \emph{(\bibinfo{series}{{CHI}
  '17})}. \bibinfo{publisher}{Association for Computing Machinery},
  \bibinfo{address}{Denver, Colorado, USA}, \bibinfo{pages}{3213--3226}.
\newblock
\showISBNx{978-1-4503-4655-9}
\urldef\tempurl%
\url{https://doi.org/10.1145/3025453.3025985}
\showDOI{\tempurl}


\bibitem[\protect\citeauthoryear{Chancellor, Pater, Clear, Gilbert, and
  De~Choudhury}{Chancellor et~al\mbox{.}}{2016}]%
        {chancellor_thyghgapp_2016}
\bibfield{author}{\bibinfo{person}{Stevie Chancellor},
  \bibinfo{person}{Jessica~Annette Pater}, \bibinfo{person}{Trustin Clear},
  \bibinfo{person}{Eric Gilbert}, {and} \bibinfo{person}{Munmun De~Choudhury}.}
  \bibinfo{year}{2016}\natexlab{}.
\newblock \showarticletitle{\#thyghgapp: {Instagram} {Content} {Moderation} and
  {Lexical} {Variation} in {Pro}-{Eating} {Disorder} {Communities}}. In
  \bibinfo{booktitle}{\emph{Proceedings of the 19th {ACM} {Conference} on
  {Computer}-{Supported} {Cooperative} {Work} \& {Social} {Computing}}}
  \emph{(\bibinfo{series}{{CSCW} '16})}. \bibinfo{publisher}{Association for
  Computing Machinery}, \bibinfo{address}{San Francisco, California, USA},
  \bibinfo{pages}{1201--1213}.
\newblock
\showISBNx{978-1-4503-3592-8}
\urldef\tempurl%
\url{https://doi.org/10.1145/2818048.2819963}
\showDOI{\tempurl}


\bibitem[\protect\citeauthoryear{Chandrasekharan, Gandhi, Mustelier, and
  Gilbert}{Chandrasekharan et~al\mbox{.}}{2019}]%
        {chandrasekharan_crossmod_2019}
\bibfield{author}{\bibinfo{person}{Eshwar Chandrasekharan},
  \bibinfo{person}{Chaitrali Gandhi}, \bibinfo{person}{Matthew~Wortley
  Mustelier}, {and} \bibinfo{person}{Eric Gilbert}.}
  \bibinfo{year}{2019}\natexlab{}.
\newblock \showarticletitle{Crossmod: {A} {Cross}-{Community} {Learning}-based
  {System} to {Assist} {Reddit} {Moderators}}.
\newblock \bibinfo{journal}{\emph{Proceedings of the ACM on Human-Computer
  Interaction}} \bibinfo{volume}{3}, \bibinfo{number}{CSCW}
  (\bibinfo{date}{Nov.} \bibinfo{year}{2019}), \bibinfo{pages}{174:1--174:30}.
\newblock
\urldef\tempurl%
\url{https://doi.org/10.1145/3359276}
\showDOI{\tempurl}


\bibitem[\protect\citeauthoryear{Chandrasekharan, Pavalanathan, Srinivasan,
  Glynn, Eisenstein, and Gilbert}{Chandrasekharan et~al\mbox{.}}{2017}]%
        {chandrasekharan_you_2017}
\bibfield{author}{\bibinfo{person}{Eshwar Chandrasekharan},
  \bibinfo{person}{Umashanthi Pavalanathan}, \bibinfo{person}{Anirudh
  Srinivasan}, \bibinfo{person}{Adam Glynn}, \bibinfo{person}{Jacob
  Eisenstein}, {and} \bibinfo{person}{Eric Gilbert}.}
  \bibinfo{year}{2017}\natexlab{}.
\newblock \showarticletitle{You {Can}'{T} {Stay} {Here}: {The} {Efficacy} of
  {Reddit}'s 2015 {Ban} {Examined} {Through} {Hate} {Speech}}.
\newblock \bibinfo{journal}{\emph{Proc. ACM Hum.-Comput. Interact.}}
  \bibinfo{volume}{1}, \bibinfo{number}{CSCW} (\bibinfo{date}{Dec.}
  \bibinfo{year}{2017}), \bibinfo{pages}{31:1--31:22}.
\newblock
\showISSN{2573-0142}
\urldef\tempurl%
\url{https://doi.org/10.1145/3134666}
\showDOI{\tempurl}


\bibitem[\protect\citeauthoryear{Chandrasekharan, Samory, Jhaver, Charvat,
  Bruckman, Lampe, Eisenstein, and Gilbert}{Chandrasekharan
  et~al\mbox{.}}{2018}]%
        {chandrasekharan_internets_2018}
\bibfield{author}{\bibinfo{person}{Eshwar Chandrasekharan},
  \bibinfo{person}{Mattia Samory}, \bibinfo{person}{Shagun Jhaver},
  \bibinfo{person}{Hunter Charvat}, \bibinfo{person}{Amy Bruckman},
  \bibinfo{person}{Cliff Lampe}, \bibinfo{person}{Jacob Eisenstein}, {and}
  \bibinfo{person}{Eric Gilbert}.} \bibinfo{year}{2018}\natexlab{}.
\newblock \showarticletitle{The {Internet}'s {Hidden} {Rules}: {An} {Empirical}
  {Study} of {Reddit} {Norm} {Violations} at {Micro}, {Meso}, and {Macro}
  {Scales}}.
\newblock \bibinfo{journal}{\emph{Proc. ACM Hum.-Comput. Interact.}}
  \bibinfo{volume}{2}, \bibinfo{number}{CSCW} (\bibinfo{date}{Nov.}
  \bibinfo{year}{2018}), \bibinfo{pages}{Article 32}.
\newblock
\showISSN{2573-0142}
\urldef\tempurl%
\url{https://doi.org/10.1145/3274301}
\showDOI{\tempurl}


\bibitem[\protect\citeauthoryear{Cheng, Bernstein, Danescu-Niculescu-Mizil, and
  Leskovec}{Cheng et~al\mbox{.}}{2017}]%
        {cheng_anyone_2017}
\bibfield{author}{\bibinfo{person}{Justin Cheng}, \bibinfo{person}{Michael
  Bernstein}, \bibinfo{person}{Cristian Danescu-Niculescu-Mizil}, {and}
  \bibinfo{person}{Jure Leskovec}.} \bibinfo{year}{2017}\natexlab{}.
\newblock \showarticletitle{Anyone {Can} {Become} a {Troll}: {Causes} of
  {Trolling} {Behavior} in {Online} {Discussions}}. In
  \bibinfo{booktitle}{\emph{Proceedings of the 2017 {ACM} {Conference} on
  {Computer} {Supported} {Cooperative} {Work} and {Social} {Computing}}}
  \emph{(\bibinfo{series}{{CSCW} '17})}. \bibinfo{publisher}{Association for
  Computing Machinery}, \bibinfo{address}{New York, NY, USA},
  \bibinfo{pages}{1217--1230}.
\newblock
\showISBNx{978-1-4503-4335-0}
\urldef\tempurl%
\url{https://doi.org/10.1145/2998181.2998213}
\showDOI{\tempurl}


\bibitem[\protect\citeauthoryear{Datta and Adar}{Datta and Adar}{2019}]%
        {datta_extracting_2019}
\bibfield{author}{\bibinfo{person}{Srayan Datta} {and} \bibinfo{person}{Eytan
  Adar}.} \bibinfo{year}{2019}\natexlab{}.
\newblock \showarticletitle{Extracting {Inter}-{Community} {Conflicts} in
  {Reddit}}.
\newblock \bibinfo{journal}{\emph{Proceedings of the International AAAI
  Conference on Web and Social Media}}  \bibinfo{volume}{13}
  (\bibinfo{date}{July} \bibinfo{year}{2019}), \bibinfo{pages}{146--157}.
\newblock
\showISSN{2334-0770}
\urldef\tempurl%
\url{https://www.aaai.org/ojs/index.php/ICWSM/article/view/3217}
\showURL{%
\tempurl}


\bibitem[\protect\citeauthoryear{DiSalvo, Sengers, and
  Brynjarsd{\'o}ttir}{DiSalvo et~al\mbox{.}}{2010}]%
        {disalvo_mapping_2010}
\bibfield{author}{\bibinfo{person}{Carl DiSalvo}, \bibinfo{person}{Phoebe
  Sengers}, {and} \bibinfo{person}{Hr{\"o}nn Brynjarsd{\'o}ttir}.}
  \bibinfo{year}{2010}\natexlab{}.
\newblock \showarticletitle{Mapping the {Landscape} of {Sustainable} {HCI}}. In
  \bibinfo{booktitle}{\emph{Proceedings of the {SIGCHI} {Conference} on {Human}
  {Factors} in {Computing} {Systems}}} \emph{(\bibinfo{series}{{CHI} '10})}.
  \bibinfo{publisher}{ACM}, \bibinfo{address}{New York, NY, USA},
  \bibinfo{pages}{1975--1984}.
\newblock
\showISBNx{978-1-60558-929-9}
\urldef\tempurl%
\url{https://doi.org/10.1145/1753326.1753625}
\showDOI{\tempurl}
\newblock
\shownote{event-place: Atlanta, Georgia, USA.}


\bibitem[\protect\citeauthoryear{Dosono and Semaan}{Dosono and Semaan}{2019}]%
        {dosono_moderation_2019}
\bibfield{author}{\bibinfo{person}{Bryan Dosono} {and} \bibinfo{person}{Bryan
  Semaan}.} \bibinfo{year}{2019}\natexlab{}.
\newblock \showarticletitle{Moderation {Practices} {As} {Emotional} {Labor} in
  {Sustaining} {Online} {Communities}: {The} {Case} of {AAPI} {Identity} {Work}
  on {Reddit}}. In \bibinfo{booktitle}{\emph{Proceedings of the 2019 {CHI}
  {Conference} on {Human} {Factors} in {Computing} {Systems}}}
  \emph{(\bibinfo{series}{{CHI} '19})}. \bibinfo{publisher}{ACM},
  \bibinfo{address}{New York, NY, USA}, \bibinfo{pages}{142:1--142:13}.
\newblock
\showISBNx{978-1-4503-5970-2}
\urldef\tempurl%
\url{https://doi.org/10.1145/3290605.3300372}
\showDOI{\tempurl}
\newblock
\shownote{event-place: Glasgow, Scotland Uk.}


\bibitem[\protect\citeauthoryear{Draper}{Draper}{2019}]%
        {draper_distributed_2019}
\bibfield{author}{\bibinfo{person}{Nora~A. Draper}.}
  \bibinfo{year}{2019}\natexlab{}.
\newblock \showarticletitle{Distributed intervention: networked content
  moderation in anonymous mobile spaces}.
\newblock \bibinfo{journal}{\emph{Feminist Media Studies}}
  \bibinfo{volume}{19}, \bibinfo{number}{5} (\bibinfo{date}{July}
  \bibinfo{year}{2019}), \bibinfo{pages}{667--683}.
\newblock
\showISSN{1468-0777}
\urldef\tempurl%
\url{https://doi.org/10.1080/14680777.2018.1458746}
\showDOI{\tempurl}
\newblock
\shownote{Publisher: Routledge \_eprint:
  https://doi.org/10.1080/14680777.2018.1458746.}


\bibitem[\protect\citeauthoryear{Duguay, Burgess, and Suzor}{Duguay
  et~al\mbox{.}}{2018}]%
        {duguay_queer_2018}
\bibfield{author}{\bibinfo{person}{Stefanie Duguay}, \bibinfo{person}{Jean
  Burgess}, {and} \bibinfo{person}{Nicolas Suzor}.}
  \bibinfo{year}{2018}\natexlab{}.
\newblock \showarticletitle{Queer women{\textquoteright}s experiences of
  patchwork platform governance on {Tinder}, {Instagram}, and {Vine}:}.
\newblock \bibinfo{journal}{\emph{Convergence}} (\bibinfo{date}{June}
  \bibinfo{year}{2018}).
\newblock
\urldef\tempurl%
\url{https://doi.org/10.1177/1354856518781530}
\showDOI{\tempurl}
\newblock
\shownote{Publisher: SAGE PublicationsSage UK: London, England.}


\bibitem[\protect\citeauthoryear{Ehrett}{Ehrett}{2016}]%
        {ehrett_e-judiciaries_2016}
\bibfield{author}{\bibinfo{person}{John~S. Ehrett}.}
  \bibinfo{year}{2016}\natexlab{}.
\newblock \showarticletitle{E-judiciaries: a model for community policing in
  cyberspace}.
\newblock \bibinfo{journal}{\emph{Information \& Communications Technology
  Law}} \bibinfo{volume}{25}, \bibinfo{number}{3} (\bibinfo{date}{Sept.}
  \bibinfo{year}{2016}), \bibinfo{pages}{272--291}.
\newblock
\showISSN{1360-0834}
\urldef\tempurl%
\url{https://doi.org/10.1080/13600834.2016.1236428}
\showDOI{\tempurl}
\newblock
\shownote{Publisher: Routledge \_eprint:
  https://doi.org/10.1080/13600834.2016.1236428.}


\bibitem[\protect\citeauthoryear{Einwiller and Kim}{Einwiller and Kim}{2020}]%
        {einwiller_how_2020}
\bibfield{author}{\bibinfo{person}{Sabine~A. Einwiller} {and}
  \bibinfo{person}{Sora Kim}.} \bibinfo{year}{2020}\natexlab{}.
\newblock \showarticletitle{How {Online} {Content} {Providers} {Moderate}
  {User}-{Generated} {Content} to {Prevent} {Harmful} {Online} {Communication}:
  {An} {Analysis} of {Policies} and {Their} {Implementation}}.
\newblock \bibinfo{journal}{\emph{Policy \& Internet}} \bibinfo{volume}{12},
  \bibinfo{number}{2} (\bibinfo{year}{2020}), \bibinfo{pages}{184--206}.
\newblock
\showISSN{1944-2866}
\urldef\tempurl%
\url{https://doi.org/10.1002/poi3.239}
\showDOI{\tempurl}
\newblock
\shownote{\_eprint: https://onlinelibrary.wiley.com/doi/pdf/10.1002/poi3.239.}


\bibitem[\protect\citeauthoryear{Facebook}{Facebook}{[n.d.]}]%
        {facebook_community_nodate}
\bibfield{author}{\bibinfo{person}{Facebook}.}
  \bibinfo{year}{[n.d.]}\natexlab{}.
\newblock \bibinfo{title}{Community {Standards}}.
\newblock
\newblock
\urldef\tempurl%
\url{https://www.facebook.com/communitystandards/}
\showURL{%
\tempurl}


\bibitem[\protect\citeauthoryear{Fan and Zhang}{Fan and Zhang}{2020}]%
        {fan_digital_2020}
\bibfield{author}{\bibinfo{person}{Jenny Fan} {and} \bibinfo{person}{Amy~X.
  Zhang}.} \bibinfo{year}{2020}\natexlab{}.
\newblock \showarticletitle{Digital {Juries}: {A} {Civics}-{Oriented}
  {Approach} to {Platform} {Governance}}. In
  \bibinfo{booktitle}{\emph{Proceedings of the 2020 {CHI} {Conference} on
  {Human} {Factors} in {Computing} {Systems}}} \emph{(\bibinfo{series}{{CHI}
  '20})}. \bibinfo{publisher}{Association for Computing Machinery},
  \bibinfo{address}{Honolulu, HI, USA}, \bibinfo{pages}{1--14}.
\newblock
\showISBNx{978-1-4503-6708-0}
\urldef\tempurl%
\url{https://doi.org/10.1145/3313831.3376293}
\showDOI{\tempurl}


\bibitem[\protect\citeauthoryear{Feuston, Taylor, and Piper}{Feuston
  et~al\mbox{.}}{2020}]%
        {feuston_conformity_2020}
\bibfield{author}{\bibinfo{person}{Jessica~L. Feuston},
  \bibinfo{person}{Alex~S. Taylor}, {and} \bibinfo{person}{Anne~Marie Piper}.}
  \bibinfo{year}{2020}\natexlab{}.
\newblock \showarticletitle{Conformity of {Eating} {Disorders} through
  {Content} {Moderation}}.
\newblock \bibinfo{journal}{\emph{Proceedings of the ACM on Human-Computer
  Interaction}} \bibinfo{volume}{4}, \bibinfo{number}{CSCW1}
  (\bibinfo{date}{May} \bibinfo{year}{2020}), \bibinfo{pages}{040:1--040:28}.
\newblock
\urldef\tempurl%
\url{https://doi.org/10.1145/3392845}
\showDOI{\tempurl}


\bibitem[\protect\citeauthoryear{Fiesler and Bruckman}{Fiesler and
  Bruckman}{2019}]%
        {fiesler_creativity_2019}
\bibfield{author}{\bibinfo{person}{Casey Fiesler} {and} \bibinfo{person}{Amy~S.
  Bruckman}.} \bibinfo{year}{2019}\natexlab{}.
\newblock \showarticletitle{Creativity, {Copyright}, and {Close}-{Knit}
  {Communities}: {A} {Case} {Study} of {Social} {Norm} {Formation} and
  {Enforcement}}.
\newblock \bibinfo{journal}{\emph{Proceedings of the ACM on Human-Computer
  Interaction}} \bibinfo{volume}{3}, \bibinfo{number}{GROUP}
  (\bibinfo{date}{Dec.} \bibinfo{year}{2019}), \bibinfo{pages}{241:1--241:24}.
\newblock
\urldef\tempurl%
\url{https://doi.org/10.1145/3361122}
\showDOI{\tempurl}


\bibitem[\protect\citeauthoryear{Fiesler, Jiang, McCann, Frye, and
  Brubaker}{Fiesler et~al\mbox{.}}{2018}]%
        {fiesler_reddit_2018}
\bibfield{author}{\bibinfo{person}{Casey Fiesler},
  \bibinfo{person}{Jialun~"Aaron" Jiang}, \bibinfo{person}{Joshua McCann},
  \bibinfo{person}{Kyle Frye}, {and} \bibinfo{person}{Jed~R. Brubaker}.}
  \bibinfo{year}{2018}\natexlab{}.
\newblock \showarticletitle{Reddit {Rules}! {Characterizing} an {Ecosystem} of
  {Governance}}. In \bibinfo{booktitle}{\emph{Twelfth {International} {AAAI}
  {Conference} on {Web} and {Social} {Media}}}.
\newblock
\urldef\tempurl%
\url{https://aaai.org/ocs/index.php/ICWSM/ICWSM18/paper/view/17898}
\showURL{%
\tempurl}


\bibitem[\protect\citeauthoryear{Gallagher and Savage}{Gallagher and
  Savage}{2016}]%
        {gallagher_comparing_2016}
\bibfield{author}{\bibinfo{person}{Silvia~Elena Gallagher} {and}
  \bibinfo{person}{Timothy Savage}.} \bibinfo{year}{2016}\natexlab{}.
\newblock \showarticletitle{Comparing learner community behavior in multiple
  presentations of a {Massive} {Open} {Online} {Course}}.
\newblock \bibinfo{journal}{\emph{Journal of Computing in Higher Education}}
  \bibinfo{volume}{28}, \bibinfo{number}{3} (\bibinfo{date}{Dec.}
  \bibinfo{year}{2016}), \bibinfo{pages}{358--369}.
\newblock
\showISSN{1867-1233}
\urldef\tempurl%
\url{https://doi.org/10.1007/s12528-016-9124-y}
\showDOI{\tempurl}


\bibitem[\protect\citeauthoryear{Gerrard}{Gerrard}{2018}]%
        {gerrard_beyond_2018}
\bibfield{author}{\bibinfo{person}{Ysabel Gerrard}.}
  \bibinfo{year}{2018}\natexlab{}.
\newblock \showarticletitle{Beyond the hashtag: {Circumventing} content
  moderation on social media}.
\newblock \bibinfo{journal}{\emph{New Media \& Society}} \bibinfo{volume}{20},
  \bibinfo{number}{12} (\bibinfo{date}{Dec.} \bibinfo{year}{2018}),
  \bibinfo{pages}{4492--4511}.
\newblock
\showISSN{1461-4448}
\urldef\tempurl%
\url{https://doi.org/10.1177/1461444818776611}
\showDOI{\tempurl}


\bibitem[\protect\citeauthoryear{Getto and Labriola}{Getto and
  Labriola}{2016}]%
        {getto_ifixit_2016}
\bibfield{author}{\bibinfo{person}{Guiseppe Getto} {and}
  \bibinfo{person}{Jack~T. Labriola}.} \bibinfo{year}{2016}\natexlab{}.
\newblock \showarticletitle{{iFixit} {Myself}: {User}-{Generated} {Content}
  {Strategy} in {\textquotedblleft}{The} {Free} {Repair} {Guide} for
  {Everything}{\textquotedblright}}.
\newblock \bibinfo{journal}{\emph{IEEE Transactions on Professional
  Communication}} \bibinfo{volume}{59}, \bibinfo{number}{1}
  (\bibinfo{date}{March} \bibinfo{year}{2016}), \bibinfo{pages}{37--55}.
\newblock
\showISSN{1558-1500}
\urldef\tempurl%
\url{https://doi.org/10.1109/TPC.2016.2527259}
\showDOI{\tempurl}
\newblock
\shownote{Conference Name: IEEE Transactions on Professional Communication.}


\bibitem[\protect\citeauthoryear{Gibson}{Gibson}{2019}]%
        {gibson_free_2019}
\bibfield{author}{\bibinfo{person}{Anna Gibson}.}
  \bibinfo{year}{2019}\natexlab{}.
\newblock \showarticletitle{Free {Speech} and {Safe} {Spaces}: {How}
  {Moderation} {Policies} {Shape} {Online} {Discussion} {Spaces}:}.
\newblock \bibinfo{journal}{\emph{Social Media + Society}}
  (\bibinfo{date}{March} \bibinfo{year}{2019}).
\newblock
\urldef\tempurl%
\url{https://doi.org/10.1177/2056305119832588}
\showDOI{\tempurl}
\newblock
\shownote{Publisher: SAGE PublicationsSage UK: London, England.}


\bibitem[\protect\citeauthoryear{Gilbert}{Gilbert}{2020}]%
        {gilbert_i_2020}
\bibfield{author}{\bibinfo{person}{Sarah~A. Gilbert}.}
  \bibinfo{year}{2020}\natexlab{}.
\newblock \showarticletitle{"{I} run the world's largest historical outreach
  project and it's on a cesspool of a website." {Moderating} a {Public}
  {Scholarship} {Site} on {Reddit}: {A} {Case} {Study} of r/{AskHistorians}}.
\newblock \bibinfo{journal}{\emph{Proceedings of the ACM on Human-Computer
  Interaction}} \bibinfo{volume}{4}, \bibinfo{number}{CSCW1}
  (\bibinfo{date}{May} \bibinfo{year}{2020}), \bibinfo{pages}{019:1--019:27}.
\newblock
\urldef\tempurl%
\url{https://doi.org/10.1145/3392822}
\showDOI{\tempurl}


\bibitem[\protect\citeauthoryear{Gillespie}{Gillespie}{2018}]%
        {gillespie_custodians_2018}
\bibfield{author}{\bibinfo{person}{Tarleton Gillespie}.}
  \bibinfo{year}{2018}\natexlab{}.
\newblock \bibinfo{booktitle}{\emph{Custodians of the {Internet}: {Platforms},
  {Content} {Moderation}, and the {Hidden} {Decisions} that {Shape} {Social}
  {Media}}}.
\newblock \bibinfo{publisher}{Yale University Press}.
\newblock
\showISBNx{978-0-300-17313-0}


\bibitem[\protect\citeauthoryear{Gillespie}{Gillespie}{2019}]%
        {gillespie_content_2019}
\bibfield{author}{\bibinfo{person}{Tarleton Gillespie}.}
  \bibinfo{year}{2019}\natexlab{}.
\newblock \bibinfo{title}{Content {Moderation}: {A} {Reading} {List}}.
\newblock
\newblock
\urldef\tempurl%
\url{https://socialmediacollective.org/reading-lists/content-moderation-reading-list/}
\showURL{%
\tempurl}


\bibitem[\protect\citeauthoryear{Gorwa, Binns, and Katzenbach}{Gorwa
  et~al\mbox{.}}{2020}]%
        {gorwa_algorithmic_2020}
\bibfield{author}{\bibinfo{person}{Robert Gorwa}, \bibinfo{person}{Reuben
  Binns}, {and} \bibinfo{person}{Christian Katzenbach}.}
  \bibinfo{year}{2020}\natexlab{}.
\newblock \showarticletitle{Algorithmic content moderation: {Technical} and
  political challenges in the automation of platform governance}.
\newblock \bibinfo{journal}{\emph{Big Data \& Society}} \bibinfo{volume}{7},
  \bibinfo{number}{1} (\bibinfo{date}{Jan.} \bibinfo{year}{2020}),
  \bibinfo{pages}{2053951719897945}.
\newblock
\showISSN{2053-9517}
\urldef\tempurl%
\url{https://doi.org/10.1177/2053951719897945}
\showDOI{\tempurl}
\newblock
\shownote{Publisher: SAGE Publications Ltd.}


\bibitem[\protect\citeauthoryear{Gray and Suzor}{Gray and Suzor}{2020}]%
        {gray_playing_2020}
\bibfield{author}{\bibinfo{person}{Joanne~E Gray} {and}
  \bibinfo{person}{Nicolas~P Suzor}.} \bibinfo{year}{2020}\natexlab{}.
\newblock \showarticletitle{Playing with machines: {Using} machine learning to
  understand automated copyright enforcement at scale}.
\newblock \bibinfo{journal}{\emph{Big Data \& Society}} \bibinfo{volume}{7},
  \bibinfo{number}{1} (\bibinfo{date}{Jan.} \bibinfo{year}{2020}),
  \bibinfo{pages}{2053951720919963}.
\newblock
\showISSN{2053-9517}
\urldef\tempurl%
\url{https://doi.org/10.1177/2053951720919963}
\showDOI{\tempurl}
\newblock
\shownote{Publisher: SAGE Publications Ltd.}


\bibitem[\protect\citeauthoryear{Grimmelmann}{Grimmelmann}{2015}]%
        {grimmelmann_virtues_2015}
\bibfield{author}{\bibinfo{person}{James Grimmelmann}.}
  \bibinfo{year}{2015}\natexlab{}.
\newblock \showarticletitle{The {Virtues} of {Moderation}}.
\newblock \bibinfo{journal}{\emph{Yale Journal of Law and Technology}}
  \bibinfo{volume}{17}, \bibinfo{number}{1} (\bibinfo{date}{Sept.}
  \bibinfo{year}{2015}).
\newblock
\urldef\tempurl%
\url{https://digitalcommons.law.yale.edu/yjolt/vol17/iss1/2}
\showURL{%
\tempurl}


\bibitem[\protect\citeauthoryear{Group}{Group}{1981}]%
        {gengle_communitree_1981}
\bibfield{author}{\bibinfo{person}{The~CommuniTree Group}.}
  \bibinfo{year}{1981}\natexlab{}.
\newblock \bibinfo{title}{Communitree}.
\newblock
\newblock


\bibitem[\protect\citeauthoryear{Grover and Mark}{Grover and Mark}{2019}]%
        {grover_detecting_2019}
\bibfield{author}{\bibinfo{person}{Ted Grover} {and} \bibinfo{person}{Gloria
  Mark}.} \bibinfo{year}{2019}\natexlab{}.
\newblock \showarticletitle{Detecting {Potential} {Warning} {Behaviors} of
  {Ideological} {Radicalization} in an {Alt}-{Right} {Subreddit}}.
\newblock \bibinfo{journal}{\emph{Proceedings of the International AAAI
  Conference on Web and Social Media}}  \bibinfo{volume}{13}
  (\bibinfo{date}{July} \bibinfo{year}{2019}), \bibinfo{pages}{193--204}.
\newblock
\showISSN{2334-0770}
\urldef\tempurl%
\url{https://www.aaai.org/ojs/index.php/ICWSM/article/view/3221}
\showURL{%
\tempurl}


\bibitem[\protect\citeauthoryear{Gurzick, White, Lutters, and Boot}{Gurzick
  et~al\mbox{.}}{2009}]%
        {gurzick_view_2009}
\bibfield{author}{\bibinfo{person}{David Gurzick}, \bibinfo{person}{Kevin~F.
  White}, \bibinfo{person}{Wayne~G. Lutters}, {and} \bibinfo{person}{Lee
  Boot}.} \bibinfo{year}{2009}\natexlab{}.
\newblock \showarticletitle{A view from {Mount} {Olympus}: the impact of
  activity tracking tools on the character and practice of moderation}. In
  \bibinfo{booktitle}{\emph{Proceedings of the {ACM} 2009 international
  conference on {Supporting} group work}} \emph{(\bibinfo{series}{{GROUP}
  '09})}. \bibinfo{publisher}{Association for Computing Machinery},
  \bibinfo{address}{Sanibel Island, Florida, USA}, \bibinfo{pages}{361--370}.
\newblock
\showISBNx{978-1-60558-500-0}
\urldef\tempurl%
\url{https://doi.org/10.1145/1531674.1531727}
\showDOI{\tempurl}


\bibitem[\protect\citeauthoryear{Heinze, Ferneley, and Child}{Heinze
  et~al\mbox{.}}{2013}]%
        {heinze_ideal_2013}
\bibfield{author}{\bibinfo{person}{Aleksej Heinze}, \bibinfo{person}{Elaine
  Ferneley}, {and} \bibinfo{person}{Paul Child}.}
  \bibinfo{year}{2013}\natexlab{}.
\newblock \showarticletitle{Ideal {Participants} in {Online} {Market}
  {Research}: {Lessons} from {Closed} {Communities}:}.
\newblock \bibinfo{journal}{\emph{International Journal of Market Research}}
  (\bibinfo{date}{Nov.} \bibinfo{year}{2013}).
\newblock
\urldef\tempurl%
\url{https://doi.org/10.2501/IJMR-2013-066}
\showDOI{\tempurl}
\newblock
\shownote{Publisher: SAGE PublicationsSage UK: London, England.}


\bibitem[\protect\citeauthoryear{Herring, Job-Sluder, Scheckler, and
  Barab}{Herring et~al\mbox{.}}{2002}]%
        {herring_searching_2002}
\bibfield{author}{\bibinfo{person}{Susan Herring}, \bibinfo{person}{Kirk
  Job-Sluder}, \bibinfo{person}{Rebecca Scheckler}, {and}
  \bibinfo{person}{Sasha Barab}.} \bibinfo{year}{2002}\natexlab{}.
\newblock \showarticletitle{Searching for {Safety} {Online}: {Managing}
  "{Trolling}" in a {Feminist} {Forum}}.
\newblock \bibinfo{journal}{\emph{The Information Society}}
  \bibinfo{volume}{18}, \bibinfo{number}{5} (\bibinfo{date}{Oct.}
  \bibinfo{year}{2002}), \bibinfo{pages}{371--384}.
\newblock
\showISSN{0197-2243}
\urldef\tempurl%
\url{https://doi.org/10.1080/01972240290108186}
\showDOI{\tempurl}
\newblock
\shownote{Publisher: Routledge \_eprint:
  https://doi.org/10.1080/01972240290108186.}


\bibitem[\protect\citeauthoryear{Holmes and Cox}{Holmes and Cox}{2011}]%
        {holmes_every_2011}
\bibfield{author}{\bibinfo{person}{Paul Holmes} {and}
  \bibinfo{person}{Andrew~M. Cox}.} \bibinfo{year}{2011}\natexlab{}.
\newblock \showarticletitle{'{Every} group carries the flavour of the admins':
  leadership on {Flickr}}.
\newblock \bibinfo{journal}{\emph{International Journal of Web Based
  Communities}} \bibinfo{volume}{7}, \bibinfo{number}{3} (\bibinfo{date}{July}
  \bibinfo{year}{2011}), \bibinfo{pages}{376--391}.
\newblock
\showISSN{1477-8394}
\urldef\tempurl%
\url{https://doi.org/10.1504/IJWBC.2011.041205}
\showDOI{\tempurl}


\bibitem[\protect\citeauthoryear{Hua, Naaman, and Ristenpart}{Hua
  et~al\mbox{.}}{2020}]%
        {hua_characterizing_2020}
\bibfield{author}{\bibinfo{person}{Yiqing Hua}, \bibinfo{person}{Mor Naaman},
  {and} \bibinfo{person}{Thomas Ristenpart}.} \bibinfo{year}{2020}\natexlab{}.
\newblock \showarticletitle{Characterizing {Twitter} {Users} {Who} {Engage} in
  {Adversarial} {Interactions} against {Political} {Candidates}}. In
  \bibinfo{booktitle}{\emph{Proceedings of the 2020 {CHI} {Conference} on
  {Human} {Factors} in {Computing} {Systems}}} \emph{(\bibinfo{series}{{CHI}
  '20})}. \bibinfo{publisher}{Association for Computing Machinery},
  \bibinfo{address}{Honolulu, HI, USA}, \bibinfo{pages}{1--13}.
\newblock
\showISBNx{978-1-4503-6708-0}
\urldef\tempurl%
\url{https://doi.org/10.1145/3313831.3376548}
\showDOI{\tempurl}


\bibitem[\protect\citeauthoryear{Jhaver, Appling, Gilbert, and Bruckman}{Jhaver
  et~al\mbox{.}}{2019a}]%
        {jhaver_did_2019}
\bibfield{author}{\bibinfo{person}{Shagun Jhaver},
  \bibinfo{person}{Darren~Scott Appling}, \bibinfo{person}{Eric Gilbert}, {and}
  \bibinfo{person}{Amy Bruckman}.} \bibinfo{year}{2019}\natexlab{a}.
\newblock \showarticletitle{"{Did} {You} {Suspect} the {Post} {Would} be
  {Removed}?": {Understanding} {User} {Reactions} to {Content} {Removals} on
  {Reddit}}.
\newblock \bibinfo{journal}{\emph{Proceedings of the ACM on Human-Computer
  Interaction}} \bibinfo{volume}{3}, \bibinfo{number}{CSCW}
  (\bibinfo{date}{Nov.} \bibinfo{year}{2019}), \bibinfo{pages}{192:1--192:33}.
\newblock
\urldef\tempurl%
\url{https://doi.org/10.1145/3359294}
\showDOI{\tempurl}


\bibitem[\protect\citeauthoryear{Jhaver, Birman, Gilbert, and Bruckman}{Jhaver
  et~al\mbox{.}}{2019b}]%
        {jhaver_human-machine_2019}
\bibfield{author}{\bibinfo{person}{Shagun Jhaver}, \bibinfo{person}{Iris
  Birman}, \bibinfo{person}{Eric Gilbert}, {and} \bibinfo{person}{Amy
  Bruckman}.} \bibinfo{year}{2019}\natexlab{b}.
\newblock \showarticletitle{Human-{Machine} {Collaboration} for {Content}
  {Regulation}: {The} {Case} of {Reddit} {Automoderator}}.
\newblock \bibinfo{journal}{\emph{ACM Trans. Comput.-Hum. Interact.}}
  \bibinfo{volume}{26}, \bibinfo{number}{5} (\bibinfo{date}{July}
  \bibinfo{year}{2019}), \bibinfo{pages}{Article 31}.
\newblock
\showISSN{1073-0516}
\urldef\tempurl%
\url{https://doi.org/10.1145/3338243}
\showDOI{\tempurl}


\bibitem[\protect\citeauthoryear{Jhaver, Bruckman, and Gilbert}{Jhaver
  et~al\mbox{.}}{2019c}]%
        {jhaver_does_2019}
\bibfield{author}{\bibinfo{person}{Shagun Jhaver}, \bibinfo{person}{Amy
  Bruckman}, {and} \bibinfo{person}{Eric Gilbert}.}
  \bibinfo{year}{2019}\natexlab{c}.
\newblock \showarticletitle{Does {Transparency} in {Moderation} {Really}
  {Matter}? {User} {Behavior} {After} {Content} {Removal} {Explanations} on
  {Reddit}}.
\newblock \bibinfo{journal}{\emph{Proceedings of the ACM on Human-Computer
  Interaction}} \bibinfo{volume}{3}, \bibinfo{number}{CSCW}
  (\bibinfo{date}{Nov.} \bibinfo{year}{2019}), \bibinfo{pages}{150:1--150:27}.
\newblock
\urldef\tempurl%
\url{https://doi.org/10.1145/3359252}
\showDOI{\tempurl}


\bibitem[\protect\citeauthoryear{Jhaver, Ghoshal, Bruckman, and Gilbert}{Jhaver
  et~al\mbox{.}}{2018}]%
        {jhaver_online_2018}
\bibfield{author}{\bibinfo{person}{Shagun Jhaver}, \bibinfo{person}{Sucheta
  Ghoshal}, \bibinfo{person}{Amy Bruckman}, {and} \bibinfo{person}{Eric
  Gilbert}.} \bibinfo{year}{2018}\natexlab{}.
\newblock \showarticletitle{Online {Harassment} and {Content} {Moderation}:
  {The} {Case} of {Blocklists}}.
\newblock \bibinfo{journal}{\emph{ACM Transactions on Computer-Human
  Interaction}} \bibinfo{volume}{25}, \bibinfo{number}{2}
  (\bibinfo{date}{March} \bibinfo{year}{2018}), \bibinfo{pages}{12:1--12:33}.
\newblock
\showISSN{1073-0516}
\urldef\tempurl%
\url{https://doi.org/10.1145/3185593}
\showDOI{\tempurl}


\bibitem[\protect\citeauthoryear{Jiang, Fiesler, and Brubaker}{Jiang
  et~al\mbox{.}}{2018}]%
        {jiang_perfect_2018}
\bibfield{author}{\bibinfo{person}{Jialun~"Aaron" Jiang},
  \bibinfo{person}{Casey Fiesler}, {and} \bibinfo{person}{Jed~R. Brubaker}.}
  \bibinfo{year}{2018}\natexlab{}.
\newblock \showarticletitle{"{The} {Perfect} {One}": {Understanding}
  {Communication} {Practices} and {Challenges} with {Animated} {GIFs}}.
\newblock \bibinfo{journal}{\emph{Proc. ACM Hum.-Comput. Interact.}}
  \bibinfo{volume}{2}, \bibinfo{number}{CSCW} (\bibinfo{year}{2018}),
  \bibinfo{pages}{Article 80}.
\newblock
\urldef\tempurl%
\url{https://doi.org/10.1145/3274349}
\showDOI{\tempurl}


\bibitem[\protect\citeauthoryear{Jiang, Kiene, Middler, Brubaker, and
  Fiesler}{Jiang et~al\mbox{.}}{2019a}]%
        {jiang_moderation_2019}
\bibfield{author}{\bibinfo{person}{Jialun~"Aaron" Jiang},
  \bibinfo{person}{Charles Kiene}, \bibinfo{person}{Skyler Middler},
  \bibinfo{person}{Jed~R. Brubaker}, {and} \bibinfo{person}{Casey Fiesler}.}
  \bibinfo{year}{2019}\natexlab{a}.
\newblock \showarticletitle{Moderation {Challenges} in {Voice}-based {Online}
  {Communities} on {Discord}}.
\newblock \bibinfo{journal}{\emph{Proc. ACM Hum.-Comput. Interact.}}
  \bibinfo{volume}{3}, \bibinfo{number}{CSCW} (\bibinfo{date}{Nov.}
  \bibinfo{year}{2019}), \bibinfo{pages}{55:1--55:23}.
\newblock
\showISSN{2573-0142}
\urldef\tempurl%
\url{https://doi.org/10.1145/3359157}
\showDOI{\tempurl}


\bibitem[\protect\citeauthoryear{Jiang, Middler, Brubaker, and Fiesler}{Jiang
  et~al\mbox{.}}{2020}]%
        {jiang_characterizing_2020}
\bibfield{author}{\bibinfo{person}{Jialun~"Aaron" Jiang},
  \bibinfo{person}{Skyler Middler}, \bibinfo{person}{Jed~R. Brubaker}, {and}
  \bibinfo{person}{Casey Fiesler}.} \bibinfo{year}{2020}\natexlab{}.
\newblock \showarticletitle{Characterizing {Community} {Guidelines} on {Social}
  {Media} {Platforms}}. In \bibinfo{booktitle}{\emph{{CSCW} 2020 {Companion}}}.
\newblock


\bibitem[\protect\citeauthoryear{Jiang, Robertson, and Wilson}{Jiang
  et~al\mbox{.}}{2019b}]%
        {jiang_bias_2019}
\bibfield{author}{\bibinfo{person}{Shan Jiang}, \bibinfo{person}{Ronald~E.
  Robertson}, {and} \bibinfo{person}{Christo Wilson}.}
  \bibinfo{year}{2019}\natexlab{b}.
\newblock \showarticletitle{Bias {Misperceived}:{The} {Role} of {Partisanship}
  and {Misinformation} in {YouTube} {Comment} {Moderation}}.
\newblock \bibinfo{journal}{\emph{Proceedings of the International AAAI
  Conference on Web and Social Media}}  \bibinfo{volume}{13}
  (\bibinfo{date}{July} \bibinfo{year}{2019}), \bibinfo{pages}{278--289}.
\newblock
\showISSN{2334-0770}
\urldef\tempurl%
\url{https://www.aaai.org/ojs/index.php/ICWSM/article/view/3229}
\showURL{%
\tempurl}


\bibitem[\protect\citeauthoryear{Juneja, Rama~Subramanian, and Mitra}{Juneja
  et~al\mbox{.}}{2020}]%
        {juneja_through_2020}
\bibfield{author}{\bibinfo{person}{Prerna Juneja}, \bibinfo{person}{Deepika
  Rama~Subramanian}, {and} \bibinfo{person}{Tanushree Mitra}.}
  \bibinfo{year}{2020}\natexlab{}.
\newblock \showarticletitle{Through the {Looking} {Glass}: {Study} of
  {Transparency} in {Reddit}'s {Moderation} {Practices}}.
\newblock \bibinfo{journal}{\emph{Proceedings of the ACM on Human-Computer
  Interaction}} \bibinfo{volume}{4}, \bibinfo{number}{GROUP}
  (\bibinfo{date}{Jan.} \bibinfo{year}{2020}), \bibinfo{pages}{17:1--17:35}.
\newblock
\urldef\tempurl%
\url{https://doi.org/10.1145/3375197}
\showDOI{\tempurl}


\bibitem[\protect\citeauthoryear{Junestr{\"o}m}{Junestr{\"o}m}{2019}]%
        {junestrom_online_2019}
\bibfield{author}{\bibinfo{person}{Amalia Junestr{\"o}m}.}
  \bibinfo{year}{2019}\natexlab{}.
\newblock \showarticletitle{Online user misconduct and an evolving
  infrastructure of practices: a practice-based study of information
  infrastructure and social practices}.
\newblock \bibinfo{journal}{\emph{Information Research: an international
  electronic journal}} \bibinfo{volume}{24}, \bibinfo{number}{1}
  (\bibinfo{date}{March} \bibinfo{year}{2019}).
\newblock
\urldef\tempurl%
\url{http://informationr.net/ir/24-1/isic2018/isic1825.html}
\showURL{%
\tempurl}
\newblock
\shownote{Library Catalog: informationr.net Publisher: University of Bor{\r
  a}s.}


\bibitem[\protect\citeauthoryear{Karunakaran and Ramakrishan}{Karunakaran and
  Ramakrishan}{2019}]%
        {karunakaran_testing_2019}
\bibfield{author}{\bibinfo{person}{Sowmya Karunakaran} {and}
  \bibinfo{person}{Rashmi Ramakrishan}.} \bibinfo{year}{2019}\natexlab{}.
\newblock \showarticletitle{Testing {Stylistic} {Interventions} to {Reduce}
  {Emotional} {Impact} of {Content} {Moderation} {Workers}}.
\newblock \bibinfo{journal}{\emph{Proceedings of the AAAI Conference on Human
  Computation and Crowdsourcing}} \bibinfo{volume}{7}, \bibinfo{number}{1}
  (\bibinfo{date}{Oct.} \bibinfo{year}{2019}), \bibinfo{pages}{50--58}.
\newblock
\urldef\tempurl%
\url{https://www.aaai.org/ojs/index.php/HCOMP/article/view/5270}
\showURL{%
\tempurl}
\newblock
\shownote{Number: 1.}


\bibitem[\protect\citeauthoryear{Kayhan and Bhattacherjee}{Kayhan and
  Bhattacherjee}{2013}]%
        {kayhan_content_2013}
\bibfield{author}{\bibinfo{person}{Varol~Onur Kayhan} {and}
  \bibinfo{person}{Anol Bhattacherjee}.} \bibinfo{year}{2013}\natexlab{}.
\newblock \showarticletitle{Content {Use} from {Websites}: {Effects} of
  {Governance} {Mechanisms}}.
\newblock \bibinfo{journal}{\emph{Journal of Computer Information Systems}}
  \bibinfo{volume}{53}, \bibinfo{number}{4} (\bibinfo{date}{June}
  \bibinfo{year}{2013}), \bibinfo{pages}{68--80}.
\newblock
\showISSN{0887-4417}
\urldef\tempurl%
\url{https://doi.org/10.1080/08874417.2013.11645652}
\showDOI{\tempurl}
\newblock
\shownote{Publisher: Taylor \& Francis \_eprint:
  https://doi.org/10.1080/08874417.2013.11645652.}


\bibitem[\protect\citeauthoryear{Keegan and Fiesler}{Keegan and
  Fiesler}{2017}]%
        {keegan_evolution_2017}
\bibfield{author}{\bibinfo{person}{Brian~C Keegan} {and} \bibinfo{person}{Casey
  Fiesler}.} \bibinfo{year}{2017}\natexlab{}.
\newblock \showarticletitle{The {Evolution} and {Consequences} of {Peer}
  {Producing} {Wikipedia} ' s {Rules}}. In
  \bibinfo{booktitle}{\emph{Proceedings of the {AAAI} {International}
  {Conference} on {Web} and {Social} {Media} ({ICWSM})}}.
  \bibinfo{address}{Montreal, Quebec, Canada}.
\newblock


\bibitem[\protect\citeauthoryear{Kiene, Jiang, and Hill}{Kiene
  et~al\mbox{.}}{2019}]%
        {kiene_technological_2019}
\bibfield{author}{\bibinfo{person}{Charles Kiene},
  \bibinfo{person}{Jialun~"Aaron" Jiang}, {and} \bibinfo{person}{Benjamin~Mako
  Hill}.} \bibinfo{year}{2019}\natexlab{}.
\newblock \showarticletitle{Technological {Frames} and {User} {Innovation}:
  {Exploring} {Technological} {Change} in {Community} {Moderation} {Teams}}.
\newblock \bibinfo{journal}{\emph{Proc. ACM Hum.-Comput. Interact.}}
  \bibinfo{volume}{3}, \bibinfo{number}{CSCW} (\bibinfo{date}{Nov.}
  \bibinfo{year}{2019}), \bibinfo{pages}{Article 44}.
\newblock
\urldef\tempurl%
\url{https://doi.org/10.1145/3359146}
\showDOI{\tempurl}


\bibitem[\protect\citeauthoryear{Kiene, Monroy-Hern{\'a}ndez, and Hill}{Kiene
  et~al\mbox{.}}{2016}]%
        {kiene_surviving_2016}
\bibfield{author}{\bibinfo{person}{Charles Kiene}, \bibinfo{person}{Andr{\'e}s
  Monroy-Hern{\'a}ndez}, {and} \bibinfo{person}{Benjamin~Mako Hill}.}
  \bibinfo{year}{2016}\natexlab{}.
\newblock \showarticletitle{Surviving an "{Eternal} {September}": {How} an
  {Online} {Community} {Managed} a {Surge} of {Newcomers}}. In
  \bibinfo{booktitle}{\emph{Proceedings of the 2016 {CHI} {Conference} on
  {Human} {Factors} in {Computing} {Systems}}} \emph{(\bibinfo{series}{{CHI}
  '16})}. \bibinfo{publisher}{ACM}, \bibinfo{address}{New York, NY, USA},
  \bibinfo{pages}{1152--1156}.
\newblock
\showISBNx{978-1-4503-3362-7}
\urldef\tempurl%
\url{https://doi.org/10.1145/2858036.2858356}
\showDOI{\tempurl}


\bibitem[\protect\citeauthoryear{Kollock and Smith}{Kollock and Smith}{1996}]%
        {kollock_managing_1996}
\bibfield{author}{\bibinfo{person}{P. Kollock} {and} \bibinfo{person}{Marc
  Smith}.} \bibinfo{year}{1996}\natexlab{}.
\newblock \showarticletitle{Managing the {Virtual} {Commons}: {Cooperation} and
  {Conflict} in {Computer} {Communities}}.
\newblock \bibinfo{journal}{\emph{Pragmatics and Beyond New Series}}
  (\bibinfo{date}{Jan.} \bibinfo{year}{1996}), \bibinfo{pages}{109--128}.
\newblock


\bibitem[\protect\citeauthoryear{Kraut, Resnick, Kiesler, Burke, Chen, Kittur,
  Konstan, Ren, and Riedl}{Kraut et~al\mbox{.}}{2011}]%
        {kraut_building_2011}
\bibfield{author}{\bibinfo{person}{Robert~E. Kraut}, \bibinfo{person}{Paul
  Resnick}, \bibinfo{person}{Sara Kiesler}, \bibinfo{person}{Moira Burke},
  \bibinfo{person}{Yan Chen}, \bibinfo{person}{Niki Kittur},
  \bibinfo{person}{Joseph Konstan}, \bibinfo{person}{Yuqing Ren}, {and}
  \bibinfo{person}{John Riedl}.} \bibinfo{year}{2011}\natexlab{}.
\newblock \bibinfo{booktitle}{\emph{Building {Successful} {Online}
  {Communities}: {Evidence}-{Based} {Social} {Design}}}.
\newblock \bibinfo{publisher}{Mit Press}.
\newblock
\showISBNx{978-0-262-01657-5}
\urldef\tempurl%
\url{https://www.jstor.org/stable/j.ctt5hhgvw}
\showURL{%
\tempurl}


\bibitem[\protect\citeauthoryear{Lampe and Johnston}{Lampe and
  Johnston}{2005}]%
        {lampe_follow_2005}
\bibfield{author}{\bibinfo{person}{Cliff Lampe} {and} \bibinfo{person}{Erik
  Johnston}.} \bibinfo{year}{2005}\natexlab{}.
\newblock \showarticletitle{Follow the ({Slash}) {Dot}: {Effects} of {Feedback}
  on {New} {Members} in an {Online} {Community}}. In
  \bibinfo{booktitle}{\emph{Proceedings of the 2005 {International} {ACM}
  {SIGGROUP} {Conference} on {Supporting} {Group} {Work}}}
  \emph{(\bibinfo{series}{{GROUP} '05})}. \bibinfo{publisher}{ACM},
  \bibinfo{address}{New York, NY, USA}, \bibinfo{pages}{11--20}.
\newblock
\showISBNx{978-1-59593-223-5}
\urldef\tempurl%
\url{https://doi.org/10.1145/1099203.1099206}
\showDOI{\tempurl}


\bibitem[\protect\citeauthoryear{Lampe and Resnick}{Lampe and Resnick}{2004}]%
        {lampe_slashdot_2004}
\bibfield{author}{\bibinfo{person}{Cliff Lampe} {and} \bibinfo{person}{Paul
  Resnick}.} \bibinfo{year}{2004}\natexlab{}.
\newblock \showarticletitle{Slash({Dot}) and {Burn}: {Distributed} {Moderation}
  in a {Large} {Online} {Conversation} {Space}}. In
  \bibinfo{booktitle}{\emph{Proceedings of the {SIGCHI} {Conference} on {Human}
  {Factors} in {Computing} {Systems}}} \emph{(\bibinfo{series}{{CHI} '04})}.
  \bibinfo{publisher}{ACM}, \bibinfo{address}{New York, NY, USA},
  \bibinfo{pages}{543--550}.
\newblock
\showISBNx{978-1-58113-702-6}
\urldef\tempurl%
\url{https://doi.org/10.1145/985692.985761}
\showDOI{\tempurl}
\newblock
\shownote{event-place: Vienna, Austria.}


\bibitem[\protect\citeauthoryear{Lampe, Zube, Lee, Park, and Johnston}{Lampe
  et~al\mbox{.}}{2014}]%
        {lampe_crowdsourcing_2014}
\bibfield{author}{\bibinfo{person}{Cliff Lampe}, \bibinfo{person}{Paul Zube},
  \bibinfo{person}{Jusil Lee}, \bibinfo{person}{Chul~Hyun Park}, {and}
  \bibinfo{person}{Erik Johnston}.} \bibinfo{year}{2014}\natexlab{}.
\newblock \showarticletitle{Crowdsourcing civility: {A} natural experiment
  examining the effects of distributed moderation in online forums}.
\newblock \bibinfo{journal}{\emph{Government Information Quarterly}}
  \bibinfo{volume}{31}, \bibinfo{number}{2} (\bibinfo{date}{April}
  \bibinfo{year}{2014}), \bibinfo{pages}{317--326}.
\newblock
\showISSN{0740-624X}
\urldef\tempurl%
\url{https://doi.org/10.1016/j.giq.2013.11.005}
\showDOI{\tempurl}


\bibitem[\protect\citeauthoryear{Liao, Pan, Zhou, and Ma}{Liao
  et~al\mbox{.}}{2010}]%
        {liao_chinese_2010}
\bibfield{author}{\bibinfo{person}{Qinying Liao}, \bibinfo{person}{Yingxin
  Pan}, \bibinfo{person}{Michelle~X. Zhou}, {and} \bibinfo{person}{Fei Ma}.}
  \bibinfo{year}{2010}\natexlab{}.
\newblock \showarticletitle{Chinese online communities: balancing
  managementcontrol and individual autonomy}. In
  \bibinfo{booktitle}{\emph{Proceedings of the {SIGCHI} {Conference} on {Human}
  {Factors} in {Computing} {Systems}}} \emph{(\bibinfo{series}{{CHI} '10})}.
  \bibinfo{publisher}{Association for Computing Machinery},
  \bibinfo{address}{Atlanta, Georgia, USA}, \bibinfo{pages}{2193--2202}.
\newblock
\showISBNx{978-1-60558-929-9}
\urldef\tempurl%
\url{https://doi.org/10.1145/1753326.1753658}
\showDOI{\tempurl}


\bibitem[\protect\citeauthoryear{Liberati, Altman, Tetzlaff, Mulrow,
  G{\o}tzsche, Ioannidis, Clarke, Devereaux, Kleijnen, and Moher}{Liberati
  et~al\mbox{.}}{2009}]%
        {liberati_prisma_2009}
\bibfield{author}{\bibinfo{person}{Alessandro Liberati},
  \bibinfo{person}{Douglas~G. Altman}, \bibinfo{person}{Jennifer Tetzlaff},
  \bibinfo{person}{Cynthia Mulrow}, \bibinfo{person}{Peter~C. G{\o}tzsche},
  \bibinfo{person}{John P.~A. Ioannidis}, \bibinfo{person}{Mike Clarke},
  \bibinfo{person}{P.~J. Devereaux}, \bibinfo{person}{Jos Kleijnen}, {and}
  \bibinfo{person}{David Moher}.} \bibinfo{year}{2009}\natexlab{}.
\newblock \showarticletitle{The {PRISMA} {Statement} for {Reporting}
  {Systematic} {Reviews} and {Meta}-{Analyses} of {Studies} {That} {Evaluate}
  {Health} {Care} {Interventions}: {Explanation} and {Elaboration}}.
\newblock \bibinfo{journal}{\emph{PLoS Medicine}} \bibinfo{volume}{6},
  \bibinfo{number}{7} (\bibinfo{date}{July} \bibinfo{year}{2009}).
\newblock
\showISSN{1549-1277}
\urldef\tempurl%
\url{https://doi.org/10.1371/journal.pmed.1000100}
\showDOI{\tempurl}


\bibitem[\protect\citeauthoryear{Luo, Hsu, Park, and Hancock}{Luo
  et~al\mbox{.}}{2020}]%
        {luo_emotional_2020}
\bibfield{author}{\bibinfo{person}{Mufan Luo}, \bibinfo{person}{Tiffany~W.
  Hsu}, \bibinfo{person}{Joon~Sung Park}, {and} \bibinfo{person}{Jeffrey~T.
  Hancock}.} \bibinfo{year}{2020}\natexlab{}.
\newblock \showarticletitle{Emotional {Amplification} {During}
  {Live}-{Streaming}: {Evidence} from {Comments} {During} and {After} {News}
  {Events}}.
\newblock \bibinfo{journal}{\emph{Proceedings of the ACM on Human-Computer
  Interaction}} \bibinfo{volume}{4}, \bibinfo{number}{CSCW1}
  (\bibinfo{date}{May} \bibinfo{year}{2020}), \bibinfo{pages}{047:1--047:19}.
\newblock
\urldef\tempurl%
\url{https://doi.org/10.1145/3392853}
\showDOI{\tempurl}


\bibitem[\protect\citeauthoryear{Matias}{Matias}{2016}]%
        {matias_going_2016}
\bibfield{author}{\bibinfo{person}{J.~Nathan Matias}.}
  \bibinfo{year}{2016}\natexlab{}.
\newblock \showarticletitle{Going {Dark}: {Social} {Factors} in {Collective}
  {Action} {Against} {Platform} {Operators} in the {Reddit} {Blackout}}. In
  \bibinfo{booktitle}{\emph{Proceedings of the 2016 {CHI} {Conference} on
  {Human} {Factors} in {Computing} {Systems}}} \emph{(\bibinfo{series}{{CHI}
  '16})}. \bibinfo{publisher}{Association for Computing Machinery},
  \bibinfo{address}{San Jose, California, USA}, \bibinfo{pages}{1138--1151}.
\newblock
\showISBNx{978-1-4503-3362-7}
\urldef\tempurl%
\url{https://doi.org/10.1145/2858036.2858391}
\showDOI{\tempurl}


\bibitem[\protect\citeauthoryear{Matias}{Matias}{2019}]%
        {matias_civic_2019}
\bibfield{author}{\bibinfo{person}{J.~Nathan Matias}.}
  \bibinfo{year}{2019}\natexlab{}.
\newblock \showarticletitle{The {Civic} {Labor} of {Volunteer} {Moderators}
  {Online}:}.
\newblock \bibinfo{journal}{\emph{Social Media + Society}}
  (\bibinfo{date}{April} \bibinfo{year}{2019}).
\newblock
\urldef\tempurl%
\url{https://doi.org/10.1177/2056305119836778}
\showDOI{\tempurl}
\newblock
\shownote{Publisher: SAGE PublicationsSage UK: London, England.}


\bibitem[\protect\citeauthoryear{Matias and Mou}{Matias and Mou}{2018}]%
        {matias_civilservant_2018}
\bibfield{author}{\bibinfo{person}{J.~Nathan Matias} {and}
  \bibinfo{person}{Merry Mou}.} \bibinfo{year}{2018}\natexlab{}.
\newblock \showarticletitle{{CivilServant}: {Community}-{Led} {Experiments} in
  {Platform} {Governance}}. In \bibinfo{booktitle}{\emph{Proceedings of the
  2018 {CHI} {Conference} on {Human} {Factors} in {Computing} {Systems}}}
  \emph{(\bibinfo{series}{{CHI} '18})}. \bibinfo{publisher}{ACM},
  \bibinfo{address}{New York, NY, USA}, \bibinfo{pages}{9:1--9:13}.
\newblock
\showISBNx{978-1-4503-5620-6}
\urldef\tempurl%
\url{https://doi.org/10.1145/3173574.3173583}
\showDOI{\tempurl}


\bibitem[\protect\citeauthoryear{Medeiros}{Medeiros}{2019}]%
        {medeiros_picketing_2019}
\bibfield{author}{\bibinfo{person}{Ben Medeiros}.}
  \bibinfo{year}{2019}\natexlab{}.
\newblock \showarticletitle{Picketing the {Virtual} {Storefront}: {Content}
  {Moderation} and {Political} {Criticism} of {Businesses} on {Yelp}}.
\newblock \bibinfo{journal}{\emph{International Journal of Communication}}
  \bibinfo{volume}{13}, \bibinfo{number}{0} (\bibinfo{date}{Sept.}
  \bibinfo{year}{2019}), \bibinfo{pages}{17}.
\newblock
\showISSN{1932-8036}
\urldef\tempurl%
\url{https://ijoc.org/index.php/ijoc/article/view/10451}
\showURL{%
\tempurl}
\newblock
\shownote{Number: 0.}


\bibitem[\protect\citeauthoryear{Newell, Jurgens, Saleem, Vala, Sassine,
  Armstrong, and Ruths}{Newell et~al\mbox{.}}{2016}]%
        {newell_user_2016}
\bibfield{author}{\bibinfo{person}{Edward Newell}, \bibinfo{person}{David
  Jurgens}, \bibinfo{person}{Haji~Mohammad Saleem}, \bibinfo{person}{Hardik
  Vala}, \bibinfo{person}{Jad Sassine}, \bibinfo{person}{Caitrin Armstrong},
  {and} \bibinfo{person}{Derek Ruths}.} \bibinfo{year}{2016}\natexlab{}.
\newblock \showarticletitle{User {Migration} in {Online} {Social} {Networks}:
  {A} {Case} {Study} on {Reddit} {During} a {Period} of {Community} {Unrest}}.
  In \bibinfo{booktitle}{\emph{Tenth {International} {AAAI} {Conference} on
  {Web} and {Social} {Media}}}.
\newblock
\urldef\tempurl%
\url{https://www.aaai.org/ocs/index.php/ICWSM/ICWSM16/paper/view/13137}
\showURL{%
\tempurl}


\bibitem[\protect\citeauthoryear{Newton}{Newton}{2019}]%
        {newton_secret_2019}
\bibfield{author}{\bibinfo{person}{Casey Newton}.}
  \bibinfo{year}{2019}\natexlab{}.
\newblock \bibinfo{title}{The secret lives of {Facebook} moderators in
  {America}}.
\newblock
\newblock
\urldef\tempurl%
\url{https://www.theverge.com/2019/2/25/18229714/cognizant-facebook-content-moderator-interviews-trauma-working-conditions-arizona}
\showURL{%
\tempurl}


\bibitem[\protect\citeauthoryear{Nurik}{Nurik}{2019}]%
        {nurik_men_2019}
\bibfield{author}{\bibinfo{person}{Chloe Nurik}.}
  \bibinfo{year}{2019}\natexlab{}.
\newblock \showarticletitle{{\textquotedblleft}{Men} {Are}
  {Scum}{\textquotedblright}: {Self}-{Regulation}, {Hate} {Speech}, and
  {Gender}-{Based} {Censorship} on {Facebook}}.
\newblock \bibinfo{journal}{\emph{International Journal of Communication}}
  \bibinfo{volume}{13}, \bibinfo{number}{0} (\bibinfo{date}{June}
  \bibinfo{year}{2019}), \bibinfo{pages}{21}.
\newblock
\showISSN{1932-8036}
\urldef\tempurl%
\url{https://ijoc.org/index.php/ijoc/article/view/9608}
\showURL{%
\tempurl}
\newblock
\shownote{Number: 0.}


\bibitem[\protect\citeauthoryear{Obar and Oeldorf-Hirsch}{Obar and
  Oeldorf-Hirsch}{2020}]%
        {obar_biggest_2020}
\bibfield{author}{\bibinfo{person}{Jonathan~A. Obar} {and}
  \bibinfo{person}{Anne Oeldorf-Hirsch}.} \bibinfo{year}{2020}\natexlab{}.
\newblock \showarticletitle{The biggest lie on the {Internet}: ignoring the
  privacy policies and terms of service policies of social networking
  services}.
\newblock \bibinfo{journal}{\emph{Information, Communication \& Society}}
  \bibinfo{volume}{23}, \bibinfo{number}{1} (\bibinfo{date}{Jan.}
  \bibinfo{year}{2020}), \bibinfo{pages}{128--147}.
\newblock
\showISSN{1369-118X}
\urldef\tempurl%
\url{https://doi.org/10.1080/1369118X.2018.1486870}
\showDOI{\tempurl}
\newblock
\shownote{Publisher: Routledge \_eprint:
  https://doi.org/10.1080/1369118X.2018.1486870.}


\bibitem[\protect\citeauthoryear{Park, Sachar, Diakopoulos, and Elmqvist}{Park
  et~al\mbox{.}}{2016}]%
        {park_supporting_2016}
\bibfield{author}{\bibinfo{person}{Deokgun Park}, \bibinfo{person}{Simranjit
  Sachar}, \bibinfo{person}{Nicholas Diakopoulos}, {and}
  \bibinfo{person}{Niklas Elmqvist}.} \bibinfo{year}{2016}\natexlab{}.
\newblock \showarticletitle{Supporting {Comment} {Moderators} in {Identifying}
  {High} {Quality} {Online} {News} {Comments}}. In
  \bibinfo{booktitle}{\emph{Proceedings of the 2016 {CHI} {Conference} on
  {Human} {Factors} in {Computing} {Systems}}} \emph{(\bibinfo{series}{{CHI}
  '16})}. \bibinfo{publisher}{Association for Computing Machinery},
  \bibinfo{address}{San Jose, California, USA}, \bibinfo{pages}{1114--1125}.
\newblock
\showISBNx{978-1-4503-3362-7}
\urldef\tempurl%
\url{https://doi.org/10.1145/2858036.2858389}
\showDOI{\tempurl}


\bibitem[\protect\citeauthoryear{Pavalanathan, Han, and
  Eisenstein}{Pavalanathan et~al\mbox{.}}{2018}]%
        {pavalanathan_mind_2018}
\bibfield{author}{\bibinfo{person}{Umashanthi Pavalanathan},
  \bibinfo{person}{Xiaochuang Han}, {and} \bibinfo{person}{Jacob Eisenstein}.}
  \bibinfo{year}{2018}\natexlab{}.
\newblock \showarticletitle{Mind {Your} {POV}: {Convergence} of {Articles} and
  {Editors} {Towards} {Wikipedia}'s {Neutrality} {Norm}}.
\newblock \bibinfo{journal}{\emph{Proceedings of the ACM on Human-Computer
  Interaction}} \bibinfo{volume}{2}, \bibinfo{number}{CSCW}
  (\bibinfo{date}{Nov.} \bibinfo{year}{2018}), \bibinfo{pages}{137:1--137:23}.
\newblock
\urldef\tempurl%
\url{https://doi.org/10.1145/3274406}
\showDOI{\tempurl}


\bibitem[\protect\citeauthoryear{Pellicone and Ahn}{Pellicone and Ahn}{2017}]%
        {pellicone_game_2017}
\bibfield{author}{\bibinfo{person}{Anthony~J. Pellicone} {and}
  \bibinfo{person}{June Ahn}.} \bibinfo{year}{2017}\natexlab{}.
\newblock \showarticletitle{The {Game} of {Performing} {Play}: {Understanding}
  {Streaming} as {Cultural} {Production}}. In
  \bibinfo{booktitle}{\emph{Proceedings of the 2017 {CHI} {Conference} on
  {Human} {Factors} in {Computing} {Systems}}} \emph{(\bibinfo{series}{{CHI}
  '17})}. \bibinfo{publisher}{Association for Computing Machinery},
  \bibinfo{address}{Denver, Colorado, USA}, \bibinfo{pages}{4863--4874}.
\newblock
\showISBNx{978-1-4503-4655-9}
\urldef\tempurl%
\url{https://doi.org/10.1145/3025453.3025854}
\showDOI{\tempurl}


\bibitem[\protect\citeauthoryear{Petri{\v c} and Petrov{\v c}i{\v c}}{Petri{\v
  c} and Petrov{\v c}i{\v c}}{2014}]%
        {petric_elements_2014}
\bibfield{author}{\bibinfo{person}{Gregor Petri{\v c}} {and}
  \bibinfo{person}{Andra{\v z} Petrov{\v c}i{\v c}}.}
  \bibinfo{year}{2014}\natexlab{}.
\newblock \showarticletitle{Elements of the management of norms and their
  effects on the sense of virtual community}.
\newblock \bibinfo{journal}{\emph{Online Information Review}}
  \bibinfo{volume}{38}, \bibinfo{number}{3} (\bibinfo{date}{Jan.}
  \bibinfo{year}{2014}), \bibinfo{pages}{436--454}.
\newblock
\showISSN{1468-4527}
\urldef\tempurl%
\url{https://doi.org/10.1108/OIR-04-2013-0083}
\showDOI{\tempurl}
\newblock
\shownote{Publisher: Emerald Group Publishing Limited.}


\bibitem[\protect\citeauthoryear{Phadke and Mitra}{Phadke and Mitra}{2020}]%
        {phadke_many_2020}
\bibfield{author}{\bibinfo{person}{Shruti Phadke} {and}
  \bibinfo{person}{Tanushree Mitra}.} \bibinfo{year}{2020}\natexlab{}.
\newblock \showarticletitle{Many {Faced} {Hate}: {A} {Cross} {Platform} {Study}
  of {Content} {Framing} and {Information} {Sharing} by {Online} {Hate}
  {Groups}}. In \bibinfo{booktitle}{\emph{Proceedings of the 2020 {CHI}
  {Conference} on {Human} {Factors} in {Computing} {Systems}}}
  \emph{(\bibinfo{series}{{CHI} '20})}. \bibinfo{publisher}{Association for
  Computing Machinery}, \bibinfo{address}{Honolulu, HI, USA},
  \bibinfo{pages}{1--13}.
\newblock
\showISBNx{978-1-4503-6708-0}
\urldef\tempurl%
\url{https://doi.org/10.1145/3313831.3376456}
\showDOI{\tempurl}


\bibitem[\protect\citeauthoryear{Potts, Small, and Trice}{Potts
  et~al\mbox{.}}{2019}]%
        {potts_boycotting_2019}
\bibfield{author}{\bibinfo{person}{Liza Potts}, \bibinfo{person}{Rebekah
  Small}, {and} \bibinfo{person}{Michael Trice}.}
  \bibinfo{year}{2019}\natexlab{}.
\newblock \showarticletitle{Boycotting the {Knowledge} {Makers}: {How} {Reddit}
  {Demonstrates} the {Rise} of {Media} {Blacklists} and {Source} {Rejection} in
  {Online} {Communities}}.
\newblock \bibinfo{journal}{\emph{IEEE Transactions on Professional
  Communication}} \bibinfo{volume}{62}, \bibinfo{number}{4}
  (\bibinfo{date}{Dec.} \bibinfo{year}{2019}), \bibinfo{pages}{351--363}.
\newblock
\showISSN{1558-1500}
\urldef\tempurl%
\url{https://doi.org/10.1109/TPC.2019.2946942}
\showDOI{\tempurl}
\newblock
\shownote{Conference Name: IEEE Transactions on Professional Communication.}


\bibitem[\protect\citeauthoryear{Proch{\'a}zka}{Proch{\'a}zka}{2019}]%
        {prochazka_making_2019}
\bibfield{author}{\bibinfo{person}{Ond{\v r}ej Proch{\'a}zka}.}
  \bibinfo{year}{2019}\natexlab{}.
\newblock \showarticletitle{Making {Sense} of {Facebook}{\textquoteright}s
  {Content} {Moderation}: {A} {Posthumanist} {Perspective} on {Communicative}
  {Competence} and {Internet} {Memes}}.
\newblock \bibinfo{journal}{\emph{Signs and Society}} \bibinfo{volume}{7},
  \bibinfo{number}{3} (\bibinfo{date}{Sept.} \bibinfo{year}{2019}),
  \bibinfo{pages}{362--397}.
\newblock
\showISSN{2326-4489}
\urldef\tempurl%
\url{https://doi.org/10.1086/704763}
\showDOI{\tempurl}
\newblock
\shownote{Publisher: The University of Chicago Press.}


\bibitem[\protect\citeauthoryear{Proferes, Jones, Gilbert, Fiesler, and
  Zimmer}{Proferes et~al\mbox{.}}{2021}]%
        {proferes_studying_2021}
\bibfield{author}{\bibinfo{person}{Nicholas Proferes}, \bibinfo{person}{Naiyan
  Jones}, \bibinfo{person}{Sarah Gilbert}, \bibinfo{person}{Casey Fiesler},
  {and} \bibinfo{person}{Michael Zimmer}.} \bibinfo{year}{2021}\natexlab{}.
\newblock \showarticletitle{Studying Reddit: A Systematic Overview of
  Disciplines, Approaches, Methods, and Ethics}.
\newblock \bibinfo{journal}{\emph{Social Media + Society}} \bibinfo{volume}{7},
  \bibinfo{number}{2} (\bibinfo{year}{2021}),
  \bibinfo{pages}{20563051211019004}.
\newblock
\urldef\tempurl%
\url{https://doi.org/10.1177/20563051211019004}
\showDOI{\tempurl}


\bibitem[\protect\citeauthoryear{Rajadesingan, Resnick, and Budak}{Rajadesingan
  et~al\mbox{.}}{2020}]%
        {rajadesingan_quick_2020}
\bibfield{author}{\bibinfo{person}{Ashwin Rajadesingan}, \bibinfo{person}{Paul
  Resnick}, {and} \bibinfo{person}{Ceren Budak}.}
  \bibinfo{year}{2020}\natexlab{}.
\newblock \showarticletitle{Quick, {Community}-{Specific} {Learning}: {How}
  {Distinctive} {Toxicity} {Norms} {Are} {Maintained} in {Political}
  {Subreddits}}.
\newblock \bibinfo{journal}{\emph{Proceedings of the International AAAI
  Conference on Web and Social Media}}  \bibinfo{volume}{14}
  (\bibinfo{date}{May} \bibinfo{year}{2020}), \bibinfo{pages}{557--568}.
\newblock
\showISSN{2334-0770}
\urldef\tempurl%
\url{https://aaai.org/ojs/index.php/ICWSM/article/view/7323}
\showURL{%
\tempurl}


\bibitem[\protect\citeauthoryear{Redmiles, Bodford, and Blackwell}{Redmiles
  et~al\mbox{.}}{2019}]%
        {redmiles_i_2019}
\bibfield{author}{\bibinfo{person}{Elissa~M. Redmiles},
  \bibinfo{person}{Jessica Bodford}, {and} \bibinfo{person}{Lindsay
  Blackwell}.} \bibinfo{year}{2019}\natexlab{}.
\newblock \showarticletitle{{\textquotedblleft}{I} {Just} {Want} to {Feel}
  {Safe}{\textquotedblright}: {A} {Diary} {Study} of {Safety} {Perceptions} on
  {Social} {Media}}.
\newblock \bibinfo{journal}{\emph{Proceedings of the International AAAI
  Conference on Web and Social Media}}  \bibinfo{volume}{13}
  (\bibinfo{date}{July} \bibinfo{year}{2019}), \bibinfo{pages}{405--416}.
\newblock
\showISSN{2334-0770}
\urldef\tempurl%
\url{https://www.aaai.org/ojs/index.php/ICWSM/article/view/3356}
\showURL{%
\tempurl}


\bibitem[\protect\citeauthoryear{Reid}{Reid}{1999}]%
        {reid_communities_1999}
\bibfield{author}{\bibinfo{person}{Reid}.} \bibinfo{year}{1999}\natexlab{}.
\newblock \bibinfo{booktitle}{\emph{Communities in {Cyberspace}}}.
\newblock \bibinfo{publisher}{Routledge}, \bibinfo{address}{London \&amp; New
  York}.
\newblock
\newblock
\shownote{Pages: 107-133.}


\bibitem[\protect\citeauthoryear{Riedl, Masullo, and Whipple}{Riedl
  et~al\mbox{.}}{2020}]%
        {riedl_downsides_2020}
\bibfield{author}{\bibinfo{person}{Martin~J. Riedl}, \bibinfo{person}{Gina~M.
  Masullo}, {and} \bibinfo{person}{Kelsey~N. Whipple}.}
  \bibinfo{year}{2020}\natexlab{}.
\newblock \showarticletitle{The downsides of digital labor: {Exploring} the
  toll incivility takes on online comment moderators}.
\newblock \bibinfo{journal}{\emph{Computers in Human Behavior}}
  \bibinfo{volume}{107} (\bibinfo{date}{June} \bibinfo{year}{2020}),
  \bibinfo{pages}{106262}.
\newblock
\showISSN{0747-5632}
\urldef\tempurl%
\url{https://doi.org/10.1016/j.chb.2020.106262}
\showDOI{\tempurl}


\bibitem[\protect\citeauthoryear{Ruckenstein and Turunen}{Ruckenstein and
  Turunen}{2020}]%
        {ruckenstein_re-humanizing_2020}
\bibfield{author}{\bibinfo{person}{Minna Ruckenstein} {and}
  \bibinfo{person}{Linda Lisa~Maria Turunen}.} \bibinfo{year}{2020}\natexlab{}.
\newblock \showarticletitle{Re-humanizing the platform: {Content} moderators
  and the logic of care}.
\newblock \bibinfo{journal}{\emph{New Media \& Society}} \bibinfo{volume}{22},
  \bibinfo{number}{6} (\bibinfo{date}{June} \bibinfo{year}{2020}),
  \bibinfo{pages}{1026--1042}.
\newblock
\showISSN{1461-4448}
\urldef\tempurl%
\url{https://doi.org/10.1177/1461444819875990}
\showDOI{\tempurl}
\newblock
\shownote{Publisher: SAGE Publications.}


\bibitem[\protect\citeauthoryear{Sarkar, Wohn, Lampe, and DeMaagd}{Sarkar
  et~al\mbox{.}}{2012}]%
        {sarkar_quantitative_2012}
\bibfield{author}{\bibinfo{person}{Chandan Sarkar}, \bibinfo{person}{Donghee
  Wohn}, \bibinfo{person}{Cliff Lampe}, {and} \bibinfo{person}{Kurt DeMaagd}.}
  \bibinfo{year}{2012}\natexlab{}.
\newblock \showarticletitle{A quantitative explanation of governance in an
  online peer-production community}. In \bibinfo{booktitle}{\emph{Proceedings
  of the {SIGCHI} {Conference} on {Human} {Factors} in {Computing} {Systems}}}
  \emph{(\bibinfo{series}{{CHI} '12})}. \bibinfo{publisher}{Association for
  Computing Machinery}, \bibinfo{address}{Austin, Texas, USA},
  \bibinfo{pages}{2939--2942}.
\newblock
\showISBNx{978-1-4503-1015-4}
\urldef\tempurl%
\url{https://doi.org/10.1145/2207676.2208701}
\showDOI{\tempurl}


\bibitem[\protect\citeauthoryear{Schoenebeck, Haimson, and
  Nakamura}{Schoenebeck et~al\mbox{.}}{2020}]%
        {schoenebeck_drawing_2020}
\bibfield{author}{\bibinfo{person}{Sarita Schoenebeck},
  \bibinfo{person}{Oliver~L Haimson}, {and} \bibinfo{person}{Lisa Nakamura}.}
  \bibinfo{year}{2020}\natexlab{}.
\newblock \showarticletitle{Drawing from justice theories to support targets of
  online harassment}.
\newblock \bibinfo{journal}{\emph{New Media \& Society}} (\bibinfo{date}{March}
  \bibinfo{year}{2020}), \bibinfo{pages}{1461444820913122}.
\newblock
\showISSN{1461-4448}
\urldef\tempurl%
\url{https://doi.org/10.1177/1461444820913122}
\showDOI{\tempurl}
\newblock
\shownote{Publisher: SAGE Publications.}


\bibitem[\protect\citeauthoryear{Seering, Kaufman, and Chancellor}{Seering
  et~al\mbox{.}}{2020}]%
        {seering_metaphors_2020}
\bibfield{author}{\bibinfo{person}{Joseph Seering}, \bibinfo{person}{Geoff
  Kaufman}, {and} \bibinfo{person}{Stevie Chancellor}.}
  \bibinfo{year}{2020}\natexlab{}.
\newblock \showarticletitle{Metaphors in moderation}.
\newblock \bibinfo{journal}{\emph{New Media \& Society}} (\bibinfo{date}{Oct.}
  \bibinfo{year}{2020}), \bibinfo{pages}{1461444820964968}.
\newblock
\showISSN{1461-4448}
\urldef\tempurl%
\url{https://doi.org/10.1177/1461444820964968}
\showDOI{\tempurl}
\newblock
\shownote{Publisher: SAGE Publications.}


\bibitem[\protect\citeauthoryear{Seering, Kraut, and Dabbish}{Seering
  et~al\mbox{.}}{2017}]%
        {seering_shaping_2017}
\bibfield{author}{\bibinfo{person}{Joseph Seering}, \bibinfo{person}{Robert
  Kraut}, {and} \bibinfo{person}{Laura Dabbish}.}
  \bibinfo{year}{2017}\natexlab{}.
\newblock \showarticletitle{Shaping pro and anti-social behavior on twitch
  through moderation and example-setting}. In
  \bibinfo{booktitle}{\emph{Proceedings of the 2017 {ACM} {Conference} on
  {Computer} {Supported} {Cooperative} {Work} and {Social} {Computing}}}
  \emph{(\bibinfo{series}{{CSCW} '17})}. \bibinfo{publisher}{ACM},
  \bibinfo{address}{New York, NY, USA}, \bibinfo{pages}{111--125}.
\newblock
\showISBNx{978-1-4503-4335-0}
\urldef\tempurl%
\url{https://doi.org/10.1145/2998181.2998277}
\showDOI{\tempurl}


\bibitem[\protect\citeauthoryear{Seering, Ng, Yao, and Kaufman}{Seering
  et~al\mbox{.}}{2018}]%
        {seering_applications_2018}
\bibfield{author}{\bibinfo{person}{Joseph Seering}, \bibinfo{person}{Felicia
  Ng}, \bibinfo{person}{Zheng Yao}, {and} \bibinfo{person}{Geoff Kaufman}.}
  \bibinfo{year}{2018}\natexlab{}.
\newblock \showarticletitle{Applications of {Social} {Identity} {Theory} to
  {Research} and {Design} in {Computer}-{Supported} {Cooperative} {Work}}.
\newblock \bibinfo{journal}{\emph{Proc. ACM Hum.-Comput. Interact.}}
  \bibinfo{volume}{2}, \bibinfo{number}{CSCW} (\bibinfo{date}{Nov.}
  \bibinfo{year}{2018}), \bibinfo{pages}{201:1--201:34}.
\newblock
\showISSN{2573-0142}
\urldef\tempurl%
\url{https://doi.org/10.1145/3274771}
\showDOI{\tempurl}


\bibitem[\protect\citeauthoryear{Seering, Wang, Yoon, and Kaufman}{Seering
  et~al\mbox{.}}{2019}]%
        {seering_moderator_2019}
\bibfield{author}{\bibinfo{person}{Joseph Seering}, \bibinfo{person}{Tony
  Wang}, \bibinfo{person}{Jina Yoon}, {and} \bibinfo{person}{Geoff Kaufman}.}
  \bibinfo{year}{2019}\natexlab{}.
\newblock \showarticletitle{Moderator engagement and community development in
  the age of algorithms}.
\newblock \bibinfo{journal}{\emph{New Media \& Society}} (\bibinfo{date}{Jan.}
  \bibinfo{year}{2019}), \bibinfo{pages}{1461444818821316}.
\newblock
\showISSN{1461-4448}
\urldef\tempurl%
\url{https://doi.org/10.1177/1461444818821316}
\showDOI{\tempurl}


\bibitem[\protect\citeauthoryear{Shen and Rose}{Shen and Rose}{2019}]%
        {shen_discourse_2019}
\bibfield{author}{\bibinfo{person}{Qinlan Shen} {and} \bibinfo{person}{Carolyn
  Rose}.} \bibinfo{year}{2019}\natexlab{}.
\newblock \showarticletitle{The {Discourse} of {Online} {Content} {Moderation}:
  {Investigating} {Polarized} {User} {Responses} to {Changes} in {Reddit}'s
  {Quarantine} {Policy}}. In \bibinfo{booktitle}{\emph{Proceedings of the
  {Third} {Workshop} on {Abusive} {Language} {Online}}}.
  \bibinfo{publisher}{Association for Computational Linguistics},
  \bibinfo{address}{Florence, Italy}, \bibinfo{pages}{58--69}.
\newblock
\urldef\tempurl%
\url{https://doi.org/10.18653/v1/W19-3507}
\showDOI{\tempurl}


\bibitem[\protect\citeauthoryear{Shieber}{Shieber}{2020}]%
        {shieber_zuckerberg_2020}
\bibfield{author}{\bibinfo{person}{Jonathan Shieber}.}
  \bibinfo{year}{2020}\natexlab{}.
\newblock \bibinfo{title}{Zuckerberg explains why {Facebook}
  won{\textquoteright}t take action on {Trump}{\textquoteright}s recent posts}.
\newblock
\newblock
\urldef\tempurl%
\url{https://social.techcrunch.com/2020/05/29/zuckerberg-explains-why-facebook-wont-take-action-on-trumps-recent-posts/}
\showURL{%
\tempurl}


\bibitem[\protect\citeauthoryear{Skousen, Safadi, Young, Karahanna, Safadi, and
  Chebib}{Skousen et~al\mbox{.}}{2020}]%
        {skousen_successful_2020}
\bibfield{author}{\bibinfo{person}{Tanner Skousen}, \bibinfo{person}{Hani
  Safadi}, \bibinfo{person}{Colleen Young}, \bibinfo{person}{Elena Karahanna},
  \bibinfo{person}{Sami Safadi}, {and} \bibinfo{person}{Fouad Chebib}.}
  \bibinfo{year}{2020}\natexlab{}.
\newblock \showarticletitle{Successful {Moderation} in {Online} {Patient}
  {Communities}: {Inductive} {Case} {Study}}.
\newblock \bibinfo{journal}{\emph{Journal of Medical Internet Research}}
  \bibinfo{volume}{22}, \bibinfo{number}{3} (\bibinfo{year}{2020}),
  \bibinfo{pages}{e15983}.
\newblock
\urldef\tempurl%
\url{https://doi.org/10.2196/15983}
\showDOI{\tempurl}
\newblock
\shownote{Company: Journal of Medical Internet Research Distributor: Journal of
  Medical Internet Research Institution: Journal of Medical Internet Research
  Label: Journal of Medical Internet Research Publisher: JMIR Publications
  Inc., Toronto, Canada.}


\bibitem[\protect\citeauthoryear{Squirrell}{Squirrell}{2019}]%
        {squirrell_platform_2019}
\bibfield{author}{\bibinfo{person}{Tim Squirrell}.}
  \bibinfo{year}{2019}\natexlab{}.
\newblock \showarticletitle{Platform dialectics: {The} relationships between
  volunteer moderators and end users on reddit}.
\newblock \bibinfo{journal}{\emph{New Media \& Society}} \bibinfo{volume}{21},
  \bibinfo{number}{9} (\bibinfo{date}{Sept.} \bibinfo{year}{2019}),
  \bibinfo{pages}{1910--1927}.
\newblock
\showISSN{1461-4448}
\urldef\tempurl%
\url{https://doi.org/10.1177/1461444819834317}
\showDOI{\tempurl}
\newblock
\shownote{Publisher: SAGE Publications.}


\bibitem[\protect\citeauthoryear{Srinivasan, Danescu-Niculescu-Mizil, Lee, and
  Tan}{Srinivasan et~al\mbox{.}}{2019}]%
        {srinivasan_content_2019}
\bibfield{author}{\bibinfo{person}{Kumar~Bhargav Srinivasan},
  \bibinfo{person}{Cristian Danescu-Niculescu-Mizil}, \bibinfo{person}{Lillian
  Lee}, {and} \bibinfo{person}{Chenhao Tan}.} \bibinfo{year}{2019}\natexlab{}.
\newblock \showarticletitle{Content {Removal} as a {Moderation} {Strategy}:
  {Compliance} and {Other} {Outcomes} in the {ChangeMyView} {Community}}.
\newblock \bibinfo{journal}{\emph{Proceedings of the ACM on Human-Computer
  Interaction}} \bibinfo{volume}{3}, \bibinfo{number}{CSCW}
  (\bibinfo{date}{Nov.} \bibinfo{year}{2019}), \bibinfo{pages}{163:1--163:21}.
\newblock
\urldef\tempurl%
\url{https://doi.org/10.1145/3359265}
\showDOI{\tempurl}


\bibitem[\protect\citeauthoryear{Suzor, West, Quodling, and York}{Suzor
  et~al\mbox{.}}{2019}]%
        {suzor_what_2019}
\bibfield{author}{\bibinfo{person}{Nicolas~P. Suzor},
  \bibinfo{person}{Sarah~Myers West}, \bibinfo{person}{Andrew Quodling}, {and}
  \bibinfo{person}{Jillian York}.} \bibinfo{year}{2019}\natexlab{}.
\newblock \showarticletitle{What {Do} {We} {Mean} {When} {We} {Talk} {About}
  {Transparency}? {Toward} {Meaningful} {Transparency} in {Commercial}
  {Content} {Moderation}}.
\newblock \bibinfo{journal}{\emph{International Journal of Communication}}
  \bibinfo{volume}{13}, \bibinfo{number}{0} (\bibinfo{date}{March}
  \bibinfo{year}{2019}), \bibinfo{pages}{18}.
\newblock
\showISSN{1932-8036}
\urldef\tempurl%
\url{https://ijoc.org/index.php/ijoc/article/view/9736}
\showURL{%
\tempurl}
\newblock
\shownote{Number: 0.}


\bibitem[\protect\citeauthoryear{Tyler, Katsaros, Meares, and Venkatesh}{Tyler
  et~al\mbox{.}}{2019}]%
        {tyler_social_2019}
\bibfield{author}{\bibinfo{person}{Tom Tyler}, \bibinfo{person}{Matt Katsaros},
  \bibinfo{person}{Tracey Meares}, {and} \bibinfo{person}{Sudhir Venkatesh}.}
  \bibinfo{year}{2019}\natexlab{}.
\newblock \showarticletitle{Social media governance: can social media companies
  motivate voluntary rule following behavior among their users?}
\newblock \bibinfo{journal}{\emph{Journal of Experimental Criminology}}
  (\bibinfo{date}{Dec.} \bibinfo{year}{2019}).
\newblock
\showISSN{1572-8315}
\urldef\tempurl%
\url{https://doi.org/10.1007/s11292-019-09392-z}
\showDOI{\tempurl}


\bibitem[\protect\citeauthoryear{Vaccaro, Sandvig, and Karahalios}{Vaccaro
  et~al\mbox{.}}{2020}]%
        {vaccaro_at_2020}
\bibfield{author}{\bibinfo{person}{Kristen Vaccaro}, \bibinfo{person}{Christian
  Sandvig}, {and} \bibinfo{person}{Karrie Karahalios}.}
  \bibinfo{year}{2020}\natexlab{}.
\newblock \showarticletitle{"{At} the {End} of the {Day} {Facebook} {Does}
  {What} {ItWants}": {How} {Users} {Experience} {Contesting} {Algorithmic}
  {Content} {Moderation}}.
\newblock \bibinfo{journal}{\emph{Proceedings of the ACM on Human-Computer
  Interaction}} \bibinfo{volume}{4}, \bibinfo{number}{CSCW2}
  (\bibinfo{date}{Oct.} \bibinfo{year}{2020}), \bibinfo{pages}{167:1--167:22}.
\newblock
\urldef\tempurl%
\url{https://doi.org/10.1145/3415238}
\showDOI{\tempurl}


\bibitem[\protect\citeauthoryear{Vashistha, Cutrell, Borriello, and
  Thies}{Vashistha et~al\mbox{.}}{2015}]%
        {vashistha_sangeet_2015}
\bibfield{author}{\bibinfo{person}{Aditya Vashistha}, \bibinfo{person}{Edward
  Cutrell}, \bibinfo{person}{Gaetano Borriello}, {and} \bibinfo{person}{William
  Thies}.} \bibinfo{year}{2015}\natexlab{}.
\newblock \showarticletitle{Sangeet {Swara}: {A} {Community}-{Moderated}
  {Voice} {Forum} in {Rural} {India}}. In \bibinfo{booktitle}{\emph{Proceedings
  of the 33rd {Annual} {ACM} {Conference} on {Human} {Factors} in {Computing}
  {Systems}}} \emph{(\bibinfo{series}{{CHI} '15})}.
  \bibinfo{publisher}{Association for Computing Machinery},
  \bibinfo{address}{New York, NY, USA}, \bibinfo{pages}{417--426}.
\newblock
\showISBNx{978-1-4503-3145-6}
\urldef\tempurl%
\url{https://doi.org/10.1145/2702123.2702191}
\showDOI{\tempurl}


\bibitem[\protect\citeauthoryear{Wang, Wang, Wang, Nika, Zheng, and Zhao}{Wang
  et~al\mbox{.}}{2014}]%
        {wang_whispers_2014}
\bibfield{author}{\bibinfo{person}{Gang Wang}, \bibinfo{person}{Bolun Wang},
  \bibinfo{person}{Tianyi Wang}, \bibinfo{person}{Ana Nika},
  \bibinfo{person}{Haitao Zheng}, {and} \bibinfo{person}{Ben~Y. Zhao}.}
  \bibinfo{year}{2014}\natexlab{}.
\newblock \showarticletitle{Whispers in the dark: analysis of an anonymous
  social network}. In \bibinfo{booktitle}{\emph{Proceedings of the 2014
  {Conference} on {Internet} {Measurement} {Conference}}}
  \emph{(\bibinfo{series}{{IMC} '14})}. \bibinfo{publisher}{Association for
  Computing Machinery}, \bibinfo{address}{Vancouver, BC, Canada},
  \bibinfo{pages}{137--150}.
\newblock
\showISBNx{978-1-4503-3213-2}
\urldef\tempurl%
\url{https://doi.org/10.1145/2663716.2663728}
\showDOI{\tempurl}


\bibitem[\protect\citeauthoryear{West}{West}{2018}]%
        {west_censored_2018}
\bibfield{author}{\bibinfo{person}{Sarah~Myers West}.}
  \bibinfo{year}{2018}\natexlab{}.
\newblock \showarticletitle{Censored, suspended, shadowbanned: {User}
  interpretations of content moderation on social media platforms}.
\newblock \bibinfo{journal}{\emph{New Media \& Society}} \bibinfo{volume}{20},
  \bibinfo{number}{11} (\bibinfo{date}{Nov.} \bibinfo{year}{2018}),
  \bibinfo{pages}{4366--4383}.
\newblock
\showISSN{1461-4448}
\urldef\tempurl%
\url{https://doi.org/10.1177/1461444818773059}
\showDOI{\tempurl}
\newblock
\shownote{Publisher: SAGE Publications.}


\bibitem[\protect\citeauthoryear{Witt, Suzor, and Huggins}{Witt
  et~al\mbox{.}}{2019}]%
        {witt_rule_2019}
\bibfield{author}{\bibinfo{person}{Alice Elizabeth~Amelia Witt},
  \bibinfo{person}{Nicolas Suzor}, {and} \bibinfo{person}{Anna Huggins}.}
  \bibinfo{year}{2019}\natexlab{}.
\newblock \showarticletitle{The rule of law on {Instagram}: {An} evaluation of
  the moderation of images depicting women's bodies}.
\newblock \bibinfo{journal}{\emph{The University of New South Wales law
  journal}} \bibinfo{volume}{42}, \bibinfo{number}{2} (\bibinfo{year}{2019}),
  \bibinfo{pages}{557--596}.
\newblock
\showISSN{0313-0096}
\urldef\tempurl%
\url{http://www.unswlawjournal.unsw.edu.au/article/the-rule-of-law-on-instagram-an-evaluation-of-the-moderation-of-images-depicting-womens-bodies/}
\showURL{%
\tempurl}
\newblock
\shownote{Number: 2 Publisher: Law School, University of New South Wales.}


\bibitem[\protect\citeauthoryear{Wohn}{Wohn}{2019}]%
        {wohn_volunteer_2019}
\bibfield{author}{\bibinfo{person}{Donghee~Yvette Wohn}.}
  \bibinfo{year}{2019}\natexlab{}.
\newblock \showarticletitle{Volunteer {Moderators} in {Twitch} {Micro}
  {Communities}: {How} {They} {Get} {Involved}, the {Roles} {They} {Play}, and
  the {Emotional} {Labor} {They} {Experience}}. In
  \bibinfo{booktitle}{\emph{Proceedings of the 2019 {CHI} {Conference} on
  {Human} {Factors} in {Computing} {Systems}}} \emph{(\bibinfo{series}{{CHI}
  '19})}. \bibinfo{publisher}{Association for Computing Machinery},
  \bibinfo{address}{Glasgow, Scotland Uk}, \bibinfo{pages}{1--13}.
\newblock
\showISBNx{978-1-4503-5970-2}
\urldef\tempurl%
\url{https://doi.org/10.1145/3290605.3300390}
\showDOI{\tempurl}


\bibitem[\protect\citeauthoryear{Yang, Kraut, Smith, Mayfield, and
  Jurafsky}{Yang et~al\mbox{.}}{2019}]%
        {yang_seekers_2019}
\bibfield{author}{\bibinfo{person}{Diyi Yang}, \bibinfo{person}{Robert~E.
  Kraut}, \bibinfo{person}{Tenbroeck Smith}, \bibinfo{person}{Elijah Mayfield},
  {and} \bibinfo{person}{Dan Jurafsky}.} \bibinfo{year}{2019}\natexlab{}.
\newblock \showarticletitle{Seekers, {Providers}, {Welcomers}, and
  {Storytellers}: {Modeling} {Social} {Roles} in {Online} {Health}
  {Communities}}. In \bibinfo{booktitle}{\emph{Proceedings of the 2019 {CHI}
  {Conference} on {Human} {Factors} in {Computing} {Systems}}}
  \emph{(\bibinfo{series}{{CHI} '19})}. \bibinfo{publisher}{Association for
  Computing Machinery}, \bibinfo{address}{New York, NY, USA},
  \bibinfo{pages}{1--14}.
\newblock
\showISBNx{978-1-4503-5970-2}
\urldef\tempurl%
\url{https://doi.org/10.1145/3290605.3300574}
\showDOI{\tempurl}


\bibitem[\protect\citeauthoryear{Zeng, Chan, and Fu}{Zeng
  et~al\mbox{.}}{2017}]%
        {zeng_how_2017}
\bibfield{author}{\bibinfo{person}{Jing Zeng}, \bibinfo{person}{Chung-hong
  Chan}, {and} \bibinfo{person}{King-wa Fu}.} \bibinfo{year}{2017}\natexlab{}.
\newblock \showarticletitle{How {Social} {Media} {Construct}
  {\textquotedblleft}{Truth}{\textquotedblright} {Around} {Crisis} {Events}:
  {Weibo}'s {Rumor} {Management} {Strategies} {After} the 2015 {Tianjin}
  {Blasts}}.
\newblock \bibinfo{journal}{\emph{Policy \& Internet}} \bibinfo{volume}{9},
  \bibinfo{number}{3} (\bibinfo{year}{2017}), \bibinfo{pages}{297--320}.
\newblock
\showISSN{1944-2866}
\urldef\tempurl%
\url{https://doi.org/10.1002/poi3.155}
\showDOI{\tempurl}
\newblock
\shownote{\_eprint: https://onlinelibrary.wiley.com/doi/pdf/10.1002/poi3.155.}


\end{thebibliography}
